\documentclass[journal]{elsarticle}

\usepackage{lineno,hyperref}
\modulolinenumbers[5]

\usepackage{amsmath, amssymb, mathbbol}

\begin{document}

\begin{frontmatter}

\title{Compressibility Analysis of Asymptotically Mean Stationary Processes}

\author{Jorge F. Silva$^*$}
\address{Information and Decision System Group}
\address{Department of Electrical Engineering, University of Chile}
\address{Av. Tupper 2007, Room 508, Santiago, Chile}


\ead{josilva@ing.uchile.cl}


\begin{abstract}
This work provides new results for the analysis of random sequences in terms of $\ell_p$-compressibility. The results characterize the degree in which a random sequence can be approximated by its best $k$-sparse version under different rates of significant coefficients (compressibility analysis). In particular, the notion of strong $\ell_p$-characterization is introduced to denote a random sequence that has a well-defined asymptotic limit (sample-wise) of its best $k$-term approximation error when a fixed rate of significant coefficients  is considered (fixed-rate analysis). The main theorem of this work shows that the rich family of asymptotically mean stationary (AMS) processes has a strong $\ell_p$-characterization. 
Furthermore, we present results that characterize and analyze the $\ell_p$-approximation error function for this family of processes. 
Adding ergodicity in the analysis of AMS processes, we introduce a theorem demonstrating that the approximation error function is constant and determined in closed-form by the stationary mean of the process. Our results and analyses contribute to the theory and understanding of discrete-time sparse processes and, on the technical side,  confirm how instrumental  the  point-wise ergodic theorem is to determine the compressibility expression of discrete-time processes even when stationarity and ergodicity assumptions are relaxed.
\end{abstract}


 \newtheorem{corollary}{\bf Corollary}
\newtheorem{theorem}{\bf Theorem}
\newtheorem{lemma}{\bf Lemma}
\newtheorem{proposition}{\bf Proposition}
\newtheorem{definition}{\bf Definition}
\newtheorem{remark}{\bf Remark}
\newtheorem{example}{\bf Example}

\end{frontmatter}


\section{Introduction}
\label{sec_intro}
Quantifying sparsity  and compressibility for random sequences has been a topic of
active research  largely motivated  by the results on sparse signal recovery and compressed 
sensing (CS) \cite{amini_2011,gribonval_2012,silva_2015,cevher_2008,amini_2014,unser_2014,unser_2014b}. Sparsity and compressibility can be understood,  in general,  
as the degree to which one can represent a random sequence (perfectly and loosely, respectively)  
by its best $k$-sparse version in the non-trivial regime when $k$ (the number of significant coefficients) is smaller than the signal or ambient dimension.  Various forms of compressibility for a random sequence have been used in signal processing problems, for instance in regression \cite{gribonval_2011}, signal reconstruction (in the classical random Gaussian linear measuring setting used in CS) \cite{gribonval_2012,silva_2015}, and inference-decision \cite{amini_2013,prasad_2013}. 
Compressibility for random sequences has also been used to analyze continuous-time processes \cite{unser_2014b,amini_2014}, for instance, in periodic generalized L\'{e}vy processes  \cite{fageot_2020}. 

A discrete time process is a well-defined infinite dimensional random object, however, the standard approach used 
to measure compressibility for finite dimensional signals (based on the rate of decay of the absolute approximation error) does not extend 
naturally for this infinite dimensional analysis. Addressing this issue,  Amini et al. \cite{amini_2011} and Gribonval et al. \cite{gribonval_2012} proposed the use of a relative approximation error analysis to measure compressibility with the objective of quantifing the rate of the best $k$-approximation error with respect to the energy of the signal when the number of significant coefficients scales at a rate proportional to the dimension of the signal.  This approach offered a meaningful way to determine the energy --- and more generally the $\ell_p$-norm --- concentration signature of independent and identically distributed (i.i.d.) processes \cite{amini_2011,gribonval_2012}. In particular, they introduced the concept of $\ell_p$-compressibility to name a random sequence that has the capacity to concentrate (with very high probability) almost all their $\ell_p$-relative energy in an arbitrary small number of coordinates (relative to the ambient dimension)  of the canonical or innovation domain.  

Two important results were presented for i.i.d. processes.  \cite[Theorem 3]{amini_2011} showed that i.i.d. processes with heavy tail distribution (including the generalized Pareto, Students`s $t$ and log-logistic) are $\ell_p$-compressible for some $\ell_p$-norms. On the other hand, \cite[Theorem 1]{amini_2011} showed that i.i.d. processes with exponentially decaying tails (such as Gaussian, Laplacian and Generalized Gaussians) are not $\ell_p$-compressible for any $\ell_p$-norm. Completing this analysis,  Silva et al. \cite{silva_2015} stipulated a necessary and sufficient condition over the process distribution to be $\ell_p$-compressible (in the sense of Amini et al.\cite[Def.6]{amini_2011}) that reduces to look at the $p$-moment of the 1D marginal stationary distribution of the process. 

Importantly, the proof of the result in  \cite{silva_2015}  was rooted in the almost sure convergence of two empirical distributions (random objects function of the process) to their respective probabilities as the number of samples goes to infinity\footnote{These almost sure convergences created a family of typical sets that was used to prove the main result in \cite[Theorem 1]{silva_2015}.}.  
This  argument offered the context to move from using the law of large numbers (to characterize i.i.d. processes) to the use of the point-wise ergodic theorem \cite{gray_2009,breiman_1968}.  Then a necessary and sufficient condition for $\ell_p$-compressibility was obtained for the family of stationary and ergodic sources under the mild assumption that the process distribution projected on one coordinate, i.e., its 1D marginal distribution on $(\mathbb{R},\mathcal{B}(\mathbb{R}))$,  has a density \cite[Theorem 1]{silva_2015}. Furthermore, for non $\ell_p$-compressible processes, Silva et al. \cite{silva_2015} provided a closed-form expression for the so called 
$\ell_p$-approximation error function, meaning that a stable asymptotic value of the 
relative  $\ell_p$-approximation error is obtained when the rate of significant coefficients is given (fixed-rate analysis).

Considering that the proof of  \cite[Theorem 1]{silva_2015} relies heavily on an almost sure (with probability one) convergence of empirical means to their respective expectations,  the idea of relaxing some of the assumptions of the process, in particular  stationarity, suggests an interesting direction in the pursuit of extending results for the analysis of $\ell_p$-compressibility for general discrete time processes. 
In this work, we extend the compressibility analysis for a family of random sequences where stationarity or ergodicity is not assumed by examining the rich family of processes with ergodic properties and,  in particular,  the important family of asymptotically mean stationary (AMS) processes \cite{gray_2009,gray_1980}. This family of processes has been studied  in the context of source coding and channel coding problems where its ergodic properties (with respect to the family of indicator functions) have been used to extend fundamental performance limits in  source and channel coding problems.  Our interest in AMS processes centers on the fact that the  $\ell_p$-characterization in \cite{silva_2015} is fundamentally rooted in a form of ergodic property over a family of indicator functions; this family of measurable functions is precisely where AMS sources have (by definition) a stable almost-sure asymptotic behavior \cite{gray_2009}. 

\subsection{Contributions of this Work}
\label{sub_sec_contributions}
Specifically, 
we apply a more refined and relevant (sample-wise) almost sure fixed-rate analysis of $\ell_p$-approximation errors,  first considered by Gribonval {\em et al.} \cite{gribonval_2012}, to the analysis of a process. Through this analysis, we determine the relationship between the rate of significant coefficients and the $\ell_p$-approximation of a process in two main results.
 Our first main result (Theorem \ref{th_main_ams&ergodic}) shows that this rate vs. approximation error has a well-defined expression function of the process distribution---in particular the stationary mean of the process---for the complete collection of AMS and ergodic processes. This result relaxes stationarity as well as some of the regularity assumptions used in \cite[Theorem 1]{silva_2015};  consequently, it is a significant extension of that result.  As a direct implication of this theorem,  we extend the dichotomy of the $\ell_p$-compressible process presented in \cite[Theorem 1]{silva_2015} to the family of AMS ergodic processes (in Theorem \ref{th_main_ams&ergodic} in Section \ref{sub_sec_lp_compressilbe}).
 
 The second main result of this work (Theorem \ref{th_main_ams}) uses the ergodic decomposition theorem (EDT) \cite{gray_2009} to extend the strong $\ell_p$-characterization to the family of AMS processes, where ergodicity and the stationarity assumptions on the process have been relaxed. Remarkably, we show that this family of processes do have a stable (almost sure) asymptotic $\ell_p$-approximation error for any given rate of significant coefficients as the block of the analysis tends to infinity.  Interestingly, this limiting value is in general a measurable (non-constant) function of the process,  which is fully determined by the so-called ergodic decomposition (ED) function that maps elements of the sample space of the process to stationary and ergodic components \cite{gray_2009}. 
 
\subsection{Organization of the Paper}
\label{sub_sec_contributions}
The rest of the paper is organized as follows. Section \ref{sec_pre} introduces notations, preliminary results and some basic elements of the $\ell_p$-compressibility analysis. In particular, Section \ref{sec_strong_lp_charact} introduces the fixed-rate almost sure approximation  error analysis that is the focus of this work. Sections \ref{sec_main_results:AMS_Ergodic} and \ref{subsec_non_ergodic_case} present the two main results of this paper for AMS processes.  The summary and  final discussion of the results are presented in Section \ref{sec_final}. To conclude, Section \ref{sub_examples} provides some context for the construction of AMS processes based on the basic principle of passing an innovation process through deterministic (coding) and random (channel) processing stages. Section \ref{sec_numerical} presents a numerical strategy to estimate the $\ell_p$-approximation error function and some examples to illustrate the main results of this work. The proofs of the two main results (Theorems \ref{th_main_ams&ergodic} and \ref{th_main_ams}) are presented in Sections \ref{proof_th_main_ams&ergodic} and \ref{proof_th_main_ams}, respectively,  while the proofs of supporting results are relegated to the Appendices.

\section{Preliminaries}
\label{sec_pre}
For any vector $x^n=(x_1,..,x_n)$ in  $\mathbb{R}^n$, let  $(x_{n,1},..,x_{n,n}) \in \mathbb{R}^n$ denote the ordered vector such that $\left| x_{n,1} \right|  \geq \left|x_{n,2} \right|   \geq \ldots  \left|x_{n,n} \right|$. For $p>0$ and $k\in \left\{1,..,n \right\}$,  let 
\begin{equation}\label{eq_sec_pre_1}
	\sigma_p(k, x^n) \equiv (\left| x_{n,k+1} \right|^p + \ldots+ \left| x_{n,n} \right|^p)^\frac{1}{p}, 
\end{equation}	
be the best $k$-term $\ell_p$-approximation error of $x^n$, in the sense that if
$$\Sigma^n_k \equiv \left\{x^n \in  \mathbb{R}^n: \sigma_p(k, x^n)=0 \right\}$$ 
is the collection of $k$-sparse signals, then $\sigma_p(k, x^n) =\min_{\tilde{x}^n\in \Sigma^n_k} \left| \left| x^n- \tilde{x}^n\right|   \right|_{\ell_p}$. 

Amini {\em et al.} \cite{amini_2011} and  Gribonval {\em et al.}  \cite{gribonval_2012} proposed  the following relative best $k$-term $\ell_p$-approximation error 
\begin{equation}\label{eq_sec_pre_2}
	\tilde{\sigma}_p(k, x^n) \equiv \frac{\sigma_p(k, x^n)}{\left|\left| x^n \right|\right|_{\ell_p}} \in [0,1], \ k\in  \left\{1,..,n\right\}, 
\end{equation}	
for the analysis of infinite sequences, with the objective of extending notions of compressibility to sequences that  have infinite $\ell_p$-norms in $\mathbb{R}^{\mathbb{N}}$.
More precisely,  let $X_1,..,X_n,...$ be a one-side random sequence with values in $(\mathbb{R}^{\mathbb{N}}, \mathcal{B}(\mathbb{R}^{\mathbb{N}}))$. 
$(X_n)_{n\geq 1}$ is fully characterized by its consistent family of finite dimensional probabilities denoted by $\mu=\left\{\mu_n \in \mathcal{P}(\mathbb{R}^n): n\geq 1\right\}$ \cite{breiman_1968}, where $X^n \equiv (X_1,..,X_n)\sim \mu_n$ for all $n\geq 1$ and $\mathcal{P}(\mathbb{R}^n)$ is the collection of probabilities on the 
space $(\mathbb{R}^n, \mathcal{B}(\mathbb{R}^n))$ \cite{breiman_1968,gray_2009}. \footnote{In other words, $\mu_n(A)=\mathbb{P}\left\{X^n\in A \right\}$ for any set $A\in \mathcal{B}(\mathbb{R}^{n})$ and therefore $\mu_n$ is a probability in the measurable space $(\mathbb{R}^{n}, \mathcal{B}(\mathbb{R}^{n}))$.}

For $d\in (0,1)$, $n\geq 1$ and $k\in \left\{1,..,n \right\}$,  let us define the following set
\begin{align}\label{eq_sec_pre_3}
	\mathcal{A}^{n,k}_d & \equiv  \left\{ x^n \in \mathbb{R}^{n}: \tilde{\sigma}_p(k, x^n) \leq d \right\}. 
\end{align}
At this point, we need to introduce the following:
\begin{definition} \label{def_general_tipicality}
For a process $(X_n)_{n\geq 1}$,  equipped with $\mu=\left\{ \mu_n, n\geq 1\right\}$, and for any $n\geq 1$,  a set 
$\mathcal{A} \in \mathcal{B}(\mathbb{R}^n)$ is said to be $\epsilon$-typical for $X^n$ (or $\mu_n$) if
\begin{align}\label{eq_sec_pre_4}
	\mu_n(\mathcal{A})\geq 1-\epsilon.
\end{align}
\end{definition}
For the following three definitions,  let $(X_n)_{n \geq 1}$ be a process equipped with a distribution $\mu=\left\{ \mu^n, n\geq 1\right\}$:
\begin{definition}\cite[Defs.5 and 6]{amini_2011} \label{def_Amini_critical_dimension}
For $\epsilon>0$, $n\geq 1$ and $d\in (0,1)$, let
	\begin{equation}\label{eq_sec_pre_5}
		\tilde{\kappa}_p(d,\epsilon,\mu_n) \equiv \min  \left\{ k \in  \left\{1,\ldots ,n \right\}: \mu_n (\mathcal{A}^{n,k}_d)\geq 1-\epsilon \right\} 
	\end{equation}
denote the critical number of terms that makes $\mathcal{A}^{n,k}_d$ in (\ref{eq_sec_pre_3}) $\epsilon$-typical for $\mu_n$ (Def. \ref{def_general_tipicality}).
\end{definition}
Using Definition \ref{def_Amini_critical_dimension}, 
we can study the asymptotic rate of innovation of $(X_n)_{n \geq 1}$ 
relative to the $\ell_p$-approximation error using  $\tilde{\kappa}_p(d,\epsilon,\mu_n)$ in (\ref{eq_sec_pre_5}):
\begin{definition}\label{def_asymtotic_rate_of_innovation}
For any $d\in (0,1)$ and $\epsilon>0$ let us define: 
\begin{align}\label{eq_sec_pre_6a}
	\tilde{r}^+_p(d,\epsilon,\mu) \equiv \lim\sup_{n \rightarrow \infty} \frac{\tilde{\kappa}_p(d,\epsilon,\mu_n)}{n},\\
	\label{eq_sec_pre_6b}
	\tilde{r}^-_p(d,\epsilon,\mu) \equiv \lim\inf_{n \rightarrow \infty} \frac{\tilde{\kappa}_p(d,\epsilon,\mu_n)}{n}.
\end{align}
\end{definition}
Alternatively to the expressions in Definition \ref{def_asymtotic_rate_of_innovation}, 
we can consider the following fixed-rate asymptotic analysis for $(X_n)_{n \geq 1}$:
\begin{definition}\cite[Defs.5 and 6]{silva_2015}
\label{def_weak_approx_error_anal}
	Let us consider $\epsilon \in (0,1)$, $r \in (0,1)$ and $d\in (0,1)$.  
	The rate-distortion pair (r,d) is said to be $\ell_p$-achievable for $(X_n)_{n\geq 1}$ with  probability $\epsilon$,  if there exists a sequence of positive integers $(k_n)_{n\geq 1}$ such that $\lim \sup_{n \rightarrow \infty} \frac{k_n}{n} \leq r$ and 
	\begin{equation}\label{eq_sec_pre_7}
		\lim \inf_{n \rightarrow \infty} \mu_n (\mathcal{A}^{n,k_n}_d) \geq 1 - \epsilon.
	\end{equation}
	Then, the rate-{approximation} error function of $(X_n)_{n\geq 1}$  with  probability $\epsilon$ is given by 
	\begin{align}\label{eq_sec_pre_8}
		&{r}_p(d, \epsilon, \mu) \equiv \nonumber\\
		&\inf  \left\{ r\in [0,1], \text{such that $(r,d)$ is $\ell_p$-achievable for $(X_n)$ with  probability $\epsilon$} \right\}.
	\end{align}
\end{definition}

In general,  it follows that  ${r}_p(d, \epsilon, \mu) \leq \tilde{r}^+_p(d,\epsilon,\mu)$ \cite[Prop. 2]{silva_2015}. 
Furthermore, for the important case of stationary and ergodic processes, it was shown in \cite[Th. 1]{silva_2015} that
\begin{equation}\label{eq_sec_pre_9}
{r}_p(d, \epsilon,\mu)= \tilde{r}^+_p(d,\epsilon,\mu)=\tilde{r}^-_p(d,\epsilon,\mu)  \text{ for all } d\in (0,1).
\end{equation}

\subsection{Revisiting the $\ell_p$-Approximation Error Analysis}
\label{sec_strong_lp_charact}
The approximation properties of a process $(X_n)_{n \geq 1}$ presented  above 
rely on a weak convergence (in probability) of the event $\mathcal{A}^{n,k}_d$ (see Defs. \ref{def_Amini_critical_dimension} and \ref{def_weak_approx_error_anal}). Here, we introduce a stronger (almost sure) convergence of the approximation error at a given rate of innovation to study a more essential asymptotic indicator of the best $k$-term $\ell_p$-approximation attributes of $(X_n)_{n\geq 1}$.  This notion will be meaningful for a large collection of processes (details presented in Section \ref{sec_ams_ergodic_sequence}),  and it will imply specific 
approximation attributes for $\mu$  in terms of  ${r}_p(d, \epsilon,\mu)$, ${r}^+_p(d, \epsilon,\mu)$, and ${r}^-_p(d, \epsilon,\mu)$.

\begin{definition}\label{strong_lp_characterization}
	A process ${\bf X}=(X_n)_{n \geq 1}$, with distribution $\mu=\left\{ \mu_n, n\geq 1\right\}$, 
	is said to have a strong  rate vs. $\ell_p$ best $k$-term approximation error characterization 
	(in short,  a strong $\ell_p$-characterization),  if 
	for any $r\in (0,1]$ and for any non-negative  sequence of 
	integers $(k_n)_{n \geq 1}$ satisfying that $\frac{k_n}{n} \longrightarrow r$: 
	\begin{itemize}
	\item [{\bf i)}] the limit
		$\lim_{n \rightarrow \infty} \tilde{\sigma}_p(k_n, X^n)$ 
	is well defined in $\mathbb{R}$ with probability one ($\mu$-almost surely), and 
	\item[{\bf ii)}] this limit only depends on $r$ and not on the sequence $(k_n)_{n \geq 1}$. 
	\end{itemize}
	Provided that ${\bf X}$ has a strong $\ell_p$-characterization, we define and denote the approximation error function of ${\bf X}$ by 
	\begin{equation}\label{eq_sec_strong_lp_charact_1}
		f_{p,\mu}({\bf X},r) \equiv \lim_{n \rightarrow \infty} \tilde{\sigma}_p(k_n, X^n), 
	\end{equation}
	which is a function of  $\bf{X}$ and $r$. 
\end{definition}
A process with a strong $\ell_p$-characterization has an almost everywhere asymptotic (with $n$) pattern for its $\ell_p$-approximation error when a finite rate of significant coefficients is considered (i.e., a fixed-rate analysis).

%
On top of the structure introduced in Definition \ref{strong_lp_characterization},  a relevant scenario to consider is when the limiting function $f_{p,\mu}({\bf X},r)$,  in (\ref{eq_sec_strong_lp_charact_1}), is constant (independent of ${\bf X}$) $\mu$-almost surely. This can be interpreted as an ergodic property of ${\bf X}$ with respect to its best-$k$ term $\ell_p$-approximation error,  reflecting a typical (almost sure) approximation attribute that is constant for the entire process\footnote{ The next section shows that $f_{p,\mu}({\bf X},r)$ is a constant function for the family of AMS and ergodic processes \cite{gray_2009}.  However,  it is not a constant function  for stationary  and AMS processes in general as presented in Section \ref{subsec_non_ergodic_case}.}. 

The following result offers a connection between $f_{p,\mu}({\bf X},r)$ and ${r}_p(d, \epsilon,\mu)$ in this very special case. 

\begin{lemma} 
\label{main_lemma}
Let us consider a process ${\bf X}=(X_n)_{n\geq 1}$ and its process distribution $\mu$. Let us assume that  ${\bf X}$ has a strong $\ell_p$-characterization (Def. \ref{strong_lp_characterization}) and that its limiting function in (\ref{eq_sec_strong_lp_charact_1}) is constant  $\mu$-almost surely, denoted by $(f_{p,\mu}(r))_{r\in (0,1]}$. 
Assume that $d_o=f_{p,\mu}(ro)$ for some $r_o\in (0,1]$ \footnote{This means that $d_o\in \left\{f_{p,\mu}(r), r\in (0,1] \right\}$.}. Then, we have the following: 
	\begin{itemize}
		\item[i)]  
		If $d_o>0$,  then $r_o>0$ and $r_o$ is the unique solution of $d_o=f_{p,\mu}(r)$ for $r\in (0,1)$.    
		Furthermore, for any $r\in (0,1)$ and any $(k_n)_{n \geq 1}$ such that $\frac{k_n}{n} \longrightarrow r$, we have that
		\begin{equation}
		\label{eq_statement_main_lemma_1}
			 \lim_{n \rightarrow \infty} \mu_n(\mathcal{A}^{n,k_n}_{d_o})= \left\{ \begin{array}{ll}   1 & \textrm{if $r > r_o$}\\
			 															  0 & \textrm{if $r < r_o$.}
 																		 \end{array} 
													      \right.
		\end{equation}	
		\item[ii)] 
		 If $d_o=0$,  then  $\forall d\in (0,1)$,  for any ${r} \geq r_o$, and any $(k_n)_{n\geq 1}$ 
		 such that $\frac{k_n}{n} \longrightarrow {r}$,  it follows that 
		\begin{equation}
		\label{eq_statement_main_lemma_2}
			 \lim_{n \rightarrow \infty} \mu_n(\mathcal{A}^{n,k_n}_d)=1.
		\end{equation}	
	\end{itemize}
\end{lemma}

This result shows that for a process with a strong  $\ell_p$-characterization and a constant rate approximation error function,  there is a $0$-$1$ phase transition on the asymptotic probability of the events $\mathcal{A}^{n,k_n}_d$ when $k_n/n \longrightarrow r$, which is governed by $(f_{p,\mu}(r))_{r\in (0,1]}$ in (\ref{eq_sec_strong_lp_charact_1}).
 More precisely, we have the following corollary, which uses 
 the fact that  the  inverse $f_{p,\mu}^{-1}(d)$ is well defined for any $d\in \left\{f_{p,\mu}(r), r\in (0,1] \right\} \setminus \left\{ 0\right\}$  (see Lemma \ref{main_lemma} i)):
\begin{corollary}\label{cor_phase_transition_lp_errors}
Under the assumptions of Lemma \ref{main_lemma},  for any $d\in \left\{f_{p,\mu}(r), r\in (0,1] \right\} \setminus \left\{ 0\right\}$, and $\epsilon>0$
\begin{equation}\label{eq_sec_strong_lp_charact_2}
	{r}_p(d, \epsilon, \mu) = \tilde{r}^+_p(d,\epsilon,\mu)=\tilde{r}^-_p(d,\epsilon,\mu) = f_{p,\mu}^{-1}(d).
\end{equation}
On the other hand,  if $f_{p,\mu}(r_o)=0$ for some $r_o\in (0,1]$, then $\forall \epsilon \in (0,1)$, $\forall d\in (0,1)$,  
\begin{equation}\label{eq_sec_strong_lp_charact_3}
	{r}_p(d, \epsilon, \mu) \leq  \tilde{r}^+_p(d,\epsilon,\mu) \leq r_o.
\end{equation}
\end{corollary}

It is worth noting  in (\ref{eq_sec_strong_lp_charact_2}) that the weak $\ell_p$-approximation error function ${r}_p(d, \epsilon, \mu)$ is independent of $\epsilon$ and fully determined by $(f_{p,\mu}^{-1}(d))_{d\in \left\{f_{p,\mu}(r), r\in (0,1] \right\}}$. 
This is consistent with the result obtained for i.i.d. in \cite{amini_2011} and stationary and ergodic processes in \cite{silva_2015}. 

The proof of Lemma \ref{main_lemma} and Corollary \ref{cor_phase_transition_lp_errors} are presented in \ref{proof_main_lemma}.

\subsection{AMS Processes}
\label{sec_ams_ergodic_sequence}
A process ${\bf X} = (X_n)_{n\geq 1}$ is fully represented by the probability space $(\mathbb{R}^{\mathbb{N}},\mathcal{B}(\mathbb{R}^{\mathbb{N}}), \mu)$, where the process distribution $\mu$  is the central object. One way to model structure and dynamics on $(X_n)_{n\geq 1}$ (and indeed on $\mu$) is through the introduction of a measurable function $T: (\mathbb{R}^{\mathbb{N}},\mathcal{B}(\mathbb{R}^{\mathbb{N}})) \rightarrow  (\mathbb{R}^{\mathbb{N}},\mathcal{B}(\mathbb{R}^{\mathbb{N}}))$. Then we have an augmented object $(\mathbb{R}^{\mathbb{N}},\mathcal{B}(\mathbb{R}^{\mathbb{N}}), \mu, T)$ to analyze (and sometimes to represent) the process $(X_n)_{n\geq 1}$.  In this work,  we focus exclusively on the standard shift operator to look at the time dynamics and invariances of a process.\footnote{For $\bar{x}\in \mathbb{R}^{\mathbb{N}}$,   $\bar{z}=T(\bar{x})$ is given by the coordinate-wise relationship $z_i=x_{i+1}$ for all $i\geq 1$ \cite{gray_2009}.}
\begin{definition}\label{def_stationarity}
For the shift operator, the process $(X_n)_{n\geq 1}$ is said to be 
stationary (relative to $T$) if for any $F\in \mathcal{B}(\mathbb{R}^{\mathbb{N}})$,  
$\mu(F)=\mu(T^{-1}(F))$. 
\end{definition}
The definitions and properties presented in this section extend this basic notion of stationarity for $(X_n)_{n\geq 1}$. To that end, we will use $(\mathbb{R}^{\mathbb{N}},\mathcal{B}(\mathbb{R}^{\mathbb{N}}), \mu, T)$ to be the underlying dynamical system representation of ${\bf X} = (X_n)_{n\geq 1}$ where $T$ is the 
shift operator.\footnote{A complete exposition of sources with ergodic properties viewed as a dynamical system is presented in \cite[Chapts. 7, 8 and 10]{gray_2009}.} 

Given the above context, let us briefly introduce the family of AMS processes that is the main object of study of this work. 
\begin{definition}\label{def_ergodic_properties}
A process ${\bf X}=(X_n)_{n \geq 1}$ 
(or its underlying dynamical system $(\mathbb{R}^{\mathbb{N}},\mathcal{B}(\mathbb{R}^{\mathbb{N}}), \mu, T)$)  is said to have an ergodic property with respect  to a measurable function $f:(\mathbb{R}^{\mathbb{N}},\mathcal{B}(\mathbb{R}^{\mathbb{N}})) \rightarrow (\mathbb{R},\mathcal{B}(\mathbb{R}))$ if the sample average of $f$,  defined by
\begin{equation}\label{eq_ams_ergodic_sequence_1}
	<f>_n({\bf X}) \equiv \frac{1}{n}\sum_{i=0}^{n-1}f(T^i({\bf X})), 
\end{equation}
converges $\mu$-almost surely as $n$ tends to infinity to a measurable function $<f>({\bf X})$ from $(\mathbb{R}^{\mathbb{N}},\mathcal{B}(\mathbb{R}^{\mathbb{N}}))$ to $(\mathbb{R},\mathcal{B}(\mathbb{R}))$ . 
\end{definition}

\begin{definition}\label{def_ergodic_properties2}
A process ${\bf X}=(X_n)_{n \geq 1}$ (or $(\mathbb{R}^{\mathbb{N}},\mathcal{B}(\mathbb{R}^{\mathbb{N}}), \mu, T)$) is said to have an ergodic property with respect to a class of measurable functions $\mathcal{M}$, if  for any $f\in \mathcal{M}$, the sequence $(<f>_n({\bf X}))_{n\geq 1}$ converges to a well-defined limit $<f>({\bf X})$,  $\mu$-almost surely. 
\end{definition}

For any $n>0$, let us define the set of arithmetic mean probabilities by
\begin{equation}\label{eq_ams_ergodic_sequence_2}
	\mu^n(\mu,F) \equiv \frac{1}{n}\sum_{i=0}^{n-1}\mu(T^{-i}(F)),
\end{equation}
for all $F\in \mathcal{B}(\mathbb{R}^{\mathbb{N}})$, where it is clear that $(\mu^n(\mu,F))_{F\in \mathcal{B}(\mathbb{R}^{\mathbb{N}})} \in \mathcal{P}(\mathbb{R}^{\mathbb{N}})$ for any $n\geq 1$.

\begin{definition}\label{def_AMS}
A process $ {\bf X} =(X_n)_{n\geq 1}$ (or $(\mathbb{R}^{\mathbb{N}},\mathcal{B}(\mathbb{R}^{\mathbb{N}}), \mu, T)$)
is said to be asymptotically mean stationary (AMS)\footnote{By definition,  if $(X_n)_{n\geq 1}$ is stationary, then it is AMS.}, if  $(\mu^n(\mu,F))_{n\geq 1}$ in (\ref{eq_ams_ergodic_sequence_2}) converges as $n$ goes to infinity for any $F\in \mathcal{B}(\mathbb{R}^{\mathbb{N}})$. This limit is denoted by $(\bar{\mu}(F))_{F\in \mathcal{B}(\mathbb{R}^{\mathbb{N}})} \in \mathcal{P}(\mathbb{R}^{\mathbb{N}})$ and is called the stationary mean of ${\bf X}$.\footnote{$\bar{\mu}$ is function of $\mu$, but for sake of simplicity this dependency will be considered implicit.}
\end{definition}

\begin{remark}
If the limit of arithmetic mean probabilities of $\mu$ exists, in the sense that $(\mu^n(\mu, F))_{n\geq 1}$ convergences in $\mathbb{R}$ as $n$ tends to infinity for any 
event $F\in \mathcal{B}(\mathbb{R}^{\mathbb{N}})$,  these values (indexed by $F\in \mathcal{B}(\mathbb{R}^{\mathbb{N}})$) induce a well-defined probability in $\mathcal{P}(\mathbb{R}^{\mathbb{N}})$ that we denoted by $\bar{\mu}$ in Definition \ref{def_AMS} \cite[Lemma 7.4]{gray_2009}. 
\end{remark}
As expected, it can be proved that if ${\bf X}$ is AMS, then $\bar{\mu}$ is a stationary probability (with respect to $T$) in the sense that $\bar{\mu}(F)=\bar{\mu}(T^{-1}(F))$ for all $F\in \mathcal{B}(\mathbb{R}^{\mathbb{N}})$. 

The following important result connects processes with ergodic properties and AMS processes:
\begin{lemma}\cite[Ths. 7.1 and 8.1]{gray_2009}
	\label{lemma_ams_aux_1}
	A necessary and sufficient condition for a process ${\bf X}= (X_n)_{n\geq 1}$ 
	to be AMS is that it has an ergodic property (Definition \ref{def_ergodic_properties2}) 
	with respect to the family of indicator functions, i.e., 
	$\left\{{\bf 1}_{F}(\cdot): F\inº \mathcal{B}(\mathbb{R}^{\mathbb{N}}) \right\}$.
\end{lemma}

\subsection{Ergodicity}
Let us first introduce a stronger version of Definition \ref{def_ergodic_properties2}.
\begin{definition}\label{def_constant_ergodic_property}
A process ${\bf X}=(X_n)_{n\geq 1}$ (or $(\mathbb{R}^{\mathbb{N}},\mathcal{B}(\mathbb{R}^{\mathbb{N}}), \mu, T)$) has a constant ergodic property over a class $\mathcal{M}$ if:  
i) ${\bf X}=(X_n)_{n\geq 1}$ has an ergodic property over $\mathcal{M}$ (Definition \ref{def_ergodic_properties2}), and 
ii) for any $f\in \mathcal{M}$, its limit $<f>({\bf X})$ is a constant function, $\mu$-almost surely.
\end{definition}

The following definition derives from the celebrated point-wise ergodic theorem for AMS sources \cite[Th. 7.5]{gray_2009}: 
\begin{definition}\label{def_ergodicity}
A process ${\bf X} = (X_n)_{n \geq 1}$ (or $(\mathbb{R}^{\mathbb{N}},\mathcal{B}(\mathbb{R}^{\mathbb{N}}), \mu, T)$) is said to be ergodic,   if the collection of invariant events (the events $F$ such that $T^{-1}(F)=F$) has $\mu$-probability 1 or 0 \cite{breiman_1968,gray_2009}.
\end{definition}

The following result connects these last definitions:
\begin{lemma}\cite[Th. 7.5 and Lem. 7.14]{gray_2009}
	\label{lemma_ams_aux_2}
	A necessary and sufficient condition for an AMS process ${\bf X} =(X_n)_{n\geq 1}$ to be ergodic is that ${\bf X}$ has a constant ergodic property  for 
	$\left\{{\bf 1}_{F}(x): F\in \mathcal{B}(\mathbb{R}^{\mathbb{N}}) \right\}$. 
\end{lemma}

In general, AMS processes are not ergodic.  In fact, the following instrumental result provides a condition for ${\bf X}$ to meet 
ergodicity that can be considered a form of a weak mixing  (asymptotic independence) condition \cite{gray_2009}.
\begin{lemma}\cite[Lem. 7.15]{gray_2009}
	\label{lemma_ams_aux_3}
	A necessary and sufficient condition for an AMS process ${\bf X} =(X_n)_{n\geq 1}$ to be ergodic is that 
	\begin{equation}\label{eq_ams_ergodic_sequence_3}
		\lim_{n \rightarrow \infty}\sum_{i=0}^{n-1}\mu(T^{-i}(F)\cup F)= \bar{\mu}(F)\mu(F)
	\end{equation}
	for all $F\in \mathcal{F}$, where $\mathcal{F}\subset \mathcal{B}(\mathbb{R}^{\mathbb{N}})$ is a subfamily that generates $\mathcal{B}(\mathbb{R}^{\mathbb{N}})$.
\end{lemma}
For the case when the process is stationary, it follows that $\bar{\mu}(F) =\mu(F)$ for all $F \in \mathcal{B}(\mathbb{R}^{\mathbb{N}})$, then the condition in (\ref{eq_ams_ergodic_sequence_3}) can be interpreted as a mixing (asymptotic independence) property  on $(X_n)_{n\geq 1}$.

Finally, we have the point-wise ergodic theorem for AMS and ergodic processes:
\begin{lemma}\cite[Th. 7.5]{gray_2009}
	\label{lemma_point_wise_ergodic_th}
	Let ${\bf X} =(X_n)_{n\geq 1}$ be an AMS and ergodic process and let $f:(\mathbb{R},\mathcal{B}(\mathbb{R})) \longrightarrow (\mathbb{R},\mathcal{B}(\mathbb{R}))$ be $\ell_1$-integrable function with respect to $\bar{\mu}_1$.  Then it follows that
	\begin{equation}\label{eq_ams_ergodic_sequence_4}
		\lim_{n \longrightarrow \infty}\frac{1}{n}\sum_{i=1}^{n} f(X_i) = \mathbb{E}_{X\sim \bar{\mu}_1} (f(X))<\infty , \mu-a.s..
	\end{equation}
\end{lemma}

\section{Strong $\ell_p$-Characterization for  AMS and Ergodic Processes}
\label{sec_main_results:AMS_Ergodic}
To present the main result of this section, we first need to introduce some notations and definitions  for the statement 
of Theorem \ref{th_main_ams&ergodic}. Let ${\bf X}=(X_n)_{n \geq 1}$ be an AMS process with stationary mean $\bar{\mu} =\left\{\bar{\mu}_n: n\geq 1 \right\}$.  If the $p$-moment 
of the 1D marginal $\bar{\mu}_1$ is well defined, i.e., $\int_{\mathbb{R}} \left| x \right|^p d\bar{\mu}_1(x) < \infty$, then we can introduce 
the following induced probability $v_p\in \mathcal{P}(\mathbb{R})$ with $v_p\ll \bar{\mu}_1$ by
\begin{equation}\label{eq_ams_ergodic_sequence_3_b}
			v_p(B) \equiv \frac{\int_{B} \left| x \right|^p d\bar{\mu}_1(x)}{\int_{\mathbb{R}} \left| x \right|^p d\bar{\mu}_1(x)}, \forall B\in \mathcal{B}(\mathbb{R}). 	
\end{equation}
In addition, let us define the following tail sets: 
\begin{equation}\label{eq_ams_ergodic_sequence_3_c}
	B_\tau \equiv (-\infty,-\tau] \cup [\tau, \infty)  \text{ and } C_\tau \equiv (-\infty, - \tau) \cup (\tau, \infty), 
\end{equation}
for any $\tau \geq 0$. With this we define the following admissible set for $\bar{\mu}_1$:
\begin{equation}\label{eq_ams_ergodic_sequence_3_d}
	\mathcal{R}^*_{\bar{\mu}_1} \equiv \left\{ \bar{\mu}_1(B_\tau), \tau \in [0,\infty) \right\}.
\end{equation}
Finally,  let $\lambda$ denote the Lebesgue measure in $(\mathbb{R}, \mathcal{B}(\mathbb{R}))$. 

\subsection{Main Result}
When an AMS process satisfies the mixing condition in (\ref{eq_ams_ergodic_sequence_3}) and, consequently,  it is ergodic,  we can state the following result:
\label{subsec_ergodic_case}
\begin{theorem}\label{th_main_ams&ergodic}
	Let ${\bf X}=(X_n)_{n \geq 1}$ (or $(\mathbb{R}^{\mathbb{N}},\mathcal{B}(\mathbb{R}^{\mathbb{N}}), \mu, T)$) 
	be an  AMS and ergodic process, 
	and let  $\bar{\mu}=\left\{\bar{\mu}_n: n\geq 1 \right\}$ be its stationary mean (Definition \ref{def_AMS}).
	Then ${\bf X}$ has a a strong $\ell_p$-characterization (Definition \ref{strong_lp_characterization}).  
	More precisely, for any $p>0$,  $r\in (0,1]$, and $(k_n)_{n\geq 1}$ satisfying that $\frac{k_n}{n} \longrightarrow r$, 
	it follows that
	\begin{equation}\label{eq_ams_ergodic_sequence_4}
		f_{p,\bar{\mu}}(r) = \lim_{n\rightarrow \infty} \tilde{\sigma}_p(k_n,X^n),  \ \mu-\text{almost surely}, 
	\end{equation}
	where $f_{p,\bar{\mu}}(r)$ is a well-defined function of the stationary mean $\bar{\mu}$. 
	Furthermore, $(f_{p,\bar{\mu}}(r))_{r\in (0,1]}$ is an exclusive function of $\bar{\mu}_1\in \mathcal{P}(\mathbb{R})$ (the 1D marginal of $\bar{\mu}$) 
	with the following characterization:
	\begin{itemize}
		\item[i)]  
		If $\int_{\mathbb{R}} \left| x \right|^p d\bar{\mu}_1(x) = \infty$,
		then it follows that  
			$f_{p,\bar{\mu}}(r)=0, \ \forall r \in (0,1]$.
		
		\item[ii)] 
		If $\int_{\mathbb{R}} \left| x \right|^p d\bar{\mu}_1(x) < \infty$
		and $\bar{\mu}_1\ll \lambda$,  then for any $r \in (0,1]$,
		\begin{equation*}
			f_{p,\bar{\mu}}(r)=\sqrt[p]{1-v_p(B_{\tau(r)})},
		\end{equation*}
		where  $\tau(r)>0$ is the unique solution of 
			$\bar{\mu}_1(B_\tau)=r$.
		
		\item[iii)] 
		If $\int_{\mathbb{R}} \left| x \right|^p d\bar{\mu}_1(x) < \infty$, 
		$\bar{\mu}_1$ is not absolutely continuous with respect to $\lambda$,\footnote{In other words, $\bar{\mu}_1$ has atomic components.} 
		we have two cases:\\  
		a) $r\in \mathcal{R}^*_{\bar{\mu}_1}$,  where it follows that
		\begin{equation*}
			f_{p,\bar{\mu}}(r)=\sqrt[p]{1-v_p(B_{\tau(r)})},
		\end{equation*}
		and $\tau(r)$ is the solution of $\bar{\mu}_1(B_\tau)=r$.\\
		b) $r\notin \mathcal{R}^*_{\bar{\mu}_1}$ where there is 
		$\tau_o>0$ such that $\bar{\mu}_1(\left\{ -\tau_o, \tau_o\right\})>0$ and 
		$r\in [\bar{\mu}_1(C_{\tau_o}),  \bar{\mu}_1({B}_{\tau_o}))$.  Then $\exists \alpha_o \in [0,1)$ such that
		\begin{equation*}
			r =  \bar{\mu}_1(C_{\tau_o}) + \alpha_o  (\bar{\mu}_1(B_{\tau_o}) - \bar{\mu}_1(C_{\tau_o})), 
		\end{equation*}
		where 
		\begin{equation*}
			f_{p,\bar{\mu}}(r)=\sqrt[p]{1-v_p(C_{\tau_o}) -\alpha_o(v_p(B_{\tau_o})- v_p(C_{\tau_o}))}.
		\end{equation*}
		In the last expression, we have that
		$$v_p(B_{\tau_o})- v_p(C_{\tau_o}) = v_p(\left\{ -\tau_o, \tau_o\right\})= \left| \tau_o\right|^p \cdot \bar{\mu}_1(\left\{ -\tau_o, \tau_o\right\}) / \left| \left| (x^p)  \right|\right|_{L_1(\bar{\mu}_1)}.$$ 
	\end{itemize}
\end{theorem}
The proof  of this result is presented in Section \ref{proof_th_main_ams&ergodic}.

\subsection{Analysis and Interpretations of Theorem \ref{th_main_ams&ergodic}}
\label{sub_sec:dis_main_result}
\begin{figure}[h]
\centering
\includegraphics[width=1.05\textwidth]{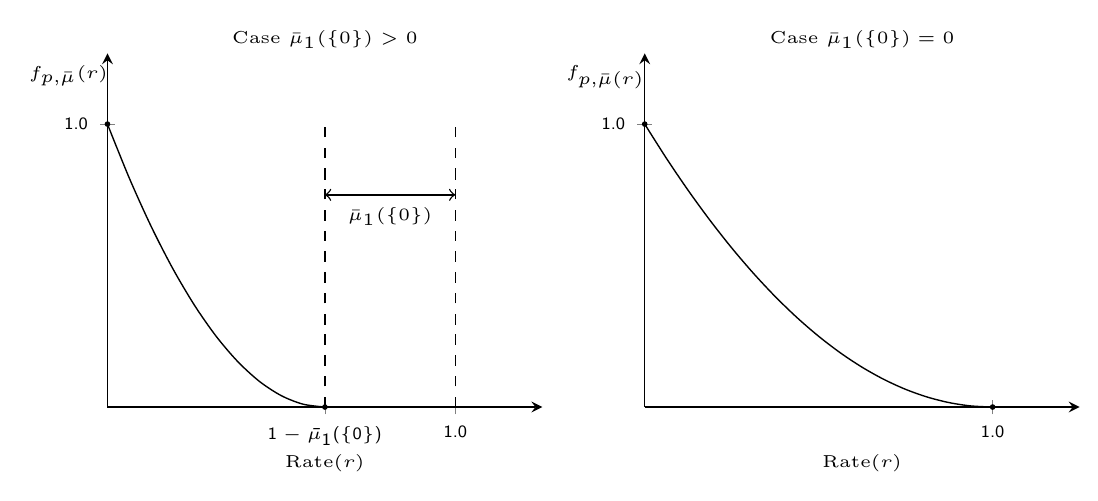}
\caption{Illustrations of the function $(f_{p,\bar{\mu}}(r))_{r\in (0,1]}$ in (\ref{eq_ams_ergodic_sequence_5}) assuming that $\int_{\mathbb{R}} \left| x \right|^p d\bar{\mu}_1(x) < \infty$. The left and right panels present the scenario where $\bar{\mu}_1(\left\{ 0\right\})\in (0,1)$ and $\bar{\mu}_1(\left\{ 0\right\})=0$, respectively.}
\label{fig_fp_structure}
\end{figure}

\begin{enumerate}
	\item The general result  in (\ref{eq_ams_ergodic_sequence_4}) shows that any ergodic AMS process has a strong $\ell_p$-characterization (Def. \ref{strong_lp_characterization}) where its point-wise (almost sure) approximation error function in (\ref{eq_sec_strong_lp_charact_1}) is completely determined by the 1D projection of its stationary mean, i.e.,  $\bar{\mu}_1\in \mathcal{P}(\mathbb{R})$. 
	
	\item Two important 
	scenarios can be highlighted. The case  
	$\int_{\mathbb{R}} \left| x \right|^p d\bar{\mu}_1(x) = \infty$
	 in which  $f_{p,\bar{\mu}}(r)=0, \ \forall r \in (0,1]$ and the case 
	 $\int_{\mathbb{R}} \left| x \right|^p d\bar{\mu}_1(x) < \infty$
	 that has a non-trivial approximation error function expressed by the following collection of (rate, distortion) pairs: 
	\begin{align}\label{eq_ams_ergodic_sequence_5}
		&\left\{(r, f_{p,\bar{\mu}}(r)), r\in (0,1] \right\}  =\left\{  \left( {\bar{\mu}_1}(B_\tau), \sqrt[p]{1-{v_p}(B_\tau)} \right), \tau\in [0,\infty) \right\}  \nonumber\\
	&\bigcup_{\tau_n\in \mathcal{Y}_{\bar{\mu}_1}} \bigcup_{\alpha \in [0,1)} \left\{ \left( \bar{\mu}_1(C_{\tau_n}) + \alpha \bar{\mu}_1(\left\{ -\tau_n,\tau_n \right\}), \sqrt[p]{1- v_p(C_{\tau_n})  - \alpha v_p(\left\{ -\tau_n,\tau_n \right\}) } \right)  \right\}
	\end{align}
	where $\mathcal{Y}_{\bar{\mu}_1}=\left\{\tau\in [0,\infty), \bar{\mu}_1(\left\{ -\tau, \tau \right\})>0 \right\}$,  which  is shown to be at most a countable set.
	It is worth noting that the expression in (\ref{eq_ams_ergodic_sequence_5}) summarizes the continuous and non-continuous result stated in ii) and iii). The details of this analysis are presented in Section \ref{proof_th_main_ams&ergodic}.
		
	\item {From the proof of Theorem \ref{th_main_ams&ergodic}, 
	the following properties for the function $(f_{p,\bar{\mu}}(r))_{r\in (0,1]}$ in (\ref{eq_ams_ergodic_sequence_4}) can be obtained:} 
	{\begin{proposition}\label{pro_properties_fp}
	Assuming that $\int_{\mathbb{R}} \left| x \right|^p d\bar{\mu}_1(x) < \infty$, 
	the function $(f_{p,\bar{\mu}}(r))_{r\in (0,1]}$ 
	is  continuous, strictly non-increasing in the domain $f_{p,\bar{\mu}}^{-1}(0,1)=(0,1-\bar{\mu}_1( \left\{0 \right\}))\subset (0,1)$,  and satisfies that $f_{p,\bar{\mu}}(r)=0$  $\forall r\in [1-\bar{\mu}(\left\{0 \right\}),1]$ and  $\lim_{r\rightarrow0}f_{p,\bar{\mu}}(r)=1$. 
	\end{proposition}}
	{The proof follows from Lemma \ref{lemma_rate_distorsion_properties} in Section \ref{proof_th_main_ams&ergodic} (and its proof in \ref{proof_lemma_rate_distorsion_properties}).} 
	
	\item {Proposition \ref{pro_properties_fp} implies that  $\forall d\in [0,1)$ there exists $r\in (0,1]$ such that  $f_{p,\bar{\mu}}(r)=d$, meaning that  $(f_{p,\bar{\mu}}(r))_{r\in (0,1]}$ achieves all values in $[0,1)$, and  $f_{p,\bar{\mu}}^{-1}(d)>0$ is well defined for any $d\in (0,1)$. Some illustrations  of $f_{p,\bar{\mu}}(\cdot)$ 
	are presented in Figure \ref{fig_fp_structure}.} 
	
	\item  
	{Under the assumption that	$\int_{\mathbb{R}} \left| x \right|^p d\bar{\mu}_1(x) < \infty$, 
	we have two important scenarios: the case when $\bar{\mu}_1(\left\{ 0\right\})\in (0,1)$ ($\bar{\mu}_1$ has atomic mass at $0$), and 
	the case when $\bar{\mu}_1(\left\{ 0\right\})=0$.
	{\bf i)} For the 
	scenario where $\bar{\mu}_1(\left\{ 0\right\})>0$, Proposition \ref{pro_properties_fp} and Theorem \ref{th_main_ams&ergodic}
	tell us that zero approximation error could be achieved at rates strictly smaller than $1$. More precisely, 
	we have that
		$f_{p,\bar{\mu}}(r)=0$  $\forall r \in [1-\bar{\mu}_1(\left\{ 0\right\}),1]$,  
	and $f_{p,\bar{\mu}}(r)>0$  $\forall r\in (0, 1-\bar{\mu}_1(\left\{ 0\right\}))$.
	{\bf ii)} On the other hand for the scenario where $\bar{\mu}_1(\left\{ 0\right\})=0$, 
	zero distortion is exclusively achieved at a rate equal to 1, meaning that $f_{p,\bar{\mu}}(r)>0$ for any $r\in (0, 1)$ (from Proposition \ref{pro_properties_fp}). 
	These two scenarios are illustrated in Figure \ref{fig_fp_structure}.}
	
	\item {Finally, we recognize two signatures for the general structure of $(f_{p,\bar{\mu}}(r))_{r\in (0,1]}$. $(f_{p,\bar{\mu}}(\cdot))$ is either a constant function equal to zero everywhere,  or it is a non-constant, strictly decreasing and continuous function from Proposition \ref{pro_properties_fp} (see Figure \ref{fig_fp_structure}).
	Indeed, we have the following dichotomy: }
	{\begin{corollary}\label{cor_dichotomy}
	$f_{p,\bar{\mu}}(r)=0$ for any $r\in (0,1)$ if, and only if,  $\int_{\mathbb{R}} \left| x \right|^p d\bar{\mu}_1(x) = \infty$.\footnote{For this last statement, we exclude the trivial process where $\bar{\mu}_1(\left\{0 \right\})=1$.}
	\end{corollary}
	The proof follows directly from Proposition \ref{pro_properties_fp} and Theorem \ref{th_main_ams&ergodic}.} 
\end{enumerate}

\subsection{$\ell_p$-compressible AMS and Ergodic Processes}
\label{sub_sec_lp_compressilbe}
{In light of Theorem \ref{th_main_ams&ergodic}, 
it is worth revisiting the  important concept of $\ell_p$-compressible processes introduced by Amini {\em et al.} \cite{amini_2011}, 
which is originally based on the weak $\ell_p$-characterization presented in Definition \ref{def_weak_approx_error_anal}. 
	\begin{definition}\cite[Def.6]{amini_2011}
	\label{def_compressible_process}
	A process $(X_n)_{n\geq 1}$, with distribution $\mu$,  is said to be $\ell_p$-compressible for $p>0$,  if
	for any $\epsilon\in (0,1)$ and $d\in (0,1)$,  $\tilde{r}^+_p(d,\epsilon,\mu)=0$.
	\end{definition}
}
	
	{Looking at Theorem \ref{th_main_ams&ergodic},  
	Lemma \ref{main_lemma}  and Corollary \ref{cor_phase_transition_lp_errors} (in particular
	the relationship expressed in Eq.(\ref{eq_sec_strong_lp_charact_2}) that connects the strong $\ell_p$-characterization in Definition \ref{strong_lp_characterization}
	with the weak $\ell_p$-characterization in (\ref{eq_sec_pre_8})), 
	the following result can be stated for the family of AMS ergodic processes:}
	{\begin{theorem}\label{cor_compressible_ams_erg_process}
		 A necessary and sufficient condition for an AMS ergodic process (with process distribution $\mu$ and stationary mean $\bar{\mu}$) to be 
		$\ell_p$-compressible (Def. \ref{def_compressible_process}) 
		is that 
			  $\int_{\mathbb{R}} \left| x \right|^p d\bar{\mu}_1(x) = \infty$.
		More specifically,  if 
		$\int_{\mathbb{R}} \left| x \right|^p d\bar{\mu}_1(x) < \infty$
		then for any $d\in (0,1)$ and $\forall \epsilon>0$, 
		$${r}_p(d, \epsilon, \mu) = \tilde{r}^+_p(d,\epsilon,\mu)=\tilde{r}^-_p(d,\epsilon,{\mu}) = f_{p,\bar{\mu}}^{-1}(d)>0,$$
		and otherwise,  i.e., 
				$\int_{\mathbb{R}} \left| x \right|^p d\bar{\mu}_1(x) = \infty$, 
		for any $d\in (0,1)$ and $\epsilon>0$, 
		$${r}_p(d, \epsilon, \mu) = \tilde{r}^+_p(d,\epsilon,\mu)=\tilde{r}^-_p(d,\epsilon,{\mu}) = 0.$$	  
	\end{theorem}
	}
	
	{Theorem \ref{cor_compressible_ams_erg_process} extends the dichotomy known in \cite[Theorem 1]{silva_2015} for ergodic and stationary processes 
	to the family of AMS and ergodic processes: i.e., a process is $\ell_p$-compressible if, and only if,  $\int_{\mathbb{R}} \left| x \right|^p d\bar{\mu}_1(x) = \infty$.}
	
	\begin{proof} 
	{Let us consider that $\int_{\mathbb{R}} \left| x \right|^p d\bar{\mu}_1(x) = \infty$.  From Theorem \ref{th_main_ams&ergodic},
	we have that $f_{p,\bar{\mu}}(r)=0$  $\forall r \in (0,1]$. Then using Corollary \ref{cor_phase_transition_lp_errors} (see  Eq(\ref{eq_sec_strong_lp_charact_3})), 
	it follows directly that $\sup_{\epsilon \in (0,1)}\sup_{d\in (0,1)} \tilde{r}^+_p(d,\epsilon,\mu) \leq r$ for any arbitrary small $r\in (0,1)$.  Consequently, we have that
	${r}_p(d, \epsilon, \mu) = \tilde{r}^+_p(d,\epsilon,\mu)=\tilde{r}^-_p(d,\epsilon, {\mu}) = 0$ for any $\epsilon \in (0,1)$ and for any $d\in (0,1)$.}
	
	{Assuming that $\int_{\mathbb{R}} \left| x \right|^p d\bar{\mu}_1(x) < \infty$, Proposition \ref{pro_properties_fp} shows  
	that $\left\{f_{p,\bar{\mu}}(r), r\in (0,1] \right\}=[0,1)$. Then using Corollary \ref{cor_phase_transition_lp_errors} (see  Eq(\ref{eq_sec_strong_lp_charact_2})), 
	we have that for any $d\in (0,1)$ and $\epsilon>0$,  ${r}_p(d, \epsilon, \mu) = \tilde{r}^+_p(d,\epsilon,\mu)=\tilde{r}^-_p(d,\epsilon,\mu) = f_{p,\bar{\mu}}^{-1}(d)$. 
	Finally, we know from the properties of $(f_{p,\bar{\mu}}(r))_{r\in (0,1]}$ stated in  Proposition \ref{pro_properties_fp} that $f_{p,\bar{\mu}}^{-1}(d)>0$ for 
	any $d\in (0,1)$}. 
	
	{To conclude, from Definition \ref{def_compressible_process} and the two previous results on $\tilde{r}^+_p(d,\epsilon,\mu)$, the main dichotomy stated in Theorem \ref{cor_compressible_ams_erg_process} is obtained.}
	\end{proof}
\subsubsection{Examples for $\bar{\mu}_1$}
\label{sub_sec_examples_lp_compressible}
{Here we show classes of distributions $\bar{\mu}_1$ where the key condition of $\ell_p$-compressibility 
stated in Theorem \ref{cor_compressible_ams_erg_process} can be tested. As this implies an isolated analysis of $\bar{\mu}_1$,  we revisit some of the conditions and examples  studied in \cite{silva_2015} to verify whether  $\int_{\mathbb{R}} \left| x \right|^p d\bar{\mu}_1(x)$ is either finite or not. Importantly, this analysis reduces to look at the  tail of $\bar{\mu}_1$ \cite{silva_2015,amini_2011}.}

{We begin covering the case of distribution (equipped with a density function) with exponential tail. \cite[Corrollary 1]{silva_2015}
shows that if for some $\gamma>0$ $\int_{\mathbb{R}} e^{\gamma  \left| x \right| } d\bar{\mu}_1(x) <\infty$ then 
$\int_{\mathbb{R}} \left| x \right|^p d\bar{\mu}_1(x) < \infty$ for any $p>0$. From this condition, it is simple to check that AMS ergodic processes with stationary 
mean ($\bar{\mu}_1$) following a Gaussian, generalized Gaussian, Laplacian and Gamma distribution (all of them with exponential tails) verify that $\mathbb{E}_{X \sim \bar{\mu}_1}(e^{\gamma\cdot \left| X \right| })<\infty$ for some $\gamma>0$ and their AMS ergodic processes are not $\ell_p$-compressible (Def. \ref{def_compressible_process}) for any $p>0$. Furthermore,   
if $\bar{\mu}_1$ is finitely supported, i.e.,   $\exists C>0$  where $\bar{\mu}_1([-C,C])=1$, then its AMS ergodic process 
is not $\ell_p$-compressible for any $p>0$.}

{On the other hand, we can consider AMS ergodic processes where $\bar{\mu}_1$ has a heavy tail distribution. For this class, 
\cite[Corrollary 2]{silva_2015} shows that if $\bar{\mu}_1 \ll \lambda$ and its density function $f_{\bar{\mu}_1}(x)=\frac{d\bar{\mu}_1}{d\lambda}(x)$
decays as $O(\left| x\right|^{-(\tau+1)})$ for some $\tau>0$  \footnote{A function $f(x)$ decays as $|x|^{-(\tau+1)}$ if there exists $x_o>0$ and $0<K_1 < K_2<\infty$  and $\lim_{x \rightarrow \infty} \frac{f(-x)}{f(x)}=\rho \in   \mathbb{R}^+ \cup   \left\{\infty \right\}$, where: if $\rho=0$,  then  $\forall x>x_o$, $K_1 x^{-(\tau+1)} \leq f(x) \leq K_2 x^{-(\tau+1)}$; 
if $\rho=\infty$,  then  $\forall x<-x_o$,  $K_1|x|^{-(\tau+1)} \leq f(x) \leq K_2|x|^{-(\tau+1)}$; 
and otherwise,  $\forall |x|>x_o$ then $K_1|x|^{-(\tau+1)} \leq f(x) \leq K_2|x|^{-(\tau+1)}$.} then
\begin{equation}\label{eq_lp_compre_cond_heavy_tail}
\int_{\mathbb{R}} \left| x \right|^p d\bar{\mu}_1(x) = \infty \text{ if, and only if, } p\geq \tau.	
\end{equation}
An example of this class of heavy tail distribution is the family of Student`s $t$-distribution with $q>0$ degrees of freedom \footnote{The pdf of a Student`s $t$-distribution with $q>0$ degrees 
of freedom is given by $f_{\bar{\mu}_1}(x)= \frac{\Gamma((q+2)/2)}{\sqrt{q\pi}} (1+ x^2/q)^{-(q+1)/2}$, where $\Gamma(\cdot)$ is the gamma function.},  
whose pdf decay (when $ \left|x \right|$ tends to infinity) as $O( \left| x \right|^{-(q+1)})$.  Consequently from  Theorem \ref{cor_compressible_ams_erg_process} 
and the condition in (\ref{eq_lp_compre_cond_heavy_tail}), an AMS ergodic process with stationary mean $\bar{\mu}_1$ following a Student`s $t$-distribution with parameter $q>0$
is $\ell_p$-compressible for any $p\geq q$ and non-$\ell_p$-compressible for $p<q$.\footnote{More examples and discussion about the verification 
of $\int_{\mathbb{R}} \left| x \right|^p d\bar{\mu}_1(x)=\infty$ can be found in \cite{silva_2015}.}}

\subsection{Estimation of $f_{p,\bar{\mu}}(r)$ from samples of ${\bf X}$}
{Theorem \ref{th_main_ams&ergodic} shows that an AMS ergodic process has a strong $\ell_p$-characterization 
that is an exclusive function of  $\bar{\mu}_1$. However, the determination of $\bar{\mu}_1$ could be a non-trivial technical task as well as the 
derivation of $(f_{p,\bar{\mu}}(r))_{r\in (0,1]}$ from the
expression presented in (\ref{eq_ams_ergodic_sequence_5}). To contextualize 
this observation and illustrate some examples,  Section \ref{sub_examples} presents results that show 
how AMS and ergodic processes can be constructed from basic processing stages. From these constructions, it is observed  
that the determination of  $\bar{\mu}_1$ could be a non-trivial task in many cases.  
Interestingly, if we can sample the process,  we could estimate $(f_{p,\bar{\mu}}(\cdot))$ (using the expression in (\ref{eq_ams_ergodic_sequence_5})) 
and numerically evaluate how $(f_{p,\bar{\mu}}(\cdot))$ behaves.  These estimations could be used to compare the compressibility pattern of different AMS and ergodic processes. 
A simple estimation strategy to infer $(f_{p,\bar{\mu}}(\cdot))$ and some numerical examples are presented in Section \ref{sec_numerical}.} 

\section{Strong $\ell_p$-Characterization for  AMS Processes}
\label{subsec_non_ergodic_case}
Relaxing the ergodic assumptions for an AMS source is the focus of this part. 
It is worth noting that the ergodic result in Theorem \ref{th_main_ams&ergodic} will be instrumental for this analysis in view of the ergodic decomposition (ED) theorem  for AMS sources nicely presented in \cite[Ths. 8.3 and 10.1]{gray_2009} and  references therein. In a nutshell, the 
ED theorem shows that the stationary mean of an AMS process (see Def. \ref{def_AMS}) can be decomposed 
as a convex combination of stationary and ergodic distributions (called the ergodic components) in $(\mathbb{R}^{\mathbb{N}},\mathcal{B}(\mathbb{R}^{\mathbb{N}}))$.  

It is important to introduce one specific aspect of this result for the statement of the following theorem.  Let us  consider an arbitrary AMS process $(X_n)_{n\geq 1}$ equipped with its process distribution $\mu \in \mathcal{P}(\mathbb{R}^{\mathbb{N}})$ and its induced stationary mean $\bar{\mu} \in \mathcal{P}(\mathbb{R}^{\mathbb{N}})$.  If we denote by $\tilde{\mathcal{P}}\subset \mathcal{P}(\mathbb{R}^{\mathbb{N}})$ the family of stationary and ergodic probabilities with respect to the shift operator, then one of the implications of the ergodic decomposition theorem \cite[Ths. 8.3 and 10.1]{gray_2009} is that there is a measurable space $(\Lambda, \mathcal{L})$ indexing this family, i.e., $\tilde{\mathcal{P}}=\left\{\mu_\lambda, \lambda\in \Lambda \right\}$. More importantly, there is 
a measurable function $\Psi: (\mathbb{R}^{\mathbb{N}},\mathcal{B}(\mathbb{R}^{\mathbb{N}})) \rightarrow (\Lambda, \mathcal{L})$ that maps points in the sequence space $\mathbb{R}^{\mathbb{N}}$ to stationary and ergodic components (more details will be given in Section \ref{proof_th_main_ams}). Then using $\Psi$, there is a probability measure $W_{\Psi}$  in $(\Lambda, \mathcal{L})$ induced by $\mu$
in the standard way,  where $\forall A \in \mathcal{L}$, we have that $W_{\Psi}(A)=\mu(\Psi^{-1}(A))$. One of the implications of the ED theorem \cite[Ths. 8.3 and 10.1]{gray_2009} is that for all $F\in \mathcal{B}(\mathbb{R}^{\mathbb{N}})$\footnote{The assumption here is that for any $F$, $\mu_{\lambda}(F) $,  as a function of $\lambda$,  is measurable from $(\Lambda, \mathcal{L})$ to $(\mathbb{R},\mathcal{B}(\mathbb{R}))$ \cite{gray_2009}.}
\begin{equation}\label{eq_subsec_non_ergodic_case_1}
	\bar{\mu}(F) = \int \mu_{\lambda}(F) \partial W_{\Psi}(\lambda).
\end{equation}
In other words, $\bar{\mu}$ can be expressed as the convex combination of stationary and ergodic components  $\left\{\mu_\lambda, \lambda\in \Lambda \right\}$, where the mixture probability on $(\Lambda, \mathcal{L})$ is induced by the decomposition function $\Psi$. This last function is universal, meaning  that  $\Psi$ is valid to decompose any stationary distribution on $(\mathbb{R}^{\mathbb{N}},\mathcal{B}(\mathbb{R}^{\mathbb{N}}))$ in stationary and ergodic components as presented in (\ref{eq_subsec_non_ergodic_case_1}).

\subsection{Main Result}
The following result uses the ED theorem for AMS sources \cite[Ths. 8.3 and 10.1]{gray_2009} and Theorem \ref{th_main_ams&ergodic} to show that AMS sources  have a strong $\ell_p$-characterization as stated in Definition \ref{strong_lp_characterization}. Furthermore, the result offers an  
expression to specify the limit $(f_{p,\mu}({\bf X},r))_{(0,1]}$ in (\ref{eq_sec_strong_lp_charact_1}). 
\begin{theorem}\label{th_main_ams}
	Let ${\bf X}=(X_n)_{n\geq 1}$ be an  AMS process with process distribution $\mu$. 
	Let us consider $\tilde{\mathcal{P}}=\left\{\mu_\lambda, \lambda\in \Lambda \right\}$ the collection of stationary and 
	ergodic probabilities and the decomposition function  $\Psi: (\mathbb{R}^{\mathbb{N}},\mathcal{B}(\mathbb{R}^{\mathbb{N}})) \rightarrow (\Lambda, \mathcal{L})$ presented in the ED theorem \cite[Th 10.1]{gray_2009}.  
	Then it follows that:
	\begin{itemize}
		\item[i)]  The process  ${\bf X}=(X_n)_{n\geq 1}$  has a strong $\ell_p$-characterization (Def. \ref{strong_lp_characterization}), where for any $r\in (0,1]$
			     and $(k_n)_{n\geq 1}$ such that $k_n/n \rightarrow r$  
			     \begin{equation} \label{eq_subsec_non_ergodic_case_2}
			     	  \lim_{n\rightarrow \infty} \tilde{\sigma}_p(k_n,X^n_1) = f_{p,\mu}({\bf X},r) =  f_{p,\mu_{\Psi({\bf X})}}(r),  \mu-\text{almost surely},
			     \end{equation}
			     where $(f_{p,\mu_{\lambda}}(r))$ has been introduced and developed in Theorem \ref{th_main_ams&ergodic}.\footnote{Note  that $\forall \lambda\in \Lambda$, $\mu_{\lambda} \in \tilde{\mathcal{P}}$ is a stationary and 
			     ergodic process.}
		\item[ii)] For any $r\in (0,1]$,  $d\in [0,1)$, and $(k_n)_{n\geq 1}$ such that $k_n/n \rightarrow r$, 
			     \begin{align} \label{eq_subsec_non_ergodic_case_3}
			     	  \lim_{n \rightarrow \infty} &\mu_n(A_d^{n,k_n}) = \int \lim_{n \rightarrow \infty} \mu^n_{\lambda} (A_d^{n,k_n}) \partial W_{\Psi}(\lambda) \nonumber\\
				  								  &=\mu \left( \left\{{\bf x}\in \mathbb{R}^{\mathbb{N}}: \mu_{\Psi({\bf x})} \text{ is $\ell_p$-compressible}  \right\} \right) \nonumber\\
												  &+\mu \left( \left\{{\bf x}\in \mathbb{R}^{\mathbb{N}}: \mu_{\Psi({\bf x})} \text{ is not $\ell_p$-compressible and } f_{p,\mu_{\Psi({\bf x})}}(r) \leq d \right\} \right).
			     \end{align}
	\end{itemize}
\end{theorem}
The proof  of this result is presented in Section \ref{proof_th_main_ams}.

\subsection{Analysis and Interpretations of Theorem \ref{th_main_ams}:} 
\begin{itemize}
	\item The first almost-sure point wise result in (\ref{eq_subsec_non_ergodic_case_2}) provides a closed-form expression for the $\ell_p$-characterization of the process ${\bf X}$ given by  $f_{p,\mu}({\bf X},r)$,  which is a function of ${\bf X}$  (not a constant function in general) through the 
	ED  function $\Psi(\cdot)$  that maps ${\bf X}$ to stationary and ergodic components in $\tilde{\mathcal{P}}$.  
	
	\item An interesting interpretation of the result in (\ref{eq_subsec_non_ergodic_case_2}),  which is a consequence of the ED theorem, is that this limiting behaviour can be seen as if one selects at $t=0$ an ergodic component $\mu_\lambda \in \tilde{\mathcal{P}}$,  and then the process evolves with the statistic of $\mu_\lambda$,  which has a strong $\ell_p$-characterization (Def. \ref{strong_lp_characterization}) given by Theorem \ref{th_main_ams&ergodic}. This  is equivalent to stating that there is one stationary ergodic component that is active all the time,  but we do not know a priori which component.  In fact, to resolve which component is active, we need to know the entire process ${\bf X}$,  as the active component $\lambda\in \Lambda$ is given by $\Psi({\bf X})$.  This interpretation has a natural connection with the standard setting used in universal source coding as clearly argued by Gray and Kieffer in \cite{gray_1980}, where it is assumed that a process is fixed and belongs to a family of process distributions from  beginning to  end, but the observer (or the designer of the coding scheme) does not know which specific distribution is active. Therefore, when observing a realization of an AMS process, what we are really observing is a realization of one (unknown a priori) stationary and ergodic component in $\tilde{\mathcal{P}}$ and, consequently,  its limiting behaviour is well defined 
	as expressed in (\ref{eq_subsec_non_ergodic_case_2}). The fact that this limit is expressed as a function of $\Psi$ can be 
	understood from the perspective that $\Psi$ is the object that chooses the active component in $\tilde{\mathcal{P}}$ from $\bf{X}$.
	
	\item An intriguing aspect of this result, which is again a consequence of the ED theorem for AMS sources, is that if we look at the limit $f_{p,\mu}({\bf X},r)$ in (\ref{eq_subsec_non_ergodic_case_2}), this is equal to $f_{p,\mu_{\Psi({\bf X})}}(r)$, which does not depend on $\mu$ explicitly as long as $\mu$ is AMS. Then,  we could say that the ED function $\Psi$ characterizes the asymptotic limit for any AMS source universally. 
	
	\item When we move to the weak $\ell_p$-characterization result expressed in (\ref{eq_subsec_non_ergodic_case_3}) (see Defs. \ref{def_Amini_critical_dimension} and \ref{def_weak_approx_error_anal}), 
	 here we can observe explicitly the role of the distribution $\mu$ in the analysis, which is consistent with the almost sure 
	 result in  (\ref{eq_subsec_non_ergodic_case_2}).  In the expression in the RHS of (\ref{eq_subsec_non_ergodic_case_3}), 
	 we note that the probability $\mu_n(A_d^{n,k_n})$ 
	 has a limit determined by the pair $(r,d)$ and the distribution $\mu$.  
	 
	 \item Complementing the previous point,  there are two clear terms in (\ref{eq_subsec_non_ergodic_case_3}): The first 
	is the probability (over $\mu$) of  the sequences that map through $\Psi(\cdot)$ to $\ell_p$-compressible  components (Def. \ref{def_compressible_process}) in $\tilde{\mathcal{P}}$.  The second 
	term is the probability of the sequences that map through $\Psi(\cdot)$ to ergodic components that are not $\ell_p$-compressible and satisfy that  its $\ell_p$-approximation error function (which is characterized in Theorem \ref{th_main_ams&ergodic}) evaluated at the rate $r$ is smaller than the distortion $d$. Note that these two events on  $(\mathbb{R}^{\mathbb{N}},\mathcal{B}(\mathbb{R}^{\mathbb{N}}))$ are distribution independent (universals) and therefore can be determined a priori (independent of $\mu$) for this weak $\ell_p$-compressibility analysis.
\end{itemize}

\section{Summary and Discussion of the Results}
\label{sec_final}
In this work,  we revisit the notion of $\ell_p$-compressibility  focusing 
on the study of the almost sure (with probability one)  limit of the $\ell_p$-relative best $k$-term approximation error when a fixed-rate of significant coefficients is considered for the analysis.   We consider the study of processes with general ergodic properties relaxing the stationarity and ergodic assumptions considered in previous work. Interestingly, we found that the family of asymptotically mean stationary (AMS) processes has an (almost-sure) stable $\ell_p$ approximation error behavior (sample-wise) when considering any arbitrary rate of 
significant coefficients per dimension of the signal. In particular, our two main results (Theorems \ref{th_main_ams&ergodic} and \ref{th_main_ams}) offer expressions for this limit, which is a function of the entire process through the known ergodic decomposition (ED) mapping used in the proof of the celebrated ED theorem. When ergodicity is added and we assume an AMS ergodic source, the $\ell_p$-approximation error function 
reduces to a closed-form expression of the stationary mean of the process.  As a direct consequence of this analysis,  we extend (in Theorem \ref{cor_compressible_ams_erg_process}) 
the dichotomy between $\ell_p$-compressibility and non $\ell_p$-compressibility observed in a previous result \cite[Th.1]{silva_2015}.

In summary, the two main theorems of this paper significantly
extend previous results in the literature on this problem that are valid under the assumption of stationarity, ergodicity and some extra regularity conditions on the process distributions. 
On the technical side, these new theorems show the important role that the general point-wise ergodic theorem and,  in particular,  the ED theorem play for the extension of the $\ell_p$-compressibility analysis to families of processes with general ergodic properties.
Finally,  from the proof of Theorem \ref{th_main_ams&ergodic}, we notice that imposing an ergodic property on the family of indicator functions is essential to obtain a stable (almost sure) result for the $\ell_p$-approximation error function,  in the way expressed in Definition \ref{strong_lp_characterization}. Consequently,  the AMS assumption (see Lemma \ref{lemma_ams_aux_1}) seems to be crucial to achieve the desired strong (almost-sure) $\ell_p$-approximation property declared in Definition \ref{strong_lp_characterization}.

\section{On the Construction and Processing of AMS Processes}
\label{sub_examples}
To conclude this paper, we provide some context to support the application of our results in Sections \ref{sec_main_results:AMS_Ergodic} and \ref{subsec_non_ergodic_case}. We consider a general generative scenario in which a process is constructed as the output of an innovation source passing through a signal processor (or coding process) and a random corruption (or channel).  This scenario permits us to observe a family of operations on a stationary and ergodic source (for example an i.i.d. source) 
that produces a process with a strong  $\ell_p$-characterization (Def. \ref{strong_lp_characterization}). For that we briefly revisit known results that guarantee that a process has stationarity and/or ergodic properties  when it is produced (deterministically or randomly) from a stationary and ergodic source.\footnote{A complete exposition can be found in \cite[Ch.2]{gray_1990_b}.}

A general way of representing a transformation of a process ${\bf X}=(X_n)_{n \geq 1}$ into another process is using the concept of a channel.  
\begin{definition}\label{def_channel}
A channel denoted by $\mathcal{C}$ is a collection of probabilities (or process distributions) in  $(\mathbb{R}^{\mathbb{N}}, \mathcal{B}(\mathbb{R}^{\mathbb{N}}))$ indexed by elements in $\mathbb{R}^{\mathbb{N}}$. More precisely,  $\mathcal{C}= \left\{v_{\bf{x}}, \bf{x}\in \mathbb{R}^{\mathbb{N}} \right\} \subset \mathcal{P}(\mathbb{R}^{\mathbb{N}})$ where for any $F\in \mathcal{B}(\mathbb{R}^{\mathbb{N}})$  $v_{\bf{x}}(F)$ is a measurable function from $(\mathbb{R}^{\mathbb{N}}, \mathcal{B}(\mathbb{R}^{\mathbb{N}}))$ to $(\mathbb{R}, \mathcal{B}(\mathbb{R}))$.
\end{definition}
Given the process distribution of $(X_n)_{n \geq 1}$, denoted by $\mu$, a channel $\mathcal{C}= \left\{v_{\bf{x}}, \bf{x}\in \mathbb{R}^{\mathbb{N}} \right\}$ induces a joint distribution in the product space $(\mathbb{R}^{\mathbb{N}}\times \mathbb{R}^{\mathbb{N}}, \mathcal{B}(\mathbb{R}^{\mathbb{N}} \times \mathbb{R}^{\mathbb{N}}))$ by
\begin{equation*}
	\mu \mathcal{C}(F\times G)= \int_{\bar{x}\in F} v_{\bar{x}}(G) d\mu(\bar{x}), \ \forall F, G \in \mathcal{B}(\mathbb{R}^{\mathbb{N}}).
\end{equation*}
The joint process distribution is denoted by $\mu \mathcal{C}$. Then a new process  ${\bf Y}=(Y_n)_{n\in \mathbb{N}}$ is obtained 
at the output of the channel when  $(X_n)_{n\in \mathbb{N}}$ is its input. If we denote the distribution of ${\bf Y}$ by $v$, this is obtained by the marginalization of $\mu \mathcal{C}$, i.e., $v(G) \equiv \mathbb{P}({\bf Y}\in G)=\mu \mathcal{C}(\mathbb{R}^{\mathbb{N}}\times G)$ for all $G\in \mathcal{B}(\mathbb{R}^{\mathbb{N}})$.  

\begin{definition}\label{def_stationary_channels}
Considering the shift operator $T$ (used to characterize stationarity and ergodic properties 
in Sec. \ref{sec_ams_ergodic_sequence}), a channel $\mathcal{C}= \left\{v_{\bf{x}}, \bf{x}\in \mathbb{R}^{\mathbb{N}} \right\}$ is said to be stationary with respect to $T$ if \cite[Sec. 2.3]{gray_1990_b}
\begin{equation*}
	v_{T(\bf{x})} (G)= v_{\bf{x}} (T^{-1}(G)), \forall \bf{x}\in \mathbb{R}^{\mathbb{N}}, \forall G\in \mathcal{B}(\mathbb{R}^{\mathbb{N}}).
\end{equation*}
\end{definition}
Then, the following result can be obtained: 
\begin{lemma}\label{lemma_stationary_channel_AMS_inputs} \cite[Lemma 2.2]{gray_1990_b}
	Let us consider an AMS process ${\bf X}$ (with stationary mean $\bar{\mu}$) as the input of a stationary channel $\mathcal{C}= \left\{v_{\bf{x}}, \bf{x}\in \mathbb{R}^{\mathbb{N}} \right\}$. Then the output process ${\bf Y}$ is AMS and its stationary mean is given by\footnote{The result shows more generally that the joint process $({\bf X}, {\bf Y})$ is AMS with respect to $T\times T$ ($T\times 
	T(\bar{x},\bar{y})=(T(\bar{x}),T(\bar{y})$), where its stationary mean is $\bar{\mu} \mathcal{C}$.}  
\begin{equation*}
\bar{v}(G)=\lim_{n \longrightarrow \infty}1/n \sum_{i=0}^{n-1} {v}(T^{-i}G) = \bar{\mu} \mathcal{C}(\mathbb{R}^{\mathbb{N}}\times G) \text{ for all } G\in \mathcal{B}(\mathbb{R}^{\mathbb{N}}).
\end{equation*}
\end{lemma}

Remarkably, Lemma \ref{lemma_stationary_channel_AMS_inputs} shows a general 
random approach to produce AMS processes from another AMS  process.  Furthermore, the result provides a closed expression for the resulting stationary mean (function of 
the stationary mean of the input $\bar{\mu}$ and the channel $\mathcal{C}$), 
which is the object that determines its strong $\ell_p$-compressibility signature from 
Theorems \ref{th_main_ams&ergodic} and \ref{th_main_ams}. 

Furthermore adding ergodicity, we highlight the following result:
\begin{lemma}\cite[Lemma 2.7]{gray_1990_b}
	\label{lemma_rand_channels_ergodic_preservation}
	If the channel $\mathcal{C}= \left\{v_{\bf{x}}, \bf{x}\in \mathbb{R}^{\mathbb{N}} \right\}$ is weakly mixing 
	in the sense that for all $\bf{x}\in \mathbb{R}^{\mathbb{N}}$ and measurable 
	events $F$, $G \in \mathcal{B}(\mathbb{R}^{\mathbb{N}})$ 
	\begin{equation*}
		\lim_{n \longrightarrow \infty} \frac{1}{n} \sum_{i=0}^{n-1} \left| v_{\bf{x}}(T^{-i}(F) \cap G) -  v_{\bf{x}}(T^{-i}(F))  v_{\bf{x}}(G)  \right|=0,
	\end{equation*}
	then if the input process is AMS and ergodic, the output of the channel is also AMS and ergodic.
\end{lemma}
We will cover two important families of channels used in many applications of statistical signal processing, source coding, and 
communications.  
\subsection{Deterministic Channels: Stationary Codes and LTI Systems}
A deterministic transformation (or measurable function) of an AMS process  ${\bf X}=(X_n)_{n\geq 1}$ can be seen as an important example of the channel framework presented above. Let us consider a measurable function $f: (\mathbb{R}^{\mathbb{N}}, \mathcal{B}(\mathbb{R}^{\mathbb{N}})) \longrightarrow  (\mathbb{R}^{\mathbb{N}}, \mathcal{B}(\mathbb{R}^{\mathbb{N}}))$  and the induced process 
${\bf Y}=f({\bf X})$, where the process distribution is $v(G)=\mu(f^{-1}(G))$ for all $G\in \mathcal{B}(\mathbb{R}^{\mathbb{N}})$. 
This form of encoding ${\bf X}$ is a special case of a channel, where $v^f_{\bf{x}}(G)=\mathbf{1}_{f^{-1}(G)}(\bf{x})$ for all ${\bf x}\in \mathbb{R}^{\mathbb{N}}$. Importantly, it follows that:
\begin{lemma}\cite{gray_1990_b}\label{lm_stationary_codings}
The deterministic channel (or coding) $\mathcal{C}^f \equiv \left\{ v^f_{\bf{x}}, \bf{x}\in \mathbb{R}^{\mathbb{N}}  \right\}$ induced by $f$ is stationary if,  and only if, $f(T({\bf x}))=T(f({\bf x}))$ for all ${\bf x}\in \mathbb{R}^{\mathbb{N}}$.
\end{lemma}
In this context, we say that $f$ produces a stationary coding of ${\bf X}$.  
\begin{corollary}\label{cor_invariant_stationary_coding}
 Any stationary coding of an AMS process produces an AMS process, where the stationary mean of ${\bf Y}$ is given by $\bar{v}(G)=\bar{\mu}(f^{-1}(G))$ for all $G\in \mathcal{B}(\mathbb{R}^{\mathbb{N}})$. 
\end{corollary}
The proof  of this result follows directly from Lemma \ref{lemma_stationary_channel_AMS_inputs}.

There is a stronger result for deterministic and stationary channels: 
\begin{lemma}\label{lemma_stationary_channel} \cite[Lemma 2.4]{gray_1990_b}
	Let us consider a deterministic and stationary channel $\mathcal{C}^f$.  If the input process to $\mathcal{C}^f$ is AMS and ergodic,  then  the output process is AMS and ergodic.
\end{lemma}

It is worth noting that a direct way of constructing stationary coding is by a scalar measurable function $\phi: (\mathbb{R}^{\mathbb{N}}, \mathcal{B}(\mathbb{R}^{\mathbb{N}})) \longrightarrow (\mathbb{R}, \mathcal{B}(\mathbb{R}))$, where given $\bf{x}\in \mathbb{R}^{\mathbb{N}}$, the output is produced by $y_n=\phi(T^{n-1}({\bf x}))$ for all $n\geq 1$.\footnote{Conversely, for any stationary code $f$ there is a function $\phi({\bf x})=\pi_1(f(\bf{x}))$ that induces $f$, where $\pi_1()$ denotes the first coordinate projection on $\mathbb{R}^{\mathbb{N}}$.} Then,  there is an infinity collection of stationary coding that preserves the AMS and ergodic characteristics of an input process.  Two emblematic cases to consider  are the finite length sliding block code where $y_n=\phi(X_{n+M},....,X_{n+D})$ with  $D>M\geq 0$ and  $\phi:\mathbb{R}^{D-M+1} \longrightarrow \mathbb{R}$, and the case when $\phi$ is a linear function, i.e.,  $\phi({\bf x})=\sum_{i\geq 1}^{D-M+1}a_i\cdot x_i$, and, consequently,  $f$ produces  a linear and time invariant (LTI) coding of ${\bf X}$.\footnote{Stationary codings play an important role in ergodic theory for the analysis of isomorphic processes \cite{gray_2009}.}

\subsection{Memoryless Channels}
\begin{definition}\label{def_memoryless_ch}
A channel $\mathcal{C}= \left\{v_{\bf{x}}, \bf{x}\in \mathbb{R}^{\mathbb{N}} \right\}$ is said to be memoryless,  
if for any  finite dimensional cylinder $\times_{i\in J}F_i \in \mathcal{B}(\mathbb{R}^{\mathbb{N}})$ 
and for any $\bf{x}\in \mathbb{R}^{\mathbb{N}}$, it follows that $v_{\bf x}(\times_{i\in J}F_i )=\prod_{i\in J} p_{x_i}(F_i)$
where $\left\{ p_x, x\in \mathbb{R} \right\} \subset \mathcal{P}(\mathbb{R})$. 
\end{definition}

Basically if $\mathcal{C}$ is memoryless, we have that $v_{\bf{x}}$ decompose as the multiplication of its marginals (memoryless) for any $\bf{x}\in \mathbb{R}^{\mathbb{N}}$. The classical example is the additive white Gaussian noise (AWGN) channel used widely in signal processing and communications where $p_x=\mathcal{N}(\mu_x, \sigma)$ is a normal distribution with the mean depending on $x$ and $\sigma>0$.  It is easy to check that memoryless channels are stationary.   Consequently, Lemma \ref{lemma_stationary_channel_AMS_inputs} tells us that a memoryless corruption of an AMS process produces an AMS process.  In addition, the mixing condition of Lemma \ref{lemma_rand_channels_ergodic_preservation} is easily verified 
for memoryless channels \cite{gray_1990_b}. Consequently,  a memoryless corruption of an AMS and ergodic process preserves 
the ergodicity of the input at the output of the channel.

Finally,  using Lemmas \ref{lemma_stationary_channel_AMS_inputs}, \ref{lemma_rand_channels_ergodic_preservation} and \ref{lemma_stationary_channel}, we have a rich collection of processing steps 
where AMS alone and AMS and ergodicity are preserved from the input to the output and, consequently,
Theorems \ref{th_main_ams&ergodic} and \ref{th_main_ams} can be adopted for compressibility analysis of these induced processes.

\subsection{Final  Examples}
\label{sub_sect_final_examples}
To conclude this section, we cover a result that offers a condition for a one-side process to be AMS 
using the following definition:
\begin{definition}\label{def_asymtotic_dominance}
	Let us consider two probabilities $\mu$ and $v$ in the measurable sequence space $(\mathbb{R}^{\mathbb{N}},\mathcal{B}(\mathbb{R}^{\mathbb{N}}))$. We say that $v$ asymptotically dominates $\mu$  with respect to the shift operator $T$ if, for any $F\in \mathcal{B}(\mathbb{R}^{\mathbb{N}})$ such that $v(F)=0$ then $\lim_{n \rightarrow  \infty} \mu(T^{-n}(F))=0$.\footnote{If $\mu \ll v$  then $v$ asymptotically dominates $\mu$. The proof of this statement follows directly from \cite[Theorem 3]{gray_1980}.}
\end{definition}

Then, we have the following result by Rechard \cite{rechard_1956} revisited in \cite[Theorem 2]{gray_1980}:
\begin{theorem}\label{th_ams_characterization} \cite{rechard_1956}
	Let ${\bf X}=(X_n)_{n\geq 1}$ be a process with process distribution $\mu$ in $(\mathbb{R}^{\mathbb{N}},\mathcal{B}(\mathbb{R}^{\mathbb{N}}))$. The process ${\bf X}$ is AMS if, and only if, there is a probability $v$  in $(\mathbb{R}^{\mathbb{N}},\mathcal{B}(\mathbb{R}^{\mathbb{N}}))$ that is stationary with respect to $T$ (Def. \ref{def_stationarity}) and asymptotically dominates $\mu$ (Def.\ref{def_asymtotic_dominance}).
\end{theorem}

Using a stationary process (or probability $v$), for example an i.i.d. process, Theorem \ref{th_ams_characterization} provides a way of constructing AMS processes. To illustrate this, the following is a selection of examples presented in \cite{gray_1980}.

\begin{example}\label{ex_1} \cite{gray_1980}
	Let $v$ be a stationary measure in $(\mathbb{R}^{\mathbb{N}},\mathcal{B}(\mathbb{R}^{\mathbb{N}}))$ and 
	let $f:(\mathbb{R}^{\mathbb{N}},\mathcal{B}(\mathbb{R}^{\mathbb{N}})) \rightarrow (\mathbb{R},\mathcal{B}(\mathbb{R})$ 
	be a non-negative integrable function with respect to $v$. If $v_f$ is the probability induced by $f$ by
	\begin{equation*}
		v_f(F) \equiv \frac{\int_{F} f(x) \delta v(x)}{\int_{\mathbb{R}^{\mathbb{N}}} f(x) \delta v(x)}, \ \forall  F\in \mathcal{B}(\mathbb{R}^{\mathbb{N}}),  
	\end{equation*}
	then $v_f$ is AMS (from Theorem \ref{th_ams_characterization} and the fact that by construction $v_f\ll v$).
	Conversely,  from the Radon-Nicodym theorem \cite{breiman_1968}, for any $\mu \ll v$ there exists
	$f(x)= \frac{\partial \mu}{\partial v}(x)$ ($v$-integrable) and, consequently, $\mu$ is AMS from Theorem \ref{th_ams_characterization}. 
\end{example}
Therefore, any integrable function $f$ with respect to a stationary probability $v$ creates an AMS process that is not stationary in general.

\begin{example}\label{ex_2} \cite{gray_1980}
	Let $v$ be a stationary measure in $(\mathbb{R}^{\mathbb{N}},\mathcal{B}(\mathbb{R}^{\mathbb{N}}))$ and 
	let $F\in \mathcal{B}(\mathbb{R}^{\mathbb{N}})$ be such that $v(F)>0$. If $\mu_F$ is the conditional distribution of $v$ given $F$ in $(\mathbb{R}^{\mathbb{N}},\mathcal{B}(\mathbb{R}^{\mathbb{N}}))$,  then $\mu_F$ is AMS (from Theorem \ref{th_ams_characterization} and the fact that by definition $\mu_F\ll v$).
\end{example}
Therefore conditioning a stationary probability on any non-trivial measurable set $F$ creates an AMS process that is not stationary in general.

\begin{example}\label{ex_3} \cite{gray_1980}
	If $\mu$ is stationary with respect to $T^N$, for some $N>1$, then $\mu$ is AMS (with respect to $T$) with stationary mean given by 
	\begin{equation}
		\bar{\mu}(F)=\frac{1}{N} \sum_{i=0}^{N-1} \mu(T^{-i}(F)) , \forall F  \in \mathcal{B}(\mathbb{R}^{\mathbb{N}}).
	\end{equation}
\end{example}
A practical case of this example is applying a $N$-block to $N$-block non-overlapping mapping (function) over a stationary process to create a process that is $T^N$-stationary by construction and, consequently, AMS. Details and examples of this block-coding construction and its use in information theory are presented in \cite{gray_1990_b}.

\section{Estimating $f_{p,\bar{\mu}}(r)$}
\label{sec_numerical}
The purpose of this last section is two-fold: First, we introduce an estimation strategy to approximate the function $f_{p,\bar{\mu}}(\cdot)$
with an arbitrary precision for an AMS and ergodic process without the need to obtain,  in closed-form,  
its invariant distribution $\bar{\mu}$ (Theorem \ref{th_main_ams&ergodic}).  Second, we illustrate the trend of the rate vs. $\ell_p$-approximation error pairs in (\ref{eq_sec_numerical_2}) for some simple cases of  $\bar{\mu}_1\in \mathcal{P}(\mathbb{R})$ (induced by Gaussian and $\alpha$-stable distributions \cite{breiman_1968,amini_2011}). 
 For this analysis, our main focus is the family of non $\ell_p$-compressible processes (see Definition \ref{def_compressible_process}
 and Theorem \ref{cor_compressible_ams_erg_process}) where in light of Theorem \ref{th_main_ams&ergodic} the behavior of $(f_{p,\bar{\mu}}(r))_{r\in (0,1]}$ is non-trivial (see Corollary \ref{cor_dichotomy}).

\subsection{Non $\ell_p$-compressible case}
\label{sub_sec_est_fp_non_compressible}
Let us consider the case of a non $\ell_p$-compressible AMS and ergodic process ${\bf X}=(X_n)_{n\geq  1}$, i.e., $\int_{\mathbb{R}} \left| x \right|^p d\bar{\mu}_1(x) < \infty$.  Let us assume for simplicity that $\bar{\mu}_1\ll \lambda$ and, consequently,  $\bar{\mu}_1$ has a density function.  In this context, Theorem \ref{th_main_ams&ergodic} part ii) shows that $f_{p,\bar{\mu}}(\cdot)$ is fully determined by the probabilities $\bar{\mu}_1$ and $v_p$  evaluated on the tail events $\left\{ B_\tau, \tau \geq 0 \right\}$ in (\ref{eq_ams_ergodic_sequence_3_c}).  More precisely,  we have that:
\begin{equation}\label{eq_sec_numerical_1}
	\left\{(r, f_{p,\bar{\mu}}(r)), r\in (0,1] \right\}  =\left\{  \left( {\bar{\mu}_1}(B_\tau), \sqrt[p]{1-{v_p}(B_\tau)} \right), \tau\in [0,\infty) \right\}. 
\end{equation}

\begin{figure}[h]
\centering
\includegraphics[width=0.85\textwidth]{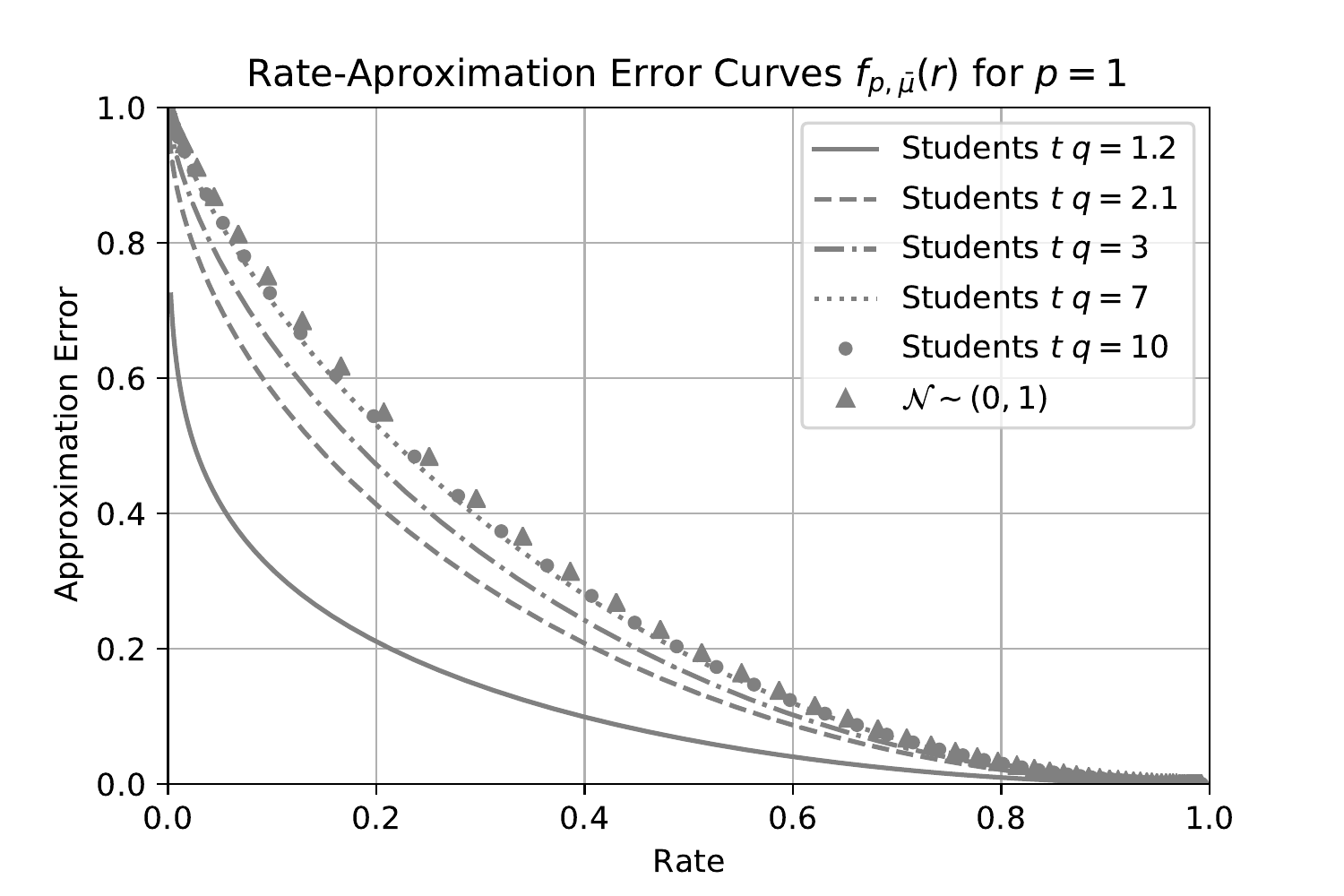}
\caption{Estimated curves of the approximation error function $f_{1,\bar{\mu}}(r)$:  for the case of an i.i.d. Gaussian process and  i.i.d. processes with Student´s $t$-distribution with $q=1.2;  2.1; 3; 7$ and  $10$.}
\label{fig_exp_1}
\end{figure}
\begin{figure}[h]
\centering
\includegraphics[width=0.85\textwidth]{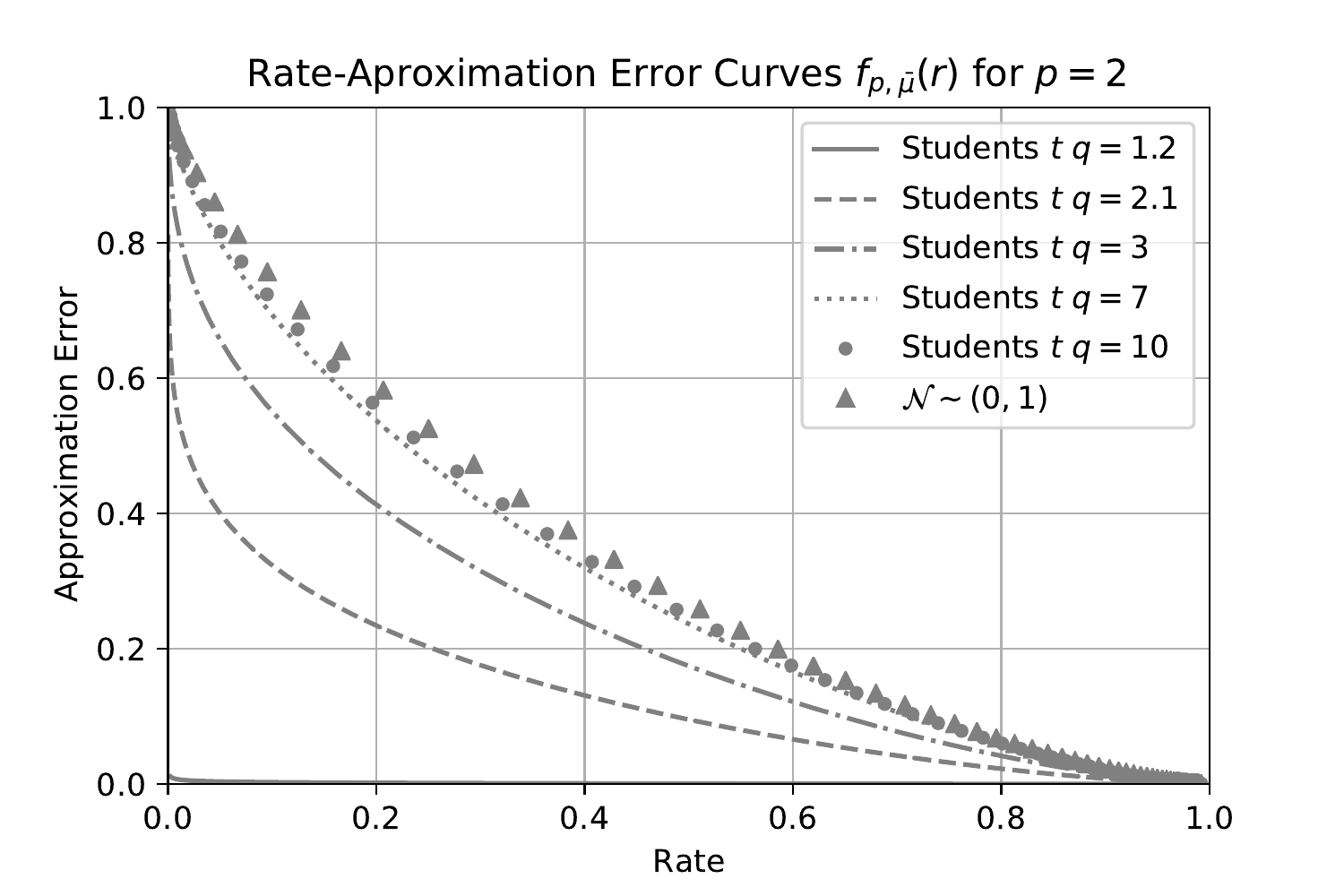}
\caption{Estimated curves of the approximation error function $f_{2,\bar{\mu}}(r)$:  for the case of an i.i.d. Gaussian process and  i.i.d. processes with Student´s $t$-distribution with $q=1.2;  2.1; 3; 7$ and  $10$.}
\label{fig_exp_2}
\end{figure}

At this point, we can use a realization of the process ${\bf X}=(X_n)_{n\geq  1}$ and the point-wise Ergodic theorem (PED) in Lemma \ref{lemma_point_wise_ergodic_th} to approximate with arbitrary precision (and almost surely) any of the pairs $\left( {\bar{\mu}_1}(B_\tau), \sqrt[p]{1-{v_p}(B_\tau)} \right)$ that define the rate vs. $\ell_p$-approximation error of ${\bf X}$. This is achieved without having to derive an expression for $\bar{\mu}$. Indeed, for any $\tau>0$, we have from the PED theorem that
\begin{equation}\label{eq_sec_numerical_2}
	\lim_{n \longrightarrow \infty} \underbrace{ \frac{1}{n}\sum_{i=1}^{n} {\bf 1}_{B_\tau}(X_i)}_{\hat{\bar{\mu}}_1(B_\tau, X^n) \equiv} = \bar{\mu}_1(B_\tau) , \mu-a.s., 
\end{equation}
\begin{equation}\label{eq_sec_numerical_3}
	\lim_{n \longrightarrow \infty}\frac{1}{n}\sum_{i=1}^{n} {\bf 1}_{B_\tau}(X_i) \cdot \left| X_i \right|^p  = \int_{B_\tau}  \left| x \right|^p d\bar{\mu}_1(x),  \mu-a.s.
\end{equation}
and,  consequently, 
\begin{equation}\label{eq_sec_numerical_4}
\lim_{n \longrightarrow \infty} \underbrace{ \frac{\sum_{i=1}^{n} {\bf 1}_{B_\tau}(X_i) \cdot \left| X_i \right|^p}{\sum_{i=1}^{n}  \left| X_i \right|^p}}_{\hat{v}_p(B,X^n) \equiv } = \frac{\int_{B_\tau}  \left| x \right|^p d\bar{\mu}_1(x)}{\left| \left| (x^p)  \right|\right|_{L_1(\bar{\mu}_1)}}= v_p(B_\tau), \mu-a.s.
\end{equation}
Therefore,  using a finite length realization of the process $X^n=(X_1,..,X_n)$,  for any $\tau>0$, we have that the empirical probabilities 
$\left( {\hat{\bar{\mu}}_1}(B_\tau,X^n), \sqrt[p]{1-\hat{{v}}_p(B_\tau, X^n)} \right)$ converge almost surely to the true 
expressions $\left( {\bar{\mu}_1}(B_\tau), \sqrt[p]{1-{v_p}(B_\tau)} \right)$ as $n$ tends to infinity. This estimation based strategy offers a  numerical way to estimate,  from a sample of the process,  the monotonic behavior of $f_{p,\bar{\mu}}(\cdot)$ (see Proposition \ref{pro_properties_fp}) using the empirical distributions in (\ref{eq_sec_numerical_4}) and (\ref{eq_sec_numerical_2}). In effect, this approach needs a strategy to sample the process and  a finite collection of threshold points  $\tau_1<\tau_2<....<\tau_L$ selected to capture the diversity of the range $[0,\infty)$. Note that as long as $L<\infty$,  the almost sure convergence of  $\left\{ \left( {\hat{\bar{\mu}}_1}(B_{\tau_i},X^n), \sqrt[p]{1-\hat{{v}}_p(B_{\tau_i}, X^n)} \right), i=1,..,L \right\} $ to $\left\{ \left( {{\bar{\mu}}_1}(B_{\tau_i}), \sqrt[p]{1-{{v}}_p(B_{\tau_i})} \right), i=1,..,L \right\} $ is guaranteed from the union bound \cite{gray_2009,breiman_1968}. In the following subsection, we illustrate this estimation strategy with some examples.

Finally, this strategy extends naturally to the more general scenario presented in Theorem \ref{th_main_ams&ergodic} part iii), for 
which the almost sure convergence result presented in (\ref{eq_pr_main_ams&ergodic_19a}) and (\ref{eq_pr_main_ams&ergodic_19b}) can be adopted (see the proof of Theorem \ref{th_main_ams&ergodic} in Section \ref{sub_sec_proof_compressible_case}).

\subsection{$\ell_p$-compressible case}
\label{sub_sec_est_fp_compressible}
{The sampling-based strategy presented in Section \ref{sub_sec_est_fp_non_compressible} can be adopted to estimate $f_{p,\bar{\mu}}(\cdot)$ when the process 
is $\ell_p$-compressible, i.e., $\int_{\mathbb{R}} \left| x \right|^p d\bar{\mu}_1(x) = \infty$. In this scenario, we know from Theorem \ref{th_main_ams&ergodic}  part i) that $f_{p,\bar{\mu}}(r)=0$ for any $r\in (0,1]$.  Although in theory, the probability $v_p$ is not well-defined  because we need that $\int_{\mathbb{R}} \left| x \right|^p d\bar{\mu}_1(x) < \infty$ (see Eq.(\ref{eq_ams_ergodic_sequence_3_b})),  we can still use the same empirical expressions $\hat{\bar{\mu}}_1(B_\tau, X^n)$ and $\hat{v}_p(B_\tau, X^n)$ presented  in (\ref{eq_sec_numerical_2}) and (\ref{eq_sec_numerical_4}).  The main difference in this case is that the PED theorem implies that
\begin{align}\label{eq_sec_numerical_4_i}
	&\lim_{n \longrightarrow \infty}\frac{1}{n}\sum_{i=1}^{n}  \left| X_i \right|^p  = \int _{\mathbb{R}} \left| x \right|^p d\bar{\mu}_1(x)= \infty,  \mu-a.s.,\\
	\label{eq_sec_numerical_4_ii}
	&\lim_{n \longrightarrow \infty}\frac{1}{n}\sum_{i=1}^{n} {\bf 1}_{B_\tau}(X_i) \cdot \left| X_i \right|^p  = \int_{B_\tau}  \left| x \right|^p d\bar{\mu}_1(x)= \infty,  \mu-a.s.,
\end{align}
using that $B_\tau=(-\infty,-\tau] \cup [\tau, \infty)$.  (\ref{eq_sec_numerical_4_i}) and  (\ref{eq_sec_numerical_4_ii})  imply 
directly that $\lim_{n \longrightarrow \infty} (1-\hat{v}_p(B_\tau, X^n))=0$, $\mu-a.s.$ for any $\tau>0$. 
Consequently, we have that the pair of empirical probabilities $\left( {\hat{\bar{\mu}}_1}(B_\tau,X^n), \sqrt[p]{1-\hat{{v}}_p(B_\tau, X^n)} \right)$ 
converge almost surely to the true expressions $\left( r_\tau={\bar{\mu}_1}(B_\tau), f_{p,\bar{\mu}}(r_\tau)= 0 \right)$ as $n$ tends to infinity. 
}

{Therefore, we can use a finite collection of threshold points  $\tau_1<\tau_2<....<\tau_L$ to sample $f_{p,\bar{\mu}}(\cdot)$, where in 
particular we have that $$\left\{ \left( {\hat{\bar{\mu}}_1}(B_{\tau_i},X^n), \sqrt[p]{1-\hat{{v}}_p(B_{\tau_i}, X^n)} \right), i=1,..,L \right\}$$ convergences 
almost surely (as $n$ tends to infinity) to the true sample points 
$$\left\{(r_i=\bar{\mu}_1(B_{\tau_i}), f_{p,\bar{\mu}}(r_i)=0):  i=1,..,L \right\}$$ of the 
zero constant function $f_{p,\bar{\mu}}(\cdot)$.} 

{Finally, looking at the trend of $\left\{ \left( {\hat{\bar{\mu}}_1}(B_{\tau_i},X^n), \sqrt[p]{1-\hat{{v}}_p(B_{\tau_i}, X^n)} \right), i=1,..,L \right\}$
as $n$ tends to infinity,  we could  have the ability to infer whether $f_{p,\bar{\mu}}(\cdot)$ is the trivial constant function equal to zero (associated 
to the case $\int_{\mathbb{R}} \left| x \right|^p d\bar{\mu}_1(x) = \infty$) or not (associated to the case $\int_{\mathbb{R}} \left| x \right|^p d\bar{\mu}_1(x) < \infty$). 
This dichotomy is observed in the next section in Figure \ref{fig_exp_2}.}

\subsection{Numerical Examples}
\label{sub_sec_examples}
We begin revisiting some of the examples used in \cite{silva_2015}, but in the context of the refined sample-wise almost sure approximation error analysis presented in this paper. We begin with  i.i.d. processes induced by different densities: from the Gaussian case that has an exponential tail,  to the case of Student`s $t$-distribution with different values for its parameter $q>0$.  The density of a Student`s $t$-distribution with  $q$ degrees of freedom is said to be heavy tailed as its tail goes to zero (with $x$) at the polynomial rate $\mathcal{O}( \left| x\right|^{-q-1})$.\footnote{ The density of a Student´s $t$-distribution with parameter $q>0$ (degrees of freedom) is: $f(x)=\frac{\Gamma((q+2)/2)}{\sqrt{q \pi}}(1+x/q)^{-\frac{q+1}{2}}$, where $\Gamma()$ denotes the Gamma function.} {From Theorem \ref{cor_compressible_ams_erg_process}}, 
we know that the Gaussian process is non $\ell_p$-compressible for any $p>0$. On the other hand, an i.i.d. process driven by a Student`s $t$-distribution with parameter $q>0$ is non $\ell_p$-compressible if, and only if,  $p<q$ {(see Eq.(\ref{eq_lp_compre_cond_heavy_tail}) in Section \ref{sub_sec_examples_lp_compressible})}.

Using the estimation strategy presented above,  and for all the aforementioned scenarios, we consider a good range of points in $\tau\in [0,\infty)$ and a sufficiently large number of samples  for each process ($n=10^5$) to obtain a precise estimation of a finite collection of rate vs. $\ell_p$-approximation error pairs in (\ref{eq_sec_numerical_1}). These points are illustrated  in Figures \ref{fig_exp_1} and \ref{fig_exp_2}.  Beginning with $p=1$, Figure \ref{fig_exp_1} presents the estimated functions $f_{1,\bar{\mu}}(r)$ for the Gaussian case and Student`s $t$-distribution with $q=1.2;  2.1; 3; 7$ and  $10$. {All the presented curves are consistent with the observation that all the selected processes are non $\ell_1$-compressible (see Section \ref{sub_sec_examples_lp_compressible}) and, consequently, for each example, the estimation of $f_{p,\bar{\mu}}(\cdot)$ should be non-zero (from Theorem \ref{th_main_ams&ergodic} part ii)), continuous, and strictly decreasing (from Proposition \ref{pro_properties_fp}).} In addition. the curves show that the Gaussian i.i.d. process (white noise)  is the least compressible and the degree of compressibility decreases from case to case as a function of how fast the tail of the density of the i.i.d process goes to zero. This observation is consistent with previous results that show that heavy tail processes are better approximated by their best sparse versions than processes with exponential tails \cite{amini_2011,silva_2015}. 

The same set of curves are obtained for $p=2$  in Figure \ref{fig_exp_2}.  {Using the analysis presented in Section \ref{sub_sec_examples_lp_compressible} (for $p=2$), we have that only one of the selected processes is $\ell_2$-compressible (the case with a Student`s $t$-distribution with $q=1.2$), and the rest are non-$\ell_2$-compressible. The estimations of $f_{p,\bar{\mu}}(\cdot)$ presented in Figure \ref{fig_exp_2} are consistent with Theorem \ref{th_main_ams&ergodic} part ii), showing non-zero functions with decreasing trends for all examples that are non-$\ell_2$-compressible. For the $\ell_2$-compressible case (Student`s $t$-distribution with $q=1.2$),  the estimated curve  matches with good precision a constant equal to zero function predicted by Theorem \ref{th_main_ams&ergodic} part i).}

\begin{figure}[h]
\centering
\includegraphics[width=0.85\textwidth]{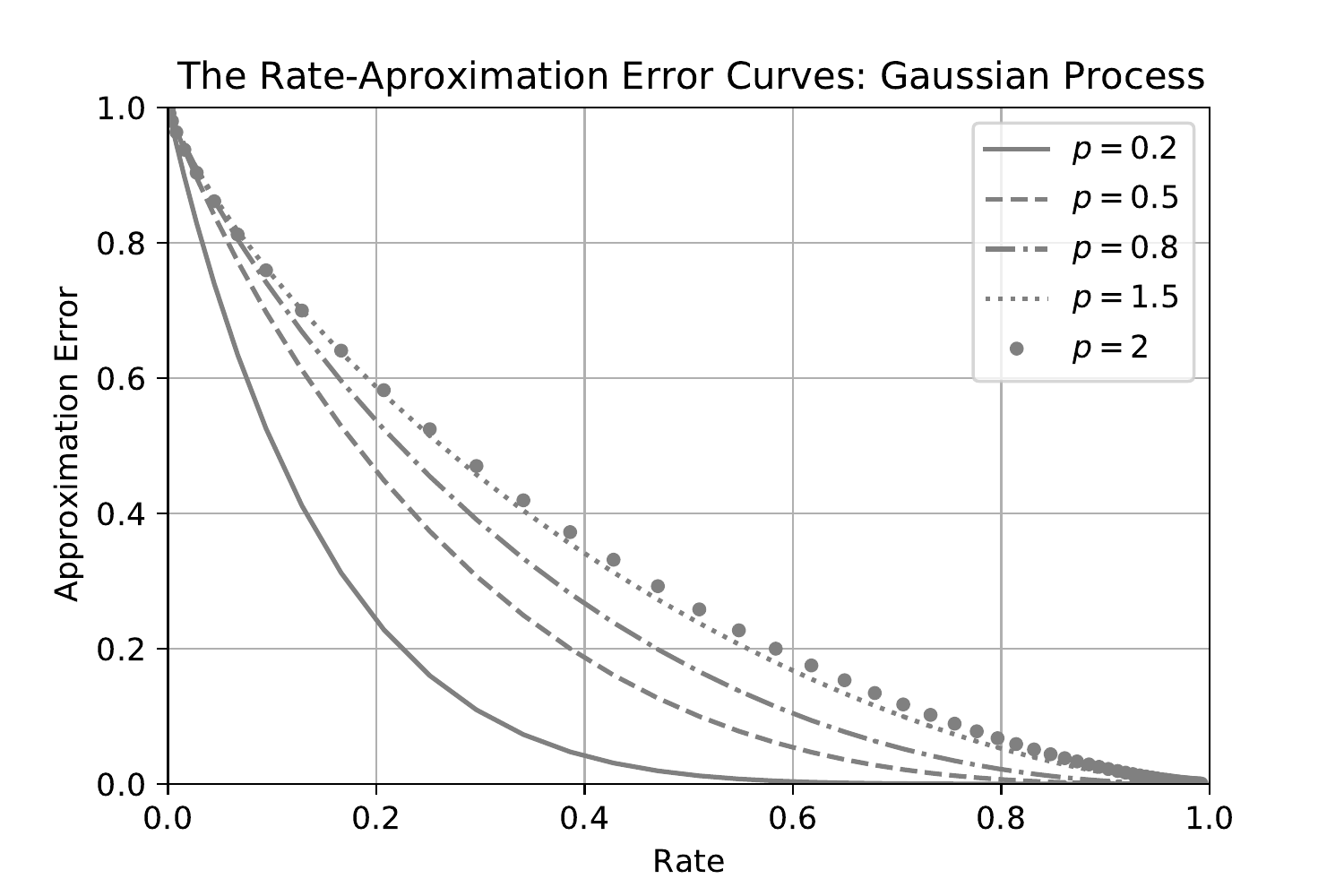}
\caption{Estimated curves of the approximation error function $(f_{p,\bar{\mu}}(r))_{r\in (0,1])}$ for the an i.i.d. Gaussian process.}
\label{fig_exp_3}
\end{figure}

To complement this analysis, Figure \ref{fig_exp_3} illustrates $f_{p,\bar{\mu}}(\cdot)$ for the same i.i.d. Gaussian process for different $p$ values ($p\in \left\{0.2, 0.5, 0.8, 1.5, 2 \right\}$). There is clear monotonic behavior of the curves by increasing the magnitude of $p$ in the analysis, where the Gaussian process becomes less compressible when $p$ increases.  {For this analysis, we know that the Gaussian i.i.d. process is non $\ell_p$-compressible for any possible value of $p$ (see Section \ref{sub_sec_examples_lp_compressible}), which is shown consistently in the estimated curves presented in Figure  \ref{fig_exp_3}.}

\begin{figure}[h]
\centering
\includegraphics[width=0.85\textwidth]{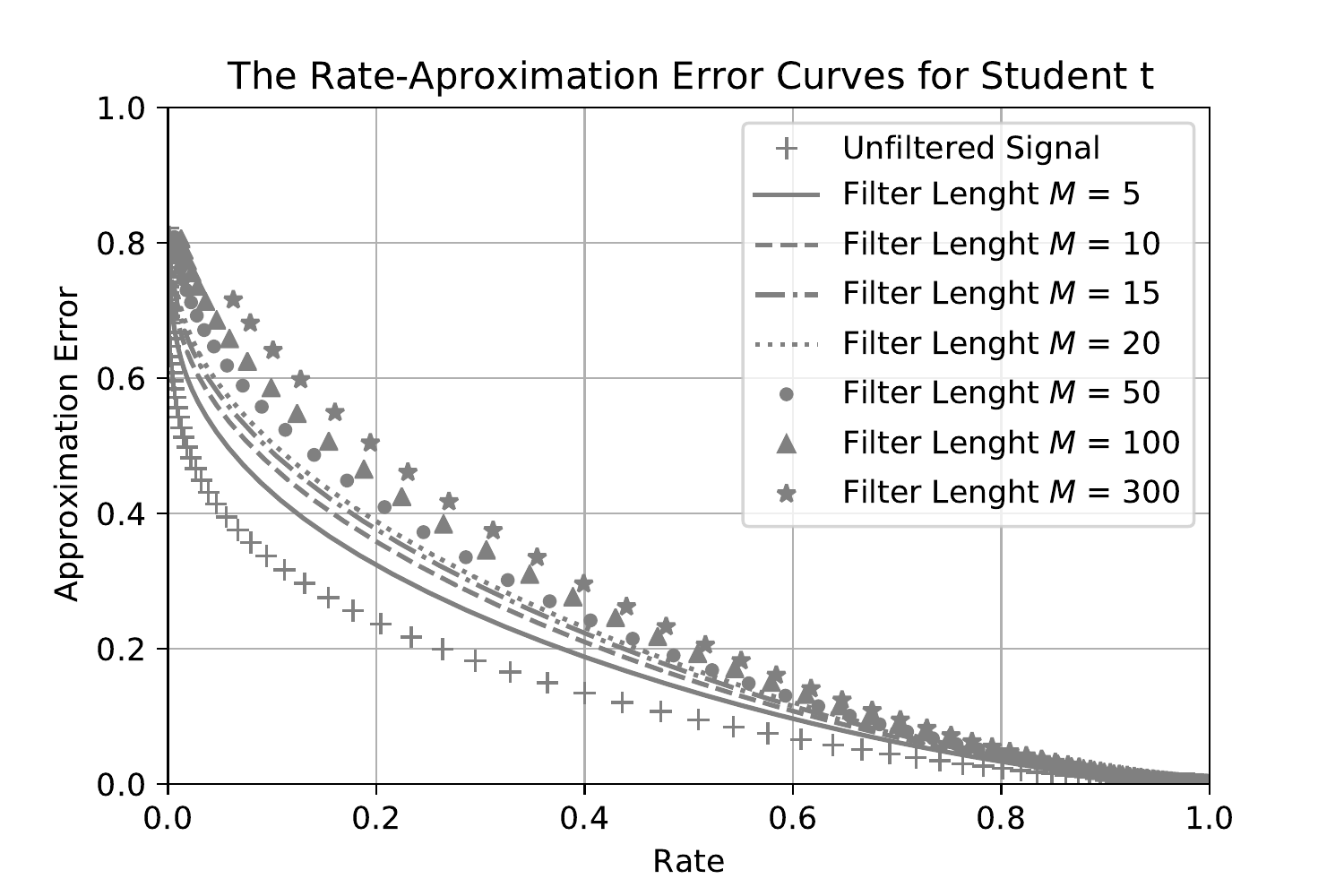}
\caption{Approximation error curves for an i.i.d. processes driven by a Student´s $t$-distribution with $q= 2.1$  and for the convolution 
of this process (innovation) with the LTI system $(a_i)^M_{i=1}=(1,1,..,1)$  for  $M\in \left\{5, 10, 15, 20, 50, 100, 300 \right\}$.}
\label{fig_exp_4}
\end{figure}

\begin{figure}[h]
\centering
\includegraphics[width=0.85\textwidth]{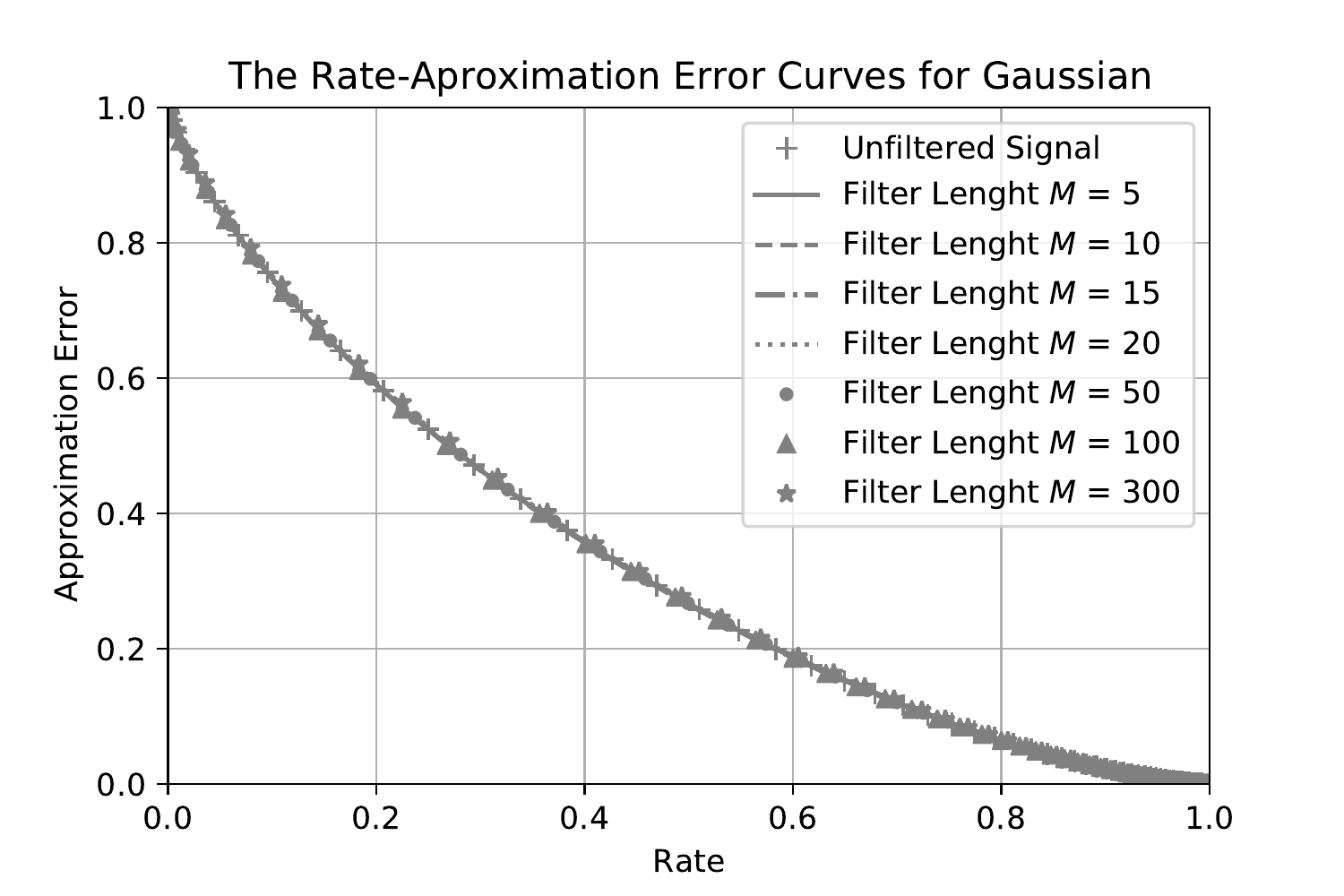}
\caption{The same set of curves presented in Figure \ref{fig_exp_4}  but using an  i.i.d. Gaussian process as the input of the LTI.}
\label{fig_exp_5}
\end{figure}

Finally,  we estimate $f_{p,\bar{\mu}}(\cdot)$ for an AMS and ergodic process that is obtained as the result of an i.i.d. innovation ${\bf X}=(X_n)_{n\geq 1}$ passing through a stationary coding (see Lemma \ref{lm_stationary_codings}).  In particular, we consider the case of a linear operator of the form $\phi({\bf x})=\sum_{i\geq 1}^{M}a_i\cdot x_{M+1-i}$. This is one of the classical LTI constructions used to model random signals in statistical signal processing \cite{gray_2004}. In this case,  the invariant distribution of the resulting process can be obtained by the product distribution of the input $(X_n)_{n\geq 1}$ and Corollary \ref{cor_invariant_stationary_coding}. However,  we use the estimation approach for which we only need a sufficiently large sample of $(X_n)_{n\geq 1}$ and  the application of convolutional  equation to reproduce $Y_n=\phi(T^{n-1}({\bf X}))$ for any $n\geq 1$. Figure \ref{fig_exp_4} shows $f_{p,\bar{\mu}}(r)$ (for {\bf X} with stationary mean $\bar{\mu}$) and $f_{p,\bar{v}}(r)$ (for ${\bf Y}=(Y_n)_{n\geq 1}$ with stationary mean $\bar{v}$) for the cases of LTI filter with $(a_i)^M_{i=1}=(1,1,..,1)$ 
and when ${\bf X}$ is i.i.d. driven by a Student`s $t$-distribution with parameter $q=2.1$ and $p=2$ (i.e., non-$\ell_p$-compressible for $p=2$).
The filter (parametrized by $(a_i)^M_{i=1}$) clearly changes the compressibility signature of the output process and, furthermore, the effect of increasing $M$  is evident in $f_{p,\bar{v}}(r)$ (the output) on its relationship with $f_{p,\bar{\mu}}(r)$ (the input). To clearly observe these changes from the input to the output, we choose the process that showed (from the previous analysis) the most compressible curve in Figure \ref{fig_exp_2}, i.e., the Student`s $t$-distribution with $q=2.1$. In the other end, Figure \ref{fig_exp_5} shows the same analysis when the input process ${\bf X}$ is an i.i.d. Gaussian process (see Figure \ref{fig_exp_2}). In this case, however, the effect of the filter $(a_i)^M_{i=1}$ and its length on the compressibility of ${\bf Y}$ is imperceptible. We observe from these two last examples (in Figures 4 and 5) that a linear stationary coding makes the output process less compressible than its input process, which is expected from the convolutional nature of the mapping, however,  when the input process has an exponential tail (white noise) the effect of linear filtering is imperceptible. 

\section{Proof of Theorem \ref{th_main_ams&ergodic}}
\label{proof_th_main_ams&ergodic}
First, we introduce a number of preliminary results,  definitions, and properties that will be essential to elaborate the main argument to prove Theorem \ref{th_main_ams&ergodic}. 
\subsection{Preliminaries}
%
For the case of AMS and ergodic sources (see Lemmas \ref{lemma_ams_aux_1} and \ref{lemma_ams_aux_3}), the ergodic theorem \cite[Th. 7.5]{gray_2009} (see Lemma \ref{lemma_point_wise_ergodic_th})  tells us that for any $\ell_1$-integrable function with respect to $\bar{\mu}_1$,  $f:(\mathbb{R},\mathcal{B}(\mathbb{R})) \longrightarrow (\mathbb{R},\mathcal{B}(\mathbb{R}))$, the sampling mean (computed with a realization of ${\bf X}$) converges with probability one (with respect to $\mu$) to the expectation of $f$ with respect to $\bar{\mu}_1$, i.e.,
\begin{equation}\label{eq_pr_main_ams&ergodic_1}
	\lim_{n \longrightarrow \infty}\frac{1}{n}\sum_{i=1}^{n} f(X_i) = \mathbb{E}_{X\sim \bar{\mu}_1} (f(X))<\infty , \mu-a.s.
\end{equation}
Therefore, we have that for any $B \in \mathcal{B}(\mathbb{R})$, 
\begin{align} \label{eq_pr_main_ams&ergodic_2a}
	&\lim_{n \longrightarrow \infty}\frac{1}{n}\sum_{i=1}^{n} {\bf 1}_{B}(X_i) = \bar{\mu}_1(B) , \mu-a.s.
\end{align}
In addition,   if $\int_{\mathbb{R}} \left| x \right|^p d\bar{\mu}_1(x) < \infty$
then for any $B \in \mathcal{B}(\mathbb{R})$, 
\begin{equation}\label{eq_pr_main_ams&ergodic_3}
	\lim_{n \longrightarrow \infty}\frac{1}{n}\sum_{i=1}^{n} {\bf 1}_{B}(X_i) \cdot \left| X_i \right|^p  = \int_{B}  \left| x \right|^p d\bar{\mu}_1(x)= \left| \left| (x^p)  \right|\right|_{L_1(\bar{\mu}_1)}, \mu-a.s.
\end{equation}
and, consequently, 
\begin{equation}\label{eq_pr_main_ams&ergodic_4}
	\lim_{n \longrightarrow \infty} \frac{\sum_{i=1}^{n} {\bf 1}_{B}(X_i) \cdot \left| X_i \right|^p}{\sum_{i=1}^{n}  \left| X_i \right|^p}  = v_p(B) = \frac{\int_{B}  \left| x \right|^p d\bar{\mu}_1(x)}{\left| \left| (x^p)  \right|\right|_{L_1(\bar{\mu}_1)}}, \mu-a.s.
\end{equation}

Let us define the tail distribution function of a probability $v$ in $(\mathbb{R}, \mathcal{B}(\mathbb{R}))$, which is an object that will play a relevant role 
in the argument to prove Theorem \ref{th_main_ams&ergodic}.
\begin{definition}\label{def_tail_function}
For $v \in \mathcal{P}(\mathbb{R})$ let us define its tail distribution function by $\phi_v(\tau) \equiv v(B_\tau)$ for all $\tau\in [0,\infty)$. 
\end{definition}

It is simple to verify that the following: 
\begin{proposition}\label{pro_tail_function_properties}
For any $v\in \mathcal{P}(\mathbb{R})$,  it follows that:
\begin{itemize}
	\item[i)] if $\tau_1>\tau_2$ then $\phi_v(\tau_1) \leq \phi_v(\tau_2)$  and  $\phi_v(\tau_1) = \phi_v(\tau_2)$ if, and only if, $v([\tau_2,\tau_1) \cup (-\tau_1,\tau_2])=0$, 
	\item[ii)] $\phi_v(0)=1$ and $\lim_{\tau \longrightarrow \infty} \phi_v(\tau)=0$, and
	\item[iii)] $(\phi_v(\tau))_{\tau\geq 0}$ is left continuous and $\phi^+_v(\tau)\equiv \lim_{t_n  \longrightarrow \tau, t_n>\tau} \phi_v(t_n)= \phi_v(\tau) - v(\left\{ \tau \right\} \cup \left\{ -\tau \right\} ).$
\end{itemize}
\end{proposition}
The proof is presented in \ref{proof_pro_tail_function_properties}.

\begin{figure}[h]
\centering
\includegraphics[width=1.0\textwidth]{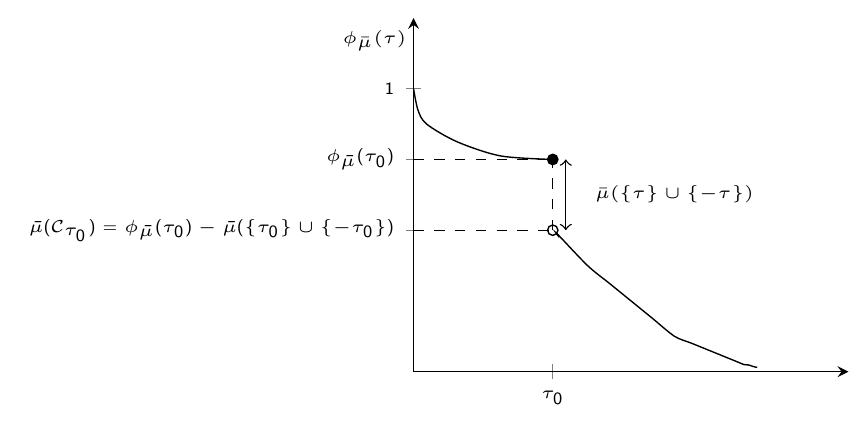}
\caption{Illustration of the tail distribution function $\phi_{\bar{\mu}}(\tau)$ of $\bar{\mu}\in \mathcal{P}(\mathbb{R})$ with a single discontinuous point at $\tau_o>0$ in $[0,\infty)$.}
\label{fig1}
\end{figure}

Therefore,  $(\phi_v(\tau))$ is a continuous function except on the points where $v$ has atomic mass (see Fig. \ref{fig1}). From a well-known result on real analysis \cite{royden_2010}, using the fact that  $(\phi_v(\tau))$ is non-decreasing, this function has at most a countable number of  discontinuities. This means that $v$ has at most a countable number of non-zero probability events on the collection $\left\{ \left\{ \tau \right\}\cup \left\{ -\tau\right\}, \tau\in [0,\infty) \right\} \subset \mathcal{B}(\mathbb{R})$ that we index and denote by $\mathcal{Y}_v \subset [0, \infty)$. 

By definition of $v_p$ (stated in (\ref{eq_ams_ergodic_sequence_3_b})), the discontinuity points of $(\phi_{\bar{\mu}_1}(\tau))$ and $(\phi_{v_p}(\tau))$  agree\footnote{There is only one exception when $\tau=0$.} from  the fact that if $\tau>0$ then $\bar{\mu}_1(\left\{ \tau \right\}\cup \left\{ -\tau\right\})=0 \Leftrightarrow  v_p(\left\{ \tau \right\}\cup \left\{ -\tau\right\})=0$. Therefore, we have that $\mathcal{Y}_{v_p} = \mathcal{Y}_{\bar{\mu}_1} \setminus \left\{ 0 \right\}$.  For the rest of the proof, 
it is relevant to consider the range of these tail functions.  These can be characterized  as follows (see Figure \ref{fig1}): 
\begin{align} \label{eq_pr_main_ams&ergodic_5a}
	\mathcal{R}^*_{\bar{\mu}_1} \equiv \left\{ \phi_{\bar{\mu_1}}(\tau), \tau\geq 0 \right\} &= (0,1] \setminus \bigcup_{\tau_n\in \mathcal{Y}_{\bar{\mu}_1}} [\bar{\mu}_1(C_{\tau_n}), \bar{\mu}_1(B_{\tau_n})),\\ 
		    \label{eq_pr_main_ams&ergodic_5b}
         \mathcal{R}^*_{v_p} \equiv  \left\{ \phi_{v_p}(\tau), \tau\geq 0 \right\} &= (0,1] \setminus \bigcup_{\tau_n\in \mathcal{Y}_{\bar{\mu}_1} \setminus \left\{ 0 \right\}} [v_p(C_{\tau_n}), v_p(B_{\tau_n})),
\end{align}
where $\mathcal{Y}_{\bar{\mu}_1}$ is either the empty set, a finite set,  or a countable set. 

With the tail functions $(\phi_{\bar{\mu}_1}(\tau))$ and $(\phi_{v_p}(\tau))$, we can introduce the collection 
\begin{align} \label{eq_pr_main_ams&ergodic_6}
	\left\{ (\phi_{\bar{\mu}_1}(\tau), \sqrt[p]{1-\phi_{v_p}(\tau)}), \tau\in [0,\infty) \right\}
\end{align}
that in its first coordinate covers the range $\mathcal{R}^*_{\bar{\mu}_1}$.  For  the non-continuous  case, i.e.,  $\left| \mathcal{Y}_{\bar{\mu}_1}  \right| >  0$, we can complete the range on the first coordinate to cover the non-achievable values $\bigcup_{\tau_n\in \mathcal{Y}_{\bar{\mu}_1}} [\bar{\mu}_1(C_{\tau_n}), \bar{\mu}_1(B_{\tau_n}))$ 
in (\ref{eq_pr_main_ams&ergodic_5a}) (Fig. \ref{fig1} illustrates this range when $\mathcal{Y}_{\bar{\mu}_1}=\left\{\tau_0 \right\}$) by the following simple extension:
\begin{align} \label{eq_pr_main_ams&ergodic_7}
	&\mathcal{F}_{\bar{\mu}_1} \equiv \left\{ (\phi_{\bar{\mu}_1}(\tau), \sqrt[p]{1-\phi_{v_p}(\tau)}), \tau\in [0,\infty) \right\}  \nonumber\\
	&\bigcup_{\tau_n\in \mathcal{Y}_{\bar{\mu}_1}} \left\{ ( \bar{\mu}_1(C_{\tau_n}) + \alpha \bar{\mu}_1(\left\{ -\tau_n, \tau_n \right\}), \sqrt[p]{1- v_p(C_{\tau_n})  - \alpha v_p(\left\{ -\tau_n,\tau_n  \right\}) }), \alpha \in [0,1) \right\}.
\end{align}
Importantly, we have the following simple result from the points in $\mathcal{F}_{\bar{\mu}_1}$: 
\begin{proposition}\label{rate_approximation_error_function}
The collection of pairs in $\mathcal{F}_{\bar{\mu}_1}$ defines a function from $(0,1]$ to $[0,1)$ that  we denote by $({f}_{\bar{\mu}_1}(r) )_{r\in (0,1]}$. 
\end{proposition}
\begin{proof}
For any $r\in (0,1]$ we have that either:  $r\in \mathcal{R}^*_{\bar{\mu}_1}$ in (\ref{eq_pr_main_ams&ergodic_5a}) for which there is a unique $\tau^*\geq 0$ such that $r=\phi_{\bar{\mu}_1}(\tau^*)$ and, consequently, there is only one $d=\sqrt[p]{1-\phi_{v_p}(\tau^*)}\in [0,1)$ such that $(r,d)\in \mathcal{F}_{\bar{\mu}_1}$, or   $r\in \bigcup_{\tau_n\in \mathcal{Y}_{\bar{\mu}_1}} [\bar{\mu}_1(C_{\tau_n}), \bar{\mu}_1(B_{\tau_n}))$,  for which there is a unique pair  $(\tau^*,\alpha^*) \in \mathcal{Y}_{\bar{\mu}_1}\times [0,1)$ such that $r=\bar{\mu}_1(C_{\tau^*}) + \alpha^* \bar{\mu}_1(\left\{ -\tau^* \right\}\cup \left\{\tau^* \right\})$ and, consequently, a unique $d=\sqrt[p]{1-\phi_{v_p}(\tau^*)  - \alpha^* v_p(\left\{ -\tau^* \right\}\cup \left\{\tau^* \right\}) }$ such that again $(r,d)\in \mathcal{F}_{\bar{\mu}_1}$. 
\end{proof}

In addition,  from the properties of  $(\phi_{\bar{\mu}_1}(\tau))$ and $(\phi_{v_p}(\tau))$, the following can be stated:
\begin{lemma}\label{lemma_rate_distorsion_properties}
The function $({f}_{\bar{\mu}_1}(r) )_{r\in (0,1]}$ induced by the set $\mathcal{F}_{\bar{\mu}_1}$ in (\ref{eq_pr_main_ams&ergodic_7}) has the following properties: 
\begin{itemize}
	\item[i)]  $({f}_{\bar{\mu}_1}(r))$ is continuous in the domain $r\in (0,1]$. 
	\item[ii)] $({f}_{\bar{\mu}_1}(r))$ is strictly decreasing in the domain $r\in {f}_{\bar{\mu}_1}^{-1}((0,1))\subset (0,1)$.
	 More precisely,  for any pair $(r_1,r_2)$ such that $0 < r_1 < r_2 \leq 1-\bar{\mu}(\left\{0 \right\})$ then  ${f}_{\bar{\mu}_1}(r_2) < {f}_{\bar{\mu}_1}(r_1)$.  
	 \item[iii)] ${f}_{\bar{\mu}_1}^{-1}((0,1))= (0,1-\bar{\mu}(\left\{0 \right\})) \subset (0,1)$. Consequently, ${f}_{\bar{\mu}_1}^{-1}((0,1))= (0,1)$
	if, and only if, $\bar{\mu}(\left\{0 \right\})=0$ (i.e., $0 \notin \mathcal{Y}_{\bar{\mu}_1}$).
	 \item[iv)] ${f}_{\bar{\mu}_1}(r)=0$  $\forall r\in [1-\bar{\mu}(\left\{0 \right\}),1]$ and $\lim_{r \rightarrow 0} {f}_{\bar{\mu}_1}(r)=1$. 
	\item[v)] The range of $({f}_{\bar{\mu}_1}(r))_{r\in (0,1]}$ is  $[0,1)$. 
\end{itemize}
\end{lemma}
The proof of this result is presented in \ref{proof_lemma_rate_distorsion_properties}.

We are in a position to prove the main result:
\subsection{Main Argument ---  Case  $\int_{\mathbb{R}} \left| x \right|^p d\bar{\mu}_1(x) < \infty$}
\label{sub_sec_proof_compressible_case}
\begin{proof} 							
Let us assume that $\int_{\mathbb{R}} \left| x \right|^p d\bar{\mu}_1(x) < \infty$.
Let us consider an arbitrary $r\in [1,0)$ and a sequence $(k_n)_{n\geq 1}$ such that $k_n/n \longrightarrow r$ as $n$ tends to infinity.  

{{\underline{Case 1: Continuous Scenario}}:}
Let us first consider the case where $r\in int(\mathcal{R}^*_{\bar{\mu}_1})$, i.e., $r\in \mathcal{R}^*_{\bar{\mu}_1} \setminus \left\{ \bar{\mu}_1(B_{\tau_n}) ,\tau_n\in \mathcal{Y}_{\bar{\mu}_1} \right\}   $,  and, consequently, there is $\tau_o$  such that $r=\phi_{\bar{\mu}_1}(\tau_o)$  being $\tau_o$ a continuous point of the tail function $\phi_{\bar{\mu}_1}(\cdot)$ (see iii) in Proposition \ref{pro_tail_function_properties}). 

Let us define  $n_\tau(x^n) \equiv \sum_{i=1}^n {\bf 1}_{B_\tau}(x_i)$, then using the (point-wise) ergodic theorem in (\ref{eq_pr_main_ams&ergodic_3}), it follows that for all $\tau\geq 0$
\begin{equation}\label{eq_pr_main_ams&ergodic_8}
	\lim_{n \rightarrow \infty}\frac{n_\tau(X^n)}{n}=\phi_{\bar{\mu}_1}(\tau), \mu-a.s.,
\end{equation}
and from (\ref{eq_pr_main_ams&ergodic_4}) and (\ref{eq_sec_pre_2})
\begin{equation}\label{eq_pr_main_ams&ergodic_9}
	\lim_{n \rightarrow \infty}  \tilde{\sigma}_p(n_\tau(X^n),X^n) = \sqrt[p]{1-\phi_{v_p}(\tau)}, \mu-a.s.
\end{equation}
In other words, we have the following family of (typical) sets: 
\begin{align} \label{eq_pr_main_ams&ergodic_10a}
	&\mathcal{A}^\tau \equiv  \left\{ (x_n)_{n\geq 0}, \lim_{n \rightarrow \infty}\frac{n_\tau(x^n)}{n}=\phi_{\bar{\mu}_1}(\tau)  \right\}\\
		     \label{eq_pr_main_ams&ergodic_10b}
	&\mathcal{B}^\tau\equiv \left\{ (x_n)_{n\geq 0},  \lim_{n \rightarrow \infty}  \tilde{\sigma}_p(n_\tau(x^n),x^n) = \sqrt[p]{1-\phi_{v_p}(\tau)} \right\},
\end{align}
satisfying that $\mu(\mathcal{A}^\tau \cap \mathcal{B}^\tau)=1$ for all $\tau\geq 0$.

Using the fact that $\phi_{\bar{\mu}_1}(\cdot)$ is continuous at $\tau_o$ and the observation that $\phi_{\bar{\mu}_1}(\cdot)$ has at most 
a countable number of discontinuities, there is $\delta \in (0,r)$ where the interval $(r-\delta,r+\delta)$ defines an open domain containing 
$\tau_o$, given by $(\tau_1,\tau_2)=\phi_{\bar{\mu}_1}^{-1}((r-\delta,r+\delta))$ where the function $\phi_{\bar{\mu}_1}(\cdot)$ is continuous (see Figure \ref{fig2}). Associated with this domain, we can consider $\left\{ \phi_{v_p}(\tau),  \tau \in (\tau_1,\tau_2) \right\}=(v_2,v_1)$ where by monotonicity $v_1=\phi_{v_p}(\tau_1) > v_2=\phi_{v_p}(\tau_2)$ (see Figure \ref{fig2} for an illustration). It is simple to show (by  the construction of $v_p$ from $\bar{\mu}_1$)\footnote{Note that for any $B\in \mathcal{B}(\mathbb{R})$ where $0 \notin B$,  $\bar{\mu}_1(B)=0$ if, and only if, $v_p(B)=0$.} that for any $\tau>0$ and $\epsilon>0$   
\begin{align}\label{eq_pr_main_ams&ergodic_11pre}
\phi_{\bar{\mu}_1}(\tau+\epsilon) < \phi_{\bar{\mu}_1}(\tau)\text{ if, and only if, } \phi_{v_p}(\tau+\epsilon) < \phi_{v_p}(\tau). 
\end{align}
Therefore,  this mutually absolute continuity property property between $\bar{\mu}_1$ and $v_p$ implies that 
\begin{align} 	\label{eq_pr_main_ams&ergodic_11a}
	&\phi_{\bar{\mu}_1}(\tau_1) >  \phi_{\bar{\mu}_1}(\tau_o)=r \Leftrightarrow  \phi_{v_p}(\tau_1)=v_1 >  \phi_{v_p}(\tau_o), \text{ and}\\
			\label{eq_pr_main_ams&ergodic_11b}
	&\phi_{\bar{\mu}_1}(\tau_2) <  \phi_{\bar{\mu}_1}(\tau_o)=r \Leftrightarrow  \phi_{v_p}(\tau_2)=v_2 <  \phi_{v_p}(\tau_o).
\end{align}
\begin{figure}[h]
\centering
\includegraphics[width=1.05\textwidth]{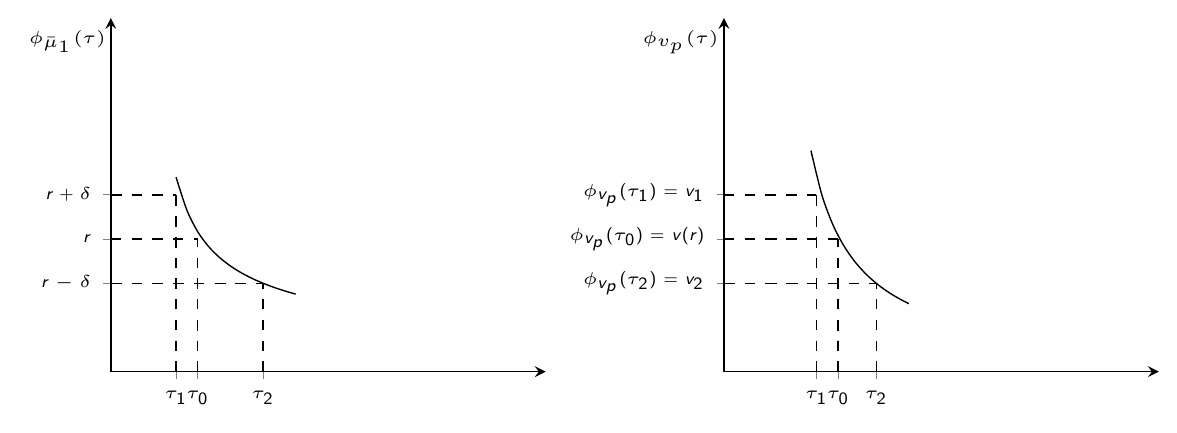}
\caption{Illustration of the tail distribution functions of $\bar{\mu}_1$ and $v_p$ at a continuous point $\tau_0$, where $r=\phi_{\bar{\mu}_1}(\tau_0)$.}
\label{fig2}
\end{figure}

We can then find $M>0$ sufficiently large such that for all $m\geq M$, $\phi_{v_p}(\tau_o)+1/m < v_1$. For any  of these $m\geq M$, 
there is $\tau_m\in (\tau_1,\tau_2)$ (from the continuity of $\phi_{v_p}(\cdot)$ in $(\tau_1,\tau_2)$) such that  $\phi_{v_p}(\tau_m) = \phi_{v_p}(\tau_o)+1/m$, where again by (\ref{eq_pr_main_ams&ergodic_11pre}) $\phi_{\bar{\mu}_1}(\tau_m) > \phi_{\bar{\mu}_1}(\tau_o)=r$. Therefore, for any $m\geq M$ and  any $(x_n)_{n\geq 1}\in \mathcal{A}^{\tau_m} \cap \mathcal{B}^{\tau_m}$,  the condition $n_{\tau_m}(x^n)>k_n$ is met eventually in $n$  as $n$ tends to infinity.  This comes from the assumption that $\lim_{n \rightarrow \infty}k_n/n = r < \phi_{\bar{\mu}_1}(\tau_m) $ and  the definition of $\mathcal{A}^{\tau_m}$ given in (\ref{eq_pr_main_ams&ergodic_10a}). Consequently,  under this context,   it follows that
\begin{align} 	\label{eq_pr_main_ams&ergodic_12}
	\tilde{\sigma}_{p}(n_{\tau_m}(x^n), x^n) \leq \tilde{\sigma}_{p}(k_n, x^n),  
\end{align}
is met eventually in $n$.
Finally,  using explicitly that $(x_n)_{n\geq 1}\in \mathcal{B}^{\tau_m}$ (see Eq.(\ref{eq_pr_main_ams&ergodic_10b})),  we have that 
\begin{align} 	\label{eq_pr_main_ams&ergodic_13}
     \sqrt[p]{1-(\phi_{v_p}(\tau_o)+1/m)} \leq \lim \inf_{n \rightarrow \infty} \tilde{\sigma}_{p}(k_n, x^n).
\end{align}
Repeating this argument,  if $(x_n)_{n\geq 1}\in \bigcap_{m\geq M} (\mathcal{A}^{\tau_m} \cap \mathcal{B}^{\tau_m})$, it follows 
from (\ref{eq_pr_main_ams&ergodic_13}) that\footnote{This is obtained by taking the supremum  ($m\geq M$) in the LHS of (\ref{eq_pr_main_ams&ergodic_13}) and using the continuity of the function $\sqrt[p]{1-x}$ in $x\in (0,1)$.}
\begin{align} 	\label{eq_pr_main_ams&ergodic_14}
    \sqrt[p]{1-\phi_{v_p}(\tau_o)} \leq \lim \inf_{n \rightarrow \infty} \tilde{\sigma}_{p}(k_n, x^n).
\end{align}
By the sigma additivity \cite{breiman_1968} and the fact that from the ergodic theorem $\mu(\mathcal{A}^{\tau_m} \cap \mathcal{B}^{\tau_m})=1$ for any $m\geq 1$,  it follows  that 
\begin{align} 	\label{eq_pr_main_ams&ergodic_15}
	\sqrt[p]{1-\phi_{v_p}(\tau_o)}  \leq \lim \inf_{n \rightarrow \infty} \tilde{\sigma}_{p}(k_n, X^n), \mu-a.s.
\end{align}

The exact  argument can be used to prove that 
\begin{align} 	\label{eq_pr_main_ams&ergodic_16}
	\lim \sup_{n \rightarrow \infty} \tilde{\sigma}_{p}(k_n, X^n) \leq \sqrt[p]{1-\phi_{v_p}(\tau_o)}, \mu-a.s..
\end{align}
by using the sequences $\tilde{\tau}_m$ such that $\phi_{v_p}(\tilde{\tau}_m) = \phi_{v_p}(\tau_o)-1/m$ for $m \geq \tilde{M}$ and 
$\tilde{M}$ sufficiently large. This part of the argument is omitted for the sake of space.
Finally, (\ref{eq_pr_main_ams&ergodic_15}) and (\ref{eq_pr_main_ams&ergodic_16}) prove the result in the continuous case.\footnote{The proof  assumes that $r<1$. The proof for the asymmetric case when $r=1\in int(\mathcal{R}^*_{\bar{\mu}_1})$, i.e., $r=1$ is a continuous point of $\phi_{\bar{\mu}_1}(\cdot)$ follows from the same argument presented above.   On the one hand,   $\tilde{\sigma}_{p}(k_n, x^n)\geq 0$ for any $k_n$ by definition. On the other hand,  the argument used to obtain (\ref{eq_pr_main_ams&ergodic_16}) follows  without any problem in this context,  implying that $\lim \sup_{n \rightarrow \infty} \tilde{\sigma}_{p}(k_n, X^n) \leq \sqrt[p]{1-\phi_{v_p}(\tau_o)}=0$ $\mu-a.s.$,  considering that $\tau_o=0$ in this case.}

{
\underline{Case 2: Discontinuous scenario:}}
Let us consider the case where $r\notin  \mathcal{R}^*_{\bar{\mu}_1}$ (see Eq.(\ref{eq_pr_main_ams&ergodic_5a})), which means that $\exists \tau_i\in \mathcal{Y}_{\bar{\mu}_1}$ such that 
\begin{align} 	\label{eq_pr_main_ams&ergodic_17}
	r\in [\bar{\mu}_1(C_{\tau_i}), \bar{\mu}_1(B_{\tau_i})), 
\end{align} 
(see the illustration in Fig. \ref{fig3}). For the moment let us assume that $r\in (\bar{\mu}_1(C_{\tau_i}), \bar{\mu}_1(B_{\tau_i}))$,\footnote{We left the case $r\in  \left\{ \bar{\mu}_1(C_{\tau_n}) : \tau_n \in \mathcal{Y}_{\bar{\mu}_1}\right\} \cup \left\{ \bar{\mu}_1(B_{\tau_n}): \tau_n \in \mathcal{Y}_{\bar{\mu}_1}\right\}$ for the mixed scenario below.} then there is a unique $\alpha_o\in (0,1)$ such that 
\begin{align} 	\label{eq_pr_main_ams&ergodic_18}
	r =  \bar{\mu}_1(C_{\tau_i}) +\alpha_o \cdot \bar{\mu}_1( \left\{ -\tau_i, \tau_i \right\}).
\end{align} 
\begin{figure}[h]
\centering
\includegraphics[width=1.0\textwidth]{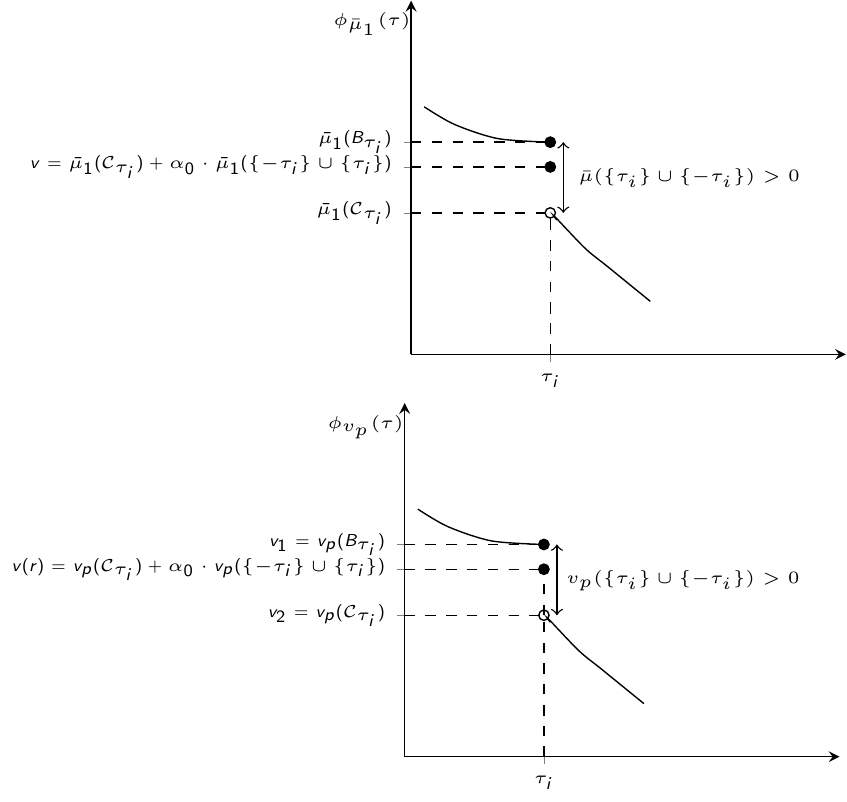}
\caption{Illustration of the tail distribution functions of $\bar{\mu}_1$ and $v_p$ at a discontinuous point $\tau_i>0$.}
\label{fig3}
\end{figure}

Here we need to use an extended version of the point-wise ergodic theorem in (\ref{eq_pr_main_ams&ergodic_1}). For that, let us introduce an  i.i.d. Bernoulli process ${\bf Y}=(Y_i)_{i\geq 1}$ of parameter $\rho\in [0,1]$,  where  $\mathbb{P}(Y_i=1)=\rho$ for all $i\geq 1$, 
that is independent of ${\bf X}=(X_n)_{n\geq 1}$. Let us denote by $\eta$ its (i.i.d) process distribution 
in ${ \left\{0,1\right\}}^\mathbb{N}$. Then,  from the ergodic theorem for AMS process in (\ref{eq_pr_main_ams&ergodic_1}) it follows,  as a natural extension of (\ref{eq_pr_main_ams&ergodic_2a}), that for all $\tau\geq 0$
\begin{align} \label{eq_pr_main_ams&ergodic_19a}
	&\lim_{n \longrightarrow \infty} \frac{1}{n}\sum_{i=1}^{n} {\bf 1}_{C_\tau}(X_i)  + \frac{1}{n}\sum_{i=1}^{n} {\bf 1}_{\left\{-\tau,  \tau \right\}}(X_i) \cdot Y_i  = \bar{\mu}_1(C_\tau) + \bar{\mu}_1( \left\{ -\tau, \tau \right\}) \rho,\\
	\label{eq_pr_main_ams&ergodic_19b}
	&\lim_{n \longrightarrow \infty}\frac{\sum_{i=1}^{n} ({\bf 1}_{C_\tau}(X_i) + {\bf 1}_{\left\{ -\tau, \tau\right\}}(X_i) Y_i) \left| X_i \right|^p}{\sum_{i=1}^{n}  \left| X_i \right|^p}  = \frac{\int_{C_\tau}  \left| x \right|^p d\bar{\mu}_1(x)  + \bar{\mu}_1( \left\{-\tau,  \tau \right\})  \tau^p  \rho}{\left| \left| (x^p)  \right|\right|_{L_1(\bar{\mu}_1)}},
\end{align}
with probability one with respect to joint process distribution of $({\bf X}, {\bf Y})$ denoted by $\mu \times \eta$.

Returning to the argument, let us consider an arbitrary $(k_n)_{\geq 1}$ such that $k_n/n \longrightarrow r$ as $n$ tends to 
infinity. Let us consider $\alpha_m \equiv \alpha_o+1/m$ and $M$ sufficiently large to make $\alpha_M<1$. 
For  any $m\geq M$, let us construct an auxiliary i.i.d. Bernoulli process ${\bf Y(\alpha_m)} \equiv (Y_i)_{i\geq 1}$, where 
$\mathbb{P}(Y_i=1)=\alpha_m$. The process distribution of ${\bf Y(\alpha_m)}$ is denoted by $\eta_{m}$. In this context,  
if we define the joint count function 
$$n_\tau(x^n,y^n)\equiv \sum_{i=1}^{n} {\bf 1}_{C_\tau}(x_i)  +\sum_{i=1}^{n} {\bf 1}_{\left\{ \tau \right\}\cup  \left\{ -\tau \right\}}(x_i) \cdot y_i$$ 
it follows from (\ref{eq_pr_main_ams&ergodic_19a}) and (\ref{eq_pr_main_ams&ergodic_19b})
that for $\tau_i$ introduced in (\ref{eq_pr_main_ams&ergodic_17}), 
\begin{align} \label{eq_pr_main_ams&ergodic_20a}
	&\lim_{n \longrightarrow \infty} \frac{n_{\tau_i}(X^n,Y^n)}{n} = \bar{\mu}_1(C_{\tau_i}) + (\alpha_o+1/m)\cdot \bar{\mu}_1( \left\{-\tau_i,  \tau_i \right\}),\\
	\label{eq_pr_main_ams&ergodic_20b}
	&\lim_{n \longrightarrow \infty} \tilde{\sigma}_p(n_{\tau_i}(X^n,Y^n),X^n)=\sqrt[p]{1-(v_p(C_{\tau_i}) - (\alpha_o+1/m)\cdot v_p( \left\{ -\tau_i, \tau_i \right\}))}
\end{align}
$\mu \times \eta_m$ -almost surely. Importantly in (\ref{eq_pr_main_ams&ergodic_20b}),  $v_p( \left\{ -\tau_i, \tau_i \right\})>0$ from the fact that $\bar{\mu}_1( \left\{ -\tau_i,  \tau_i \right\})>0$.\footnote{Here we assume that $\tau_i>0$. The important sparse case when $\tau_i=0$ will be treated below.}  

Let us consider an arbitrary (typical) sequence $((x_n)_{n\geq1}, (y_n)_{n\geq1})$ satisfying the limiting  conditions in (\ref{eq_pr_main_ams&ergodic_20a}) and (\ref{eq_pr_main_ams&ergodic_20b}).  From (\ref{eq_pr_main_ams&ergodic_20a}), it follows that the condition
$n_\tau(x^n,y^n)> k_n$ happens eventually in $n$ as $k_n/n\longrightarrow r =\bar{\mu}_1(C_{\tau_i}) + \alpha_o\cdot \bar{\mu}_1( \left\{ -\tau_i, \tau_i \right\}) < \bar{\mu}_1(C_{\tau_i}) + (\alpha_o+1/m)\cdot \bar{\mu}_1( \left\{ -\tau_i,  \tau_i \right\})$ by construction. Therefore, the condition
\begin{align} \label{eq_pr_main_ams&ergodic_21}
	\tilde{\sigma}_p(n_\tau(x^n,y^n),x^n) \leq  \tilde{\sigma}_p(k_n,x^n) \text{ holds eventually in $n$}.   
\end{align}
The left hand side of (\ref{eq_pr_main_ams&ergodic_21})  converges to $\sqrt[p]{1-(v_p(C_{\tau_i}) - (\alpha_o+1/m)\cdot v_p( \left\{ -\tau_i, \tau_i \right\}))}$ as $n$ tends to infinity by the construction of  $((x_n)_{n\geq1}, (y_n)_{n\geq1})$ and (\ref{eq_pr_main_ams&ergodic_20b}). Finally,  by the almost sure convergence in (\ref{eq_pr_main_ams&ergodic_20a}) and (\ref{eq_pr_main_ams&ergodic_20b}),  it follows that
\begin{align} \label{eq_pr_main_ams&ergodic_22}
	\lim \inf_{n \longrightarrow \infty}  \tilde{\sigma}_p(k_n,X^n) \geq  \sqrt[p]{1-(v_p(C_{\tau_i}) - (\alpha_o+1/m)\cdot v_p( \left\{ -\tau_i, \tau_i \right\}))}, 
\end{align}
$\mu$- almost surely.\footnote{We remove the dependency on $\eta_m$, as both terms in (\ref{eq_pr_main_ams&ergodic_22}) (in the limit)  turn out to be independent of the auxiliary process $(Y_i)_{i\geq 1}$.} 

Let us denote by $\mathcal{D}^{\tau_m} \equiv \left\{ (x_n)_{n\geq 1} \text{, where (\ref{eq_pr_main_ams&ergodic_22}) holds}\right\}$. From (\ref{eq_pr_main_ams&ergodic_22}), $\mu(\mathcal{D}^{\tau_m})=1$ and by sigma-additivity \cite{breiman_1968},  it follows that $\mu(\cap_{m\geq M}\mathcal{D}^{\tau_m})=1$, which implies that 
\begin{align} \label{eq_pr_main_ams&ergodic_23}
	\lim \inf_{n \longrightarrow \infty}  \tilde{\sigma}_p(k_n,X^n) \geq  \sqrt[p]{1-(v_p(C_{\tau_i}) - \alpha_o \cdot v_p( \left\{ -\tau_i, \tau_i \right\}))}, \mu-a.s.
\end{align}

To conclude, an equivalent (symmetric) argument can be used to prove  that
\begin{align} \label{eq_pr_main_ams&ergodic_24}
	\lim \sup_{n \longrightarrow \infty}  \tilde{\sigma}_p(k_n,X^n) \leq  \sqrt[p]{1-(v_p(C_{\tau_i}) - \alpha_o \cdot v_p( \left\{-\tau_i,  \tau_i \right\}))}, \mu-a.s., 
\end{align}
using $\tilde{\alpha}_m \equiv \alpha_o-1/m$ and $\tilde{M}$ sufficiently large to make $\tilde{\alpha}_{\tilde{M}}>0$. For the sake of space,  the proof of this part is omitted. This concludes the result in this case.  

{
 \underline{Case 3: Mixed scenario:}}
Here we consider the scenario where $$r\in  \left\{ \bar{\mu}_1(B_{\tau_n}), \bar{\mu}_1(C_{\tau_n}) : \tau_n \in \mathcal{Y}_{\bar{\mu}_1}\right\}.$$ 
The proof reduces to the same procedure presented above in the continuous and discontinues scenarios,  but adopted in a mixed form. 
A  sketch with the  basic steps of the argument 
is provided here as no new technical elements are needed. 

For $r= \bar{\mu}_1(B_{\tau_i})$ for  $\tau_i \in \mathcal{Y}_{\bar{\mu}_1}$ and $\tau_i\neq 0$, the same argument adopted in the continuous case (to obtain (\ref{eq_pr_main_ams&ergodic_15})) can be adopted here to obtain that
\begin{align} 	\label{eq_pr_main_ams&ergodic_25}
	\lim \inf_{n \rightarrow \infty} \tilde{\sigma}_{p}(k_n, X^n)  \geq \sqrt[p]{1-\phi_{v_p}(\tau_i)}, \mu-a.s.,
\end{align}
for any sequence $(k_n)_{n\geq 1}$ such that $k_n/n \longrightarrow \bar{\mu}_1(B_{\tau_i})$.
For the other inequality, the strategy with the auxiliary Bernoulli process presented in the proof of the discontinuous case 
can be adopted considering $\alpha_o=1$ and $\tilde{\alpha}_m=1-1/m$ for $m$ sufficiently large. Then, a 
result equivalent to (\ref{eq_pr_main_ams&ergodic_24}) is obtained,  meaning in this specific context that
\begin{align} 	\label{eq_pr_main_ams&ergodic_26}
	\lim \sup_{n \rightarrow \infty} \tilde{\sigma}_{p}(k_n, X^n)  \leq \sqrt[p]{1-\phi_{v_p}(\tau_i)}, \mu-a.s.
\end{align}

For $r= \bar{\mu}_1(C_{\tau_i})$ for  $\tau_i \in \mathcal{Y}_{\bar{\mu}_1}$ and $\tau_i\neq 0$, 
the same argument with the auxiliary Bernoulli process used to obtain  (\ref{eq_pr_main_ams&ergodic_23})
can be adopted here, considering $\alpha_o=0$ and $\alpha_m=1/m$ for $m$ sufficiently large, to obtain that
\begin{align} \label{eq_pr_main_ams&ergodic_27}
	\lim \inf_{n \longrightarrow \infty}  \tilde{\sigma}_p(k_n,X^n) \geq  \sqrt[p]{1- v_p(C_{\tau_i}) }, \mu-a.s.,
\end{align}
for any sequence $(k_n)_{n\geq 1}$ such that $k_n/n \longrightarrow \bar{\mu}_1(C_{\tau_i})$.
For the other inequality, the argument of the continuous case proposed to obtain (\ref{eq_pr_main_ams&ergodic_16}) can be 
adopted here (with no differences) to obtain that
\begin{align} \label{eq_pr_main_ams&ergodic_28}
	\lim \sup_{n \longrightarrow \infty}  \tilde{\sigma}_p(k_n,X^n) \leq  \sqrt[p]{1- v_p(C_{\tau_i}) }, \mu-a.s.
\end{align}

{\em 
\underline{Case 4: The case when $0\in \mathcal{Y}_{\bar{\mu}_1}$:}
}
This case deserves a special treatment because it offers some insights about a property of the function $(f_{p,\bar{\mu}}(r))_{r\in (0,1]}$ when $\bar{\mu}_1(\left\{ 0\right\})>0$, i.e., $\bar{\mu}_1$ has atomic mass at $0$. 

Let us consider the case that $\bar{\mu}_1(\left\{ 0\right\})=\rho_o>0$, then $\phi^+_{\bar{\mu}_1}(0) = \lim_{\tau \rightarrow 0} \phi_{\bar{\mu}_1}(\tau) = \bar{\mu}_1(C_{0})=1-\rho_o\in (0,1)$ (see, the illustration in Fig. \ref{fig4}). On the other hand, we have that $v_p(\left\{ 0\right\})=\frac{0^p\cdot \bar{\mu}_1(\left\{ 0\right\})}{\left| \left| (x^p)  \right|\right|_{L_1(\bar{\mu}_1)}}=0$. Therefore,  $(\phi_{v_p}(\tau))$ is continuous at $\tau=0$ (Fig. \ref{fig4}). From the fact that $\mathcal{Y}_{\bar{\mu}_1}$ is at most a countable set, there is $\tau_1>0$ with $\phi_{\bar{\mu}_1}(\tau_1)<1-\rho_o$ where $\phi_{\bar{\mu}_1}(\cdot)$ is continuous in $(0,\tau_1)$ and, consequently, so is $\phi_{v_p}(\cdot)$ in $(0,\tau_1)$ (from Proposition \ref{join_dependency_of_tail_functions} in 
\ref{proof_lemma_rate_distorsion_properties}). If we consider the range of $\phi_{v_p}(\cdot)$ in this continuous domain,  we have that $\left\{\phi_{v_p}(\tau), \tau\in (0,\tau_1) \right\}=(v_1,1)$ where $v_1=\phi_{v_p}(\tau_1)<1$.
\begin{figure}[h]
\centering
\includegraphics[width=1.05\textwidth]{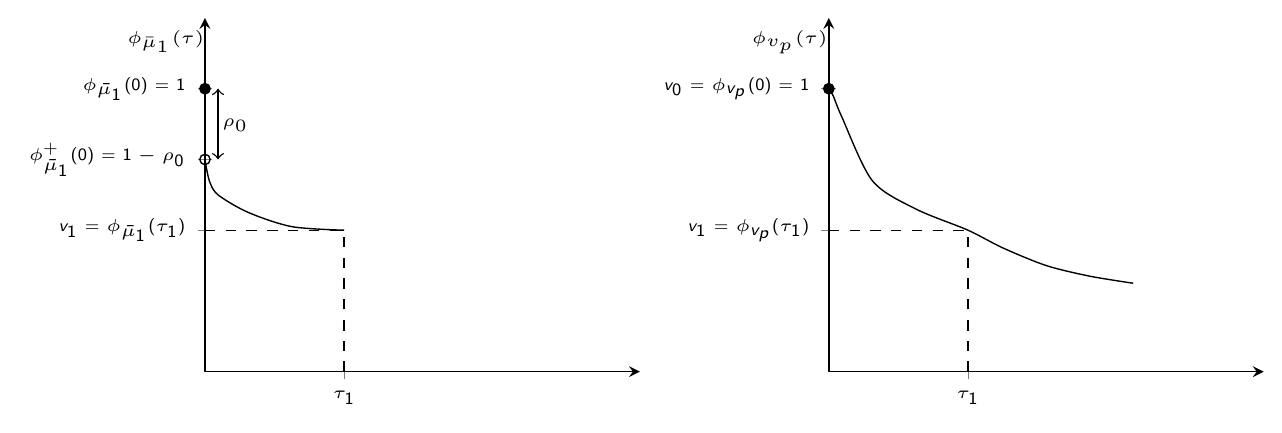}
\caption{Illustration of the tail distribution functions of $\bar{\mu}_1$ and $v_p$ in the sparse case where 
${\bar{\mu}_1}(\left\{ 0 \right\})=\rho_0>0$.}
\label{fig4}
\end{figure}

Here,  we adopt the same argument used in the continuous scenario to obtain the upper bound in (\ref{eq_pr_main_ams&ergodic_16}). Let us consider an arbitrary sequence $(k_n)_{n\geq 1}$, such that 
$k_n/n \longrightarrow 1-\rho_o$ with $n$. By the continuity of $\phi_{v_p}(\tau)$ in $(0,\tau_1)$ for 
any $m$ sufficiently large such that $1-\frac{1}{m}<v_1$, there is $\tau_m>0$ such that $\phi_{v_p}(\tau_m)=1-\frac{1}{m}$. For  any of these $\tau_m$,  it follows that $\phi_{\bar{\mu}_1}(\tau_m)<1-\rho_o$.\footnote{This from the fact that if $\phi_{v_p}(\tau) < \phi_{v_p}(\tilde{\tau})$ then $\phi_{\bar{\mu}_1}(\tau) < \phi_{\bar{\mu}_1}(\tilde{\tau})$ from the definition of $v_p$.} Then, we can consider  the set of typical sequences defined in (\ref{eq_pr_main_ams&ergodic_10a}) and (\ref{eq_pr_main_ams&ergodic_10b}), where 
if $(x_n)_{n\geq 1}\in \mathcal{A}^{\tau_m}\cap \mathcal{B}^{\tau_m}$  then  eventually in $n$ it follows that $k_n > n_{\tau_m}(x^n)$  (from the fact that  $\phi_{\bar{\mu}_1}(\tau_m)< 1-\rho_o$) and, consequently, 
\begin{align} \label{eq_pr_main_ams&ergodic_29}
	\lim \sup_{n \longrightarrow \infty}  \tilde{\sigma}_p(k_n,x^n) \leq  \sqrt[p]{1/m}.
\end{align}
This last result follows from the definition of $\mathcal{B}^{\tau_m}$ and the construction of $\tau_m$ (i.e., $\phi_{v_p}(\tau_m)=1-\frac{1}{m}$). Then if $(x_n)_{n\geq 1}\in \bigcap_{m\geq M} (\mathcal{A}^{\tau_m}\cap \mathcal{B}^{\tau_m})$, where $M>0$ is set such that $1-\frac{1}{M}<v_1$,  then
\begin{align} \label{eq_pr_main_ams&ergodic_30}
	\lim \sup_{n \longrightarrow \infty}  \tilde{\sigma}_p(k_n,x^n) \leq 0.
\end{align}
Finally,  from the (point-wise) ergodic theorem for AMS sources in (\ref{eq_pr_main_ams&ergodic_1}), it follows that $\mu(\bigcap_{m\geq M} \mathcal{A}^{\tau_m}\cap \mathcal{B}^{\tau_m})=1$, meaning from (\ref{eq_pr_main_ams&ergodic_30}) that $\lim_{n \longrightarrow \infty}  \tilde{\sigma}_p(k_n,X^n) = 0$, $\mu$-almost surely.  

The last observation to conclude this part is that if $(\tilde{k}_n)$ dominates $(k_n)$,  in the sense that  $\tilde{k}_n\geq k_n$ eventually, then from definition $\tilde{\sigma}_p(\tilde{k}_n,x^n) \leq \tilde{\sigma}_p({k}_n,x^n)$ for all $x^n$. Therefore from (\ref{eq_pr_main_ams&ergodic_30}),  for any $r\in [1-\rho_o,1]$ and for any $(k_n)$ such that $k_n/n \longrightarrow r$, it follows that 
\begin{align} \label{eq_pr_main_ams&ergodic_30b}
	\lim_{n \longrightarrow \infty}  \tilde{\sigma}_p(k_n,X^n) = 0, \ \mu-a.s.
\end{align}
Then we obtain in this case that $\forall r \in [1-\bar{\mu}_1(\left\{ 0\right\}),1]$
\begin{equation}\label{eq_ams_ergodic_sequence_6}
		f_{p,\bar{\mu}}(r)=0,
	\end{equation}
while $f_{p,\bar{\mu}}(r)>0$ if $r\in (0, 1-\bar{\mu}_1(\left\{ 0\right\}))$. 
\end{proof} 							

\begin{remark}  The result in (\ref{eq_pr_main_ams&ergodic_30b}) is consistent with the statement of Theorem \ref{th_main_ams&ergodic} because  if $r\in [1-\rho_o,1]$, then it can be written as $r=\bar{\mu}_1((0,\infty)) + \alpha \cdot \bar{\mu}_1(\left\{ 0\right\})$ for some $\alpha\in [0,1]$ where  $f_{p,\bar{\mu}}(r)=\sqrt[p]{1-v_p((0,\infty)) -\alpha \cdot v_p(\left\{ 0\right\})}=0$.
\end{remark}

\subsection{Main Argument --- Case $\int_{\mathbb{R}} \left| x \right|^p d\bar{\mu}_1(x) = \infty$}
\begin{proof} 							
When 
$\int_{\mathbb{R}} \left| x \right|^p d\bar{\mu}_1(x) = \infty$, 
it follows that  $\forall \tau\geq 0$, 
\begin{equation}\label{eq_pr_main_ams&ergodic_31}
	\lim_{n \longrightarrow \infty} \frac{\sum_{i=1}^{n} (1-{\bf 1}_{B_\tau}(X_i)) \cdot \left| X_i \right|^p}{\sum_{i=1}^{n}  \left| X_i \right|^p}  = 0, \mu-a.s..
\end{equation}
this from the (point-wise) ergodic theorem in (\ref{eq_pr_main_ams&ergodic_1}) and the fact that 
$\int_{\mathbb{R}}  \left| x \right|^p d\bar{\mu}_1(x)=\infty$. Then,  from (\ref{eq_pr_main_ams&ergodic_2a}) and (\ref{eq_pr_main_ams&ergodic_31}), it follows in this case that 
\begin{align} \label{eq_pr_main_ams&ergodic_32}
	&\lim_{n \rightarrow \infty}\frac{n_\tau(X^n)}{n}=\phi_{\bar{\mu}_1}(\tau), \mu-a.s.,\\
	&\lim_{n \rightarrow \infty}  \tilde{\sigma}_p(n_\tau(X^n),X^n) =0, \mu-a.s.
\end{align}
for all $\tau\geq 0$. Again  we can consider $\mathcal{A}^\tau =  \left\{ (x_n)_{n\geq 1}, \lim_{n \rightarrow \infty}\frac{n_\tau(x^n)}{n}=\phi_{\bar{\mu}_1}(\tau)  \right\}$ and 
$$\mathcal{B}^\tau = \left\{ (x_n)_{n\geq 1},  \lim_{n \rightarrow \infty}  \tilde{\sigma}_p(n_\tau(x^n),x^n) = 0 \right\},$$ 
where $\mu(\mathcal{A}^\tau \cap \mathcal{B}^\tau)=1$ for all $\tau$.  

Let us fix $r\in (0,1]$ and $(k_n)_{n\geq 1}$ such that $k_n/n \longrightarrow r$. We can consider $\bar{r}<r$ and $\tau_o$,  such  that $\phi_{\bar{\mu}_1}(\tau_o)=\bar{r}$.  Then, for any $(x_n)_{n\geq 1} \in \mathcal{A}^{\tau_o}\cap \mathcal{B}^{\tau_o}$, it follows that the condition $k_n> n_{\tau_o}(x^n)$ happens eventually in $n$ (from the fact that $r>\bar{r}$ and the definition of $\mathcal{A}^{\tau_o}$), therefore eventually we have that $\tilde{\sigma}_p(n_\tau(x^n),x^n) \geq \tilde{\sigma}_p(k_n,x^n)$. Finally, from the definition of  $\mathcal{B}^{\tau_o}$, $\lim_{n \rightarrow \infty}  \tilde{\sigma}_p(k_n,x^n) = 0$.  The proof concludes
noting that $\mu(\mathcal{A}^{\tau_o} \cap \mathcal{B}^{\tau_o})=1$.
\end{proof}

\section{Proof of Theorem \ref{th_main_ams}} 
\label{proof_th_main_ams}
	First, we introduce formally the ergodic decomposition (ED) theorem: 
	\begin{theorem}\label{th_ed}\cite[Th. 10.1]{gray_2009}
		Let ${\bf X}=(X_n)_{n\geq 1}$ be an  AMS process characterized by $(\mathbb{R}^{\mathbb{N}},\mathcal{B}(\mathbb{R}^{\mathbb{N}}), \mu)$.  Then there is a measurable space given by $(\Lambda, \mathcal{L})$ that parametrizes the family of 
		stationary and ergodic distribution, i.e., $\tilde{\mathcal{P}}=\left\{\mu_\lambda, \lambda\in \Lambda \right\}$, and a
		measurable function $\Psi: (\mathbb{R}^{\mathbb{N}},\mathcal{B}(\mathbb{R}^{\mathbb{N}})) \rightarrow (\Lambda, \mathcal{L})$ such that: 
		\begin{itemize}
			\item[i)] $\Psi$ is invariant with respect to $T$, i.e., $\Psi({\bf x})=\Psi(T({\bf x}))$ for all ${\bf x} \in \mathbb{R}^{\mathbb{N}}$.
			\item[ii)] Using the stationary mean $\bar{\mu}$ of ${\bf X}$ and its induced probability in $(\Lambda, \mathcal{L})$, denoted by $W_{\Psi}$, it follows that  $\forall F\in \mathcal{B}(\mathbb{R}^{\mathbb{N}})$
			\begin{equation}\label{eq_proof_th_main_ams_1}
				\bar{\mu}(F) = \int \mu_{\lambda}(F) \partial W_{\Psi}(\lambda).
			\end{equation}
			\item[iii)] Finally,\footnote{This result can be interpreted as a more sophisticated re-statement of the point-wise ergodic theorem for AMS sources under the assumption of a standard space, which is the case for $(\mathbb{R}^{\mathbb{N}}, \mathcal{B}(\mathbb{R}^{\mathbb{N}}))$. Details and the interpretations of this result are presented in \cite[Chs. 8 and 10]{gray_2009}.} for any $L_1(\bar{\mu})$-integrable and measurable function $f: (\mathbb{R}^{\mathbb{N}},\mathcal{B}(\mathbb{R}^{\mathbb{N}})) \rightarrow (\mathbb{R}, \mathcal{B}(\mathbb{R}))$, 
			\begin{equation}\label{eq_proof_th_main_ams_2}
				\lim_{n \rightarrow \infty} \frac{1}{n} \sum_{i=0}^{n-1} f(T^i({\bf X}))  = \mathbb{E}_{{\bf Z} \sim \mu_{\Psi({\bf X})}} \left( f({\bf Z}) \right),\  \mu-\text{almost surely},
			\end{equation}
			where ${\bf Z}$ in (\ref{eq_proof_th_main_ams_2}) 
			denotes a stationary  and ergodic process in $(\mathbb{R}^{\mathbb{N}},\mathcal{B}(\mathbb{R}^{\mathbb{N}}))$ 
			with process distribution given by $\Psi({\bf X}) \in \Lambda$.
		\end{itemize}
	\end{theorem}
	\begin{proof}
	
	Let us first prove the almost sure sample-wise convergence in (\ref{eq_subsec_non_ergodic_case_2}). For $r\in (0,1]$
	and $(k_n)_{n\geq 1}$ such that $k_n/n \rightarrow r$, we need to study the limit of the following random object $Y_n = \tilde{\sigma}_p(k_n, X^n_1)$.  As in the proof of Theorem \ref{th_main_ams&ergodic}, we consider the tail events  
	\begin{align}\label{eq_proof_th_main_ams_3}
		B_\tau = (-\infty, - \tau] \cup [\tau, \infty) \text{ and } 
		C_\tau = (-\infty, - \tau) \cup (\tau, \infty)
	\end{align}
	for $\tau \geq 0$. From Theorem \ref{th_ed}, it follows that for any $B \in \mathcal{B}(\mathbb{R})$, 
	\begin{align} \label{eq_proof_th_main_ams_4}
		&\lim_{n \longrightarrow \infty}\frac{1}{n}\sum_{i=1}^{n} {\bf 1}_{B}(X_i) = \mathbb{E}_{{\bf Z} \sim \mu_{\Psi({\bf X})}} ({\bf 1}_{B}(Z_1))  = \mu_{1,\Psi({\bf X})}(B) , \mu-a.s.
	\end{align}
	where ${\bf Z}=(Z_i)_{i\geq 1}$ and $\mu_{1,\Psi({\bf X})}$ denotes the probability of $Z_1$ (the 1D marginalization of the process distribution $\mu_{\Psi({\bf X})}$) in $(\mathbb{R}, \mathcal{B}(\mathbb{R}))$.
	In addition, from Theorem \ref{th_ed}  we have that
	\begin{equation}\label{eq_proof_th_main_ams_5}
		\lim_{n \longrightarrow \infty} \frac{\sum_{i=1}^{n} {\bf 1}_{B_\tau}(X_i) \cdot \left| X_i \right|^p}{\sum_{i=1}^{n}  \left| X_i \right|^p}  = \xi_p({\bf X},B_\tau), \  \mu-a.s.,
	\end{equation}
	where 
	\begin{equation}\label{eq_proof_th_main_ams_6}
		\xi_p({\bf X},B_\tau) \equiv   
					\left\{ \begin{array}{ll}
					\frac{\int_{B_\tau}  \left| x \right|^p d\mu_{1,\Psi({\bf X})}(x)}{\left| \left| (x^p)  \right|\right|_{L_1(\mu_{1,\Psi({\bf X})})}} &\text{ if $(x^p)_{x\in \mathbb{R}} \in L_1(\mu_{1,\Psi({\bf X})})$}\\
		 			 1 &\text{ if $(x^p)_{x\in \mathbb{R}} \notin L_1(\mu_{1,\Psi({\bf X})})$}
		 			\end{array} \right..
	\end{equation}
	From the results in (\ref{eq_proof_th_main_ams_4}) and (\ref{eq_proof_th_main_ams_5}), we can proceed with the same arguments 
	used  in the proof of Theorem \ref{th_main_ams&ergodic} to obtain that\footnote{We omit the argument here as it is redundant, following directly the structure presented in Section \ref{proof_th_main_ams&ergodic}.}
	\begin{equation}\label{eq_proof_th_main_ams_7}
		\lim_{n \longrightarrow  \infty} \tilde{\sigma}_p(k_n, X^n_1) =  
														f_{p,\mu_{\Psi({\bf X})}}(r), \ \mu-\text{almost surely},
	\end{equation}
	where $f_{p,\mu_{\Psi({\bf X})}}(r)$ is the almost-sure asymptotic limit of the stationary and ergodic component 
	$\mu_{\Psi({\bf X})}\in \tilde{\mathcal{P}}$ stated in (\ref{eq_ams_ergodic_sequence_4}) and elaborated in  the statement of Theorem \ref{th_main_ams&ergodic}. This proves the first part of the result. 
	
	For the second part, we consider again $r\in (0,1]$ and $(k_n)_{n\geq 1 }$ such that $k_n/n \rightarrow r$.  
	Let us denote the almost sure limit in (\ref{eq_proof_th_main_ams_7}) by $f_{p}({\bf X},r)$\footnote{We omit the dependency on $\mu$ in the notation because this limit  (as a random variable of $\mathbf{X}$) is  independent of $\mu$.}, 
	which is in general a random variable from $(\mathbb{R}^{\mathbb{N}},\mathcal{B}(\mathbb{R}^{\mathbb{N}}))$
	to $(\mathbb{R},\mathcal{B}(\mathbb{R}))$. For an arbitrary $d\in [0,1)$, we need to analyze the asymtotic limit 
	of $\mu_n(A_d^{n,k_n})$. By additivity, we decompose this probability in two terms: 
	\begin{align}\label{eq_proof_th_main_ams_8}
		\mu_n(A_d^{n,k_n}) =& \mu \left( \left\{ \bar{x}=(x_i)_{i\geq 1} \in \mathbb{R}^{\mathbb{N}}: \tilde{\sigma}_p(k_n, x^n_1) \leq d,   f_{p}(\bar{x},r) \leq d \right\} \right)\nonumber\\ 
						+& \mu \left( \left\{ \bar{x} \in \mathbb{R}^{\mathbb{N}}: \tilde{\sigma}_p(k_n, x^n_1) \leq d,   f_{p}(\bar{x},r) > d \right\} \right), 
 	\end{align}
	For the first term (from left to right) in the RHS of (\ref{eq_proof_th_main_ams_8}), 
	we can consider the following bounds
	\begin{align}\label{eq_proof_th_main_ams_9}
		\mu \left( \left\{ \bar{x}: f_{p}(\bar{x},r) \leq d \right\} \right) &\geq \mu \left( \left\{ \bar{x}: \tilde{\sigma}_p(k_n, x^n_1) \leq
		 d,   f_{p}(\bar{x},r) \leq d \right\} \right) \geq \nonumber\\
		 &\mu \left( \left\{ \bar{x}: \tilde{\sigma}_p(k_n, x^n_1) \leq f_{p,\mu}(\bar{x},r),   f_{p}(\bar{x},r) \leq d \right\} \right). 
	\end{align}
	The lower and upper bounds in (\ref{eq_proof_th_main_ams_9}) have the same asymptotic limit, i.e.,  
	\begin{align}\label{eq_proof_th_main_ams_10}
	\lim_{n \longrightarrow \infty} \mu \left( \left\{ \bar{x}: \tilde{\sigma}_p(k_n, x^n_1) \leq f_{p}(\bar{x},r),   f_{p}(\bar{x},r) \leq d \right\} \right)= \lim_{n \longrightarrow \infty} \mu \left( \left\{ \bar{x}: f_{p}(\bar{x},r) \leq d \right\} \right).
	\end{align}
	This can be shown by the following equality
	\begin{align}\label{eq_proof_th_main_ams_10b}
	\mu \left( \left\{ \bar{x}: f_{p}(\bar{x},r) \leq d \right\} \right) &= \mu \left( \left\{ \bar{x}: \tilde{\sigma}_p(k_n, x^n_1) \leq f_{p}(\bar{x},r),   f_{p}(\bar{x},r) \leq d \right\} \right) \nonumber\\
	&+ \mu \left( \left\{ \bar{x}: \tilde{\sigma}_p(k_n, x^n_1) > f_{p}(\bar{x},r),   f_{p}(\bar{x},r) \leq d \right\} \right), 
	\end{align}
	and  $\mu \left( \left\{ \bar{x}: \tilde{\sigma}_p(k_n, x^n_1) > f_{p}(\bar{x},r),   f_{p}(\bar{x},r) \leq d \right\} \right) \leq \mu \left( \left\{ \bar{x}: \tilde{\sigma}_p(k_n, x^n_1) > f_{p}(\bar{x},r) \right\} \right)$, where the almost sure convergence of $\tilde{\sigma}_p(k_n, X^n_1)$ to $f_{p}({\bf X},r)$ in (\ref{eq_proof_th_main_ams_7}) implies that 
	$$\lim_{n \longrightarrow \infty} \mu \left( \left\{ \bar{x}: \tilde{\sigma}_p(k_n, x^n_1) > f_{p}(\bar{x},r),   f_{p}(\bar{x},r) \leq d \right\} \right)=0$$ 
	obtaining the result in (\ref{eq_proof_th_main_ams_10}).  
	
	Consequently, we have from (\ref{eq_proof_th_main_ams_9}) that 
	\begin{align}\label{eq_proof_th_main_ams_11}
		&\lim_{n \longrightarrow \infty} \mu \left( \left\{ \bar{x} \in \mathbb{R}^{\mathbb{N}}: \tilde{\sigma}_p(k_n, x^n_1) \leq d,   f_{p}(\bar{x},r) \leq d \right\} \right)=\nonumber\\ 
		&\lim_{n \longrightarrow \infty} \mu \left( \left\{ \bar{x} \in \mathbb{R}^{\mathbb{N}}: f_{p}(\bar{x},r) \leq d \right\} \right).
	\end{align}
	For the second term in the RHS of (\ref{eq_proof_th_main_ams_8}), it is simple to verify that 
	$$\mu \left( \left\{ \bar{x} \in \mathbb{R}^{\mathbb{N}}: \tilde{\sigma}_p(k_n, x^n_1) \leq d,   f_{p}(\bar{x},r) > d \right\} \right) 
	\leq \mu \left( \left\{ \bar{x} \in \mathbb{R}^{\mathbb{N}}: \tilde{\sigma}_p(k_n, x^n_1) <  f_{p}(\bar{x},r)  \right\} \right),$$
	then the almost sure convergence in (\ref{eq_proof_th_main_ams_7}) implies that  
	$$\lim_{n \longrightarrow \infty}  \mu \left( \left\{ \bar{x} \in \mathbb{R}^{\mathbb{N}}: \tilde{\sigma}_p(k_n, x^n_1) \leq d,   f_{p}(\bar{x},r) > d \right\} \right) =0.$$  
	Putting this result in (\ref{eq_proof_th_main_ams_8}) and using (\ref{eq_proof_th_main_ams_11}), it follows that
	\begin{align}\label{eq_proof_th_main_ams_12}
		\lim_{n \longrightarrow \infty} \mu_n(A_d^{n,k_n}) =  \lim_{n \longrightarrow \infty} \mu \left( \left\{ \bar{x} \in \mathbb{R}^{\mathbb{N}}: f_{p}(\bar{x},r) \leq d \right\} \right), 
	\end{align}
	which concludes the argument. 
	
	Finally to obtain the specific statement presented in (\ref{eq_subsec_non_ergodic_case_3}), 
	we first note that $f_{p}(\bar{x},r) = f_{p,\mu_{\Psi(\bar{x}})}(r)$ for all $\bar{x}\in \mathbb{R}^{\mathbb{N}}$, where $(f_{p,\mu_\lambda}(r))_{r\in (0,1]}$ is the expression that has been fully characterized in Theorem \ref{th_main_ams&ergodic} for any $\mu_\lambda\in \tilde{\mathcal{P}}$.  In addition,  we can use Theorem \ref{th_main_ams&ergodic} i) stating that when $\mu_\lambda$ is $\ell_p$-compressible, meaning that $(x^p)_{x\in \mathbb{R}} \notin L_1({\mu_\lambda}_1)$, then $f_{p,{\mu_\lambda}}(r)=0$ for all $r\in (0,1]$. Therefore, all the stationary and ergodic components $\mu_{\Psi(\bar{x})}$ that are $\ell_p$-compressible satisfy that $f_{p}(\bar{x},r)=f_{p,\mu_{\Psi(\bar{x}})}(r) \leq d$ independent of the pair $(r,d)$, this observation explains the first term in the expression presented in (\ref{eq_subsec_non_ergodic_case_3}). 
\end{proof}

\section{Acknowledgment}
This material is based on work supported by grants of CONICYT-Chile, Fondecyt 1210315 and the Advanced Center for Electrical and Electronic Engineering, Basal Project FB0008.  I want to thank the two anonymous reviewers for providing valuable comments and suggestions that helped to improve the technical quality and organization of this paper. The author thanks Professor Martin Adams for providing valuable comments about the organization and presentation of this paper. The author thanks Sebastian Espinosa. Felipe Cordova and Mario Vicuna for helping with the figures and the simulations presented in this work. Finally, I thank Diane Greenstein for editing and proofreading all this material.

\appendix
\section{Proof of Lemma \ref{main_lemma} (and Corollary \ref{cor_phase_transition_lp_errors})}
\label{proof_main_lemma}
First,  some properties of $(f_{p,\mu}(r))_{r\in (0,1]}$ will be needed.
\begin{proposition}\label{pro_monotonic_properties_lp_approx_func}
It follows that:
\begin{itemize} 
	\item If $0<r_1<r_2\leq 1$, then  $f_{p,\mu}(r_2)\leq f_{p,\mu}(r_1)$.
	\item	
	If $0<r_1<r_2\leq 1$ and $f_{p,\mu}(r_2)= f_{p,\mu}(r_1)$, then 
	$f_{p,\mu}(r_2)= f_{p,\mu}(r_1)=0$.
\end{itemize}
\end{proposition}
The proof of this result derives directly from the definition of $\tilde{\sigma}_p(k_n, X^n)$ and some basic inequalities.
\footnote{This result is revisited and proved (including additional properties) 
in Lemma \ref{lemma_rate_distorsion_properties},  Section \ref{proof_th_main_ams&ergodic}.} 
From Proposition \ref{pro_monotonic_properties_lp_approx_func},  $f_{p,\mu}(\cdot)$ is strictly monotonic and injective in the domain $f_{p,\mu}^{-1}((0,1))$. Therefore,  $f_{p,\mu}^{-1}(d)$ is well defined for any $d\in \left\{f_{p,\mu}(r), r\in (0,1] \right\} \setminus \left\{ 0\right\}$. 

\begin{proof}
Let us first consider the case $d\in \left\{f_{p,\mu}(r), r\in (0,1] \right\} \setminus \left\{ 0\right\}$ assuming for a moment that this set is non-empty.  Then,  there exists $r_o\in (0,1)$ such that $d=f_{p,\mu}(r_o)$, where by the strict monotonicity of $f_{p,\mu}(\cdot)$, we have that 
$f_{p,\mu}(r_2)<d<f_{p,\mu}(r_1)$ for any $r_1<r_o<r_2\leq 1$.  On the other hand, using the 
convergence of the approximation error  to the function $f_{p,\mu}(r)$ in (\ref{eq_sec_strong_lp_charact_1}) and the definition of $\mathcal{A}^{n,k}_d$ in (\ref{eq_sec_pre_3}), it follows  that for any $r\in (0,1)$,  $k_n$ with $k_n/n \longrightarrow r$, and $\epsilon>0$
\begin{equation}\label{eq_main_lemma_1}
	\lim_{n \rightarrow \infty} \mu_n(\mathcal{A}^{n,k_n}_{f_{p,\mu}(r)+\epsilon})=1 \ \text{and}
\end{equation}
\begin{equation}\label{eq_main_lemma_2}
	\lim_{n \rightarrow \infty} \mu_n(\mathcal{A}^{n,k_n}_{f_{p,\mu}(r)-\epsilon})=0.
\end{equation}
Then assuming that $k_n/n \longrightarrow r_o$,  if $r>r_o$,  then $f_{p,\mu}(r) < d$, and from (\ref{eq_main_lemma_1}) we obtain that $\lim_{n \rightarrow \infty} \mu_n(\mathcal{A}^{n,k_n}_{d})=1$. On the other hand,   if $r<r_o$, then  $f_{p,\mu}(r) > d$, and  from (\ref{eq_main_lemma_2}) we obtain that $\lim_{n \rightarrow \infty} \mu_n(\mathcal{A}^{n,k_n}_{d})=0$.  This proves (\ref{eq_statement_main_lemma_1}).

\begin{remark}\label{rm_1_main_lemma} (for Corollary \ref{cor_phase_transition_lp_errors})
Adopting the definition in  (\ref{eq_sec_pre_8}) and setting $\epsilon>0$, it follows 
from (\ref{eq_main_lemma_1}) and (\ref{eq_main_lemma_2}) that for any arbitrary small $\delta>0$, $r_o-\delta \leq {r}_p(d, \epsilon, \mu)\leq  r_o+ \delta$, and, consequently, 
	${r}_p(d, \epsilon, \mu) = r_o= f_{p,\mu}^{-1}(d)$.
%
Furthermore, adopting the definition of $\tilde{\kappa}_p(d,\epsilon,\mu^n)$ in (\ref{eq_sec_pre_5}) with a fixed $\epsilon>0$ and its asymptotic limits (with $n$) in (\ref{eq_sec_pre_6a}) and (\ref{eq_sec_pre_6b}), it follows from (\ref{eq_main_lemma_1}) and (\ref{eq_main_lemma_2}) that 
for any arbitrary small $\delta>0$,  $r_o-\delta \leq \tilde{r}^-_p(d,\epsilon,\mu) \leq \tilde{r}^+_p(d,\epsilon,\mu)  \leq  r_o+ \delta$, and, 
consequently, $\tilde{r}^-_p(d,\epsilon,\mu) =\tilde{r}^+_p(d,\epsilon,\mu)  = r_o= f_{p,\mu}^{-1}(d)$.
\end{remark}

Concerning the second part of the result, let us assume $r_o\in (0,1)$ such that $f_{p,\mu}(r_o)=0$. 
From the convergence in (\ref{eq_sec_strong_lp_charact_1}) assumed in this result, we have that
if $(k_n)_{n\geq 1}$ is such that $k_n/n \longrightarrow r_o$,  then
\begin{equation}\label{eq_main_lemma_3}
	\lim_{n \rightarrow \infty} \tilde{\sigma}_p(k_n, X^n)=0, \mu-a.s.
\end{equation}
Then adopting $\mathcal{A}^{n,k}_d$ in (\ref{eq_sec_pre_3}), it follows 
from (\ref{eq_main_lemma_3}) that
for any $d>0$
\begin{equation}\label{eq_main_lemma_4}
	 \lim_{n \rightarrow \infty} \mu_n(\mathcal{A}^{n,k_n}_d)=1, 
\end{equation}
which proves (\ref{eq_statement_main_lemma_2}).
\begin{remark}\label{rm_2_main_lemma} (for Corollary \ref{cor_phase_transition_lp_errors})
	Using the definition of $\tilde{\kappa}_p(d,\epsilon,\mu^n)$ in (\ref{eq_sec_pre_5}), from (\ref{eq_main_lemma_4}) it is clear that for any $\epsilon>0$,  $\tilde{\kappa}_p(d,\epsilon,\mu^n) \leq k_n$ eventually (in $n$).  From (\ref{eq_sec_pre_6a}), this last inequality implies that $\tilde{r}^+_p(d,\epsilon,\mu) \leq r_o$.
\end{remark}
\end{proof}

\section{Proof of Lemma \ref{lemma_rate_distorsion_properties}}
\label{proof_lemma_rate_distorsion_properties}
	The following properties of the tail functions (that define $\mathcal{F}_{\bar{\mu}_1}$ in (\ref{eq_pr_main_ams&ergodic_7}))  will be used: 
	\begin{proposition}\label{join_dependency_of_tail_functions}
		\begin{itemize}
			\item $\mathcal{Y}_{v_p} = \mathcal{Y}_{\bar{\mu}_1}\setminus \left\{0 \right\}$, meaning that for all $\tau>0$, $(\phi_{\bar{\mu}_1}(\cdot))$ is continuous at $\tau$ if, and only if, $(\phi_{v_p}(\cdot))$ is continuous at $\tau$.
			\item $\forall \tau_1>\tau_2>0$, $\phi_{\bar{\mu}_1}(\tau_1)=\phi_{\bar{\mu}_1}(\tau_2)$ if, and only if, $\phi_{v_p}(\tau_1)=\phi_{v_p}(\tau_2)$.
		\end{itemize} 
	\end{proposition}
	The proof of this result is presented in  \ref{proof_join_dependency_of_tail_functions}.
	
\begin{proof}
	{\bf Proof of i)}: Let us first show that $({f}_{\bar{\mu}_1}(\cdot))$ is continuous in $(0,1]$.  It is 
	sufficient  to  prove continuity on the function $\tilde{f}_{\bar{\mu}_1}(r) \equiv 1- {f}_{\bar{\mu}_1}(r)^p$, which is induced by the following simpler relationship:\footnote{This from the continuity of the function $g(x)=\sqrt[p]{1-x}$ in $x\in [0,1]$.} 
	\begin{align} \label{eq_proof_lemma_rate_distorsion_properties_1a}
 		&\mathcal{\tilde{F}}_{\bar{\mu}_1} \equiv \left\{ (\phi_{\bar{\mu}_1}(\tau), {\phi_{v_p}(\tau)}), \tau\in [0,\infty) \right\}  \\
		\label{eq_proof_lemma_rate_distorsion_properties_1b}
		&\bigcup_{\tau_n\in \mathcal{Y}_{\bar{\mu}_1}} \left\{ ( \bar{\mu}_1(C_{\tau_n}) + \alpha \bar{\mu}_1(\left\{ -\tau_n,\tau_n \right\}), {v_p(C_{\tau_n})  + \alpha v_p(\left\{ -\tau_n,\tau_n \right\}) }), \alpha \in [0,1) \right\}.
	\end{align}
	 There are three distinct scenarios to consider:  
	 \begin{itemize}
	 \item Let us first focus on the case when $r\in \mathcal{R}^*_{\bar{\mu}_1} \setminus \left\{ \bar{\mu}_1(B_{\tau_n}), \tau_n\in \mathcal{Y}_{\bar{\mu}_1} \right\}$ (see, Eq.(\ref{eq_pr_main_ams&ergodic_5a})). Under this assumption there exists $\tau_o\in [0,\infty)\setminus \mathcal{Y}_{\bar{\mu}_1}$ (in the domain where $\phi_{\bar{\mu}_1}(\cdot)$ is continuous) where $r=\phi_{\bar{\mu}_1}(\tau_o)$. From Proposition \ref{join_dependency_of_tail_functions},  $\phi_{v_p}(\cdot)$ is also continuous at $\tau_o$ where by construction in (\ref{eq_proof_lemma_rate_distorsion_properties_1a}) $\tilde{f}_{\bar{\mu}_1}(r)=\phi_{v_p}(\tau_o)$. Let us consider an 
	arbitrary $\epsilon>0$. From the continuity of $\phi_{v_p}(\cdot)$ at $\tau_o$,  there exists $\delta>0$ such that 
	$\left\{\phi_{v_p}(\tau), \tau\in B_{\delta}(\tau_o) \right\} \subset B_{\epsilon}(\tilde{f}_{\bar{\mu}_1}(r))$.\footnote{$B_{\epsilon}(x)\equiv (x-\epsilon, x+\epsilon)\subset \mathbb{R}$ denotes the open ball of radius $\epsilon>0$ centered at $x\in \mathbb{R}$.} Without loss of generality,  we can assume that $\phi_{v_p}(\tau_o-\delta) > \tilde{f}_{\bar{\mu}_1}(r)= \phi_{v_p}(\tau_o) > \phi_{v_p}(\tau_o +\delta)$. Then from Proposition \ref{join_dependency_of_tail_functions},  it follows that $\phi_{\bar{\mu}_1}(\tau_o-\delta) > r=\phi_{\bar{\mu}_1}(\tau_o) > \phi_{\bar{\mu}_1}(\tau_o +\delta)$. Then,  there exists $\bar{\delta}>0$ such that $B_{\bar{\delta}}(r) \subset \left\{\phi_{\bar{\mu}_1}(\tau), \tau\in B_{\delta}(\tau_o)\right\}$.  Therefore from (\ref{eq_proof_lemma_rate_distorsion_properties_1a}), we have that for any $\bar{r}\in B_{\bar{\delta}}(r)$, there exists $\tau_{\bar{r}} \in B_{\delta}(\tau_o)$ where $\bar{r}=\phi_{\bar{\mu}_1}(\tau_{\bar{r}})$ and, consequently, $\tilde{f}_{\bar{\mu}_1}(\bar{r})=\phi_{v_p}(\tau_{\bar{r}})\in B_{\epsilon}(\tilde{f}_{\mu_1}(r))$, which concludes the argument in this case.
	 \item Let us assume that $r\in \bigcup_{\tau_n\in \mathcal{Y}_{\bar{\mu}_1}} (\bar{\mu}_1(C_{\tau_n}), \bar{\mu}_1(B_{\tau_n}))$ (see, Eq.(\ref{eq_pr_main_ams&ergodic_5a})).  Then there is $\tau_n \in \mathcal{Y}_{\bar{\mu}_1}$ and a unique $\alpha_o\in (0,1)$ such that 
	 $r=\bar{\mu}_1(C_{\tau_n}) + \alpha_o \cdot \bar{\mu}_1(B_{\tau_n}\setminus C_{\tau_n})$ and, consequently, 
	 $\tilde{f}_{\bar{\mu}_1}(r)=v_p(C_{\tau_n}) + \alpha_o \cdot v_p(B_{\tau_n}\setminus C_{\tau_n})$ from (\ref{eq_proof_lemma_rate_distorsion_properties_1b}). Without loss of generality, let us consider $\epsilon>0$ small  enough 
	 such that $B_{\epsilon}(\tilde{f}_{\bar{\mu}_1}(r)) \subset (v_p(C_{\tau_n}),v_p(B_{\tau_n}))$. Then from the continuity 
	 of the affine function $g(\alpha) \equiv v_p(C_{\tau_n}) + \alpha \cdot v_p(B_{\tau_n}\setminus C_{\tau_n})$ in $(0,1)$, 
	 there exists  $\delta>0$ (function of $\epsilon$) such that  $\left\{ g(\alpha), \alpha \in B_{\delta}(\alpha_o) \right\}  \subset B_{\epsilon}(\tilde{f}_{\bar{\mu}_1}(r))$. Therefore for any $\bar{r} \in \left\{ \bar{\mu}_1(C_{\tau_n}) + \alpha \cdot \bar{\mu}_1(B_{\tau_n}), \alpha\in (\alpha_o-\delta,\alpha_o + \delta) \right\}$,  $\tilde{f}_{\bar{\mu}_1}(\bar{r}) \in B_{\epsilon}(\tilde{f}_{\bar{\mu}_1}(r))$ from the construction in (\ref{eq_proof_lemma_rate_distorsion_properties_1b}).  Finally fixing $\bar{\delta}=\delta \cdot \bar{\mu}_1(B_{\tau_n}\setminus C_{\tau_n})$,  we have that $\left\{ \tilde{f}_{\bar{\mu}_1}(\bar{r}), \bar{r} \in B_{\bar{\delta}}(r) \right\} \subset B_{\epsilon}(\tilde{f}_{\bar{\mu}_1}({r}))$, which concludes the argument in this case. 
	 \item Finally,  we need to consider the case where $r\in \left\{ \bar{\mu}_1(C_{\tau_n}), \tau_n\in \mathcal{Y}_{\bar{\mu}_1} \right\}\cup \left\{ \bar{\mu}_1(B_{\tau_n}), \tau_n\in \mathcal{Y}_{\bar{\mu}_1} \right\}$. The argument mixed the steps already presented in the two previous scenarios,  and for the sake of space it is omitted here as no new technical elements are needed.   
	\end{itemize}
	
	{\bf Proof of iv)}: Using the fact that $\lim_{\tau \rightarrow \infty} \phi_{\bar{\mu}_1}(\tau)=0$  and $\lim_{\tau \rightarrow \infty} \phi_{v_p}(\tau)=0$, from the construction of ${f}_{\bar{\mu}_1}(r)$ in (\ref{eq_pr_main_ams&ergodic_7}) we have that $\lim_{r \rightarrow 0} {f}_{\bar{\mu}_1}(r)=1$. For the other condition,  let us consider $r=1-\bar{\mu}_1(\left\{ 0 \right\})$. There are two cases. The simplest case to analyze is when $0 \notin \mathcal{Y}_{\bar{\mu}_1}$. In this case, we are only looking at the point $r=1$. This point is achieved at $\tau=0$, i.e., $r=\phi_{\bar{\mu}_1}(\tau=0)=1$, which is mapped to ${f}_{\bar{\mu}_1}(1)=\sqrt[p]{1-\phi_{v_p}(\tau=0)}=0$. When $0 \in \mathcal{Y}_{\bar{\mu}_1}$, 
	then from (\ref{eq_pr_main_ams&ergodic_7}) we can focus on the  following range of pairs determining ${f}_{\bar{\mu}_1}(r)$:
	$$\left\{ ( \bar{\mu}_1(C_{0}) + \alpha \bar{\mu}_1(\left\{ 0 \right\}), \sqrt[p]{1- v_p(C_{0})  - \alpha \cdot  v_p(\left\{ 0  \right\}) }), \alpha \in [0,1) \right\}$$
	 looking at $\tau=0\in \mathcal{Y}_{\bar{\mu}_1}$.  We know that $0 \notin \mathcal{Y}_{v_p}$ then $v_p(\left\{ 0  \right\})=0$ and 
	 $v_p(C_{0})=v_p(B_{0})=1$. On the other hand,  $ \bar{\mu}_1(C_{0}) =  \bar{\mu}_1(B_{0}) - \bar{\mu}_1(\left\{ 0 \right\})= 1- \bar{\mu}_1(\left\{ 0 \right\})$. 
	 Therefore, by exploring all the values of $\alpha\in [0,1)$, we have that ${f}_{\bar{\mu}_1}(r)=0$ for all $r\in [1- \bar{\mu}_1(\left\{ 0 \right\}),1)$.
	
	{\bf Proof of ii)}: Let us consider $r_2>r_1$ and assume that both belong to $\mathcal{R}^*_{\bar{\mu}_1}$ in (\ref{eq_pr_main_ams&ergodic_5a}). This means 
	that there exist $ \tau_1 > \tau_2\geq 0$ such that $r_1=\phi_{\bar{\mu}_1}(\tau_1)$ and $r_2=\phi_{\bar{\mu}_1}(\tau_1)$. Then 
	$\phi_{v_p}(\tau_1) < \phi_{v_p}(\tau_2)$ from Proposition \ref{join_dependency_of_tail_functions}, which implies the result 
	by the construction of $f_{\bar{\mu}_1}(\cdot)$ in (\ref{eq_pr_main_ams&ergodic_7}).  Another important scenario to cover 
	is the case when $r_1=\bar{\mu}_1(C_{\tau_n}) + \alpha_1 \bar{\mu}_1(B_{\tau_n} \setminus C_{\tau_n})$ and  $r_2=\bar{\mu}_1(C_{\tau_n}) + \alpha_2 \bar{\mu}_1(B_{\tau_n} \setminus C_{\tau_n})$
	with $\alpha_2>\alpha_1$,  $\tau_n \in \mathcal{Y}_{\bar{\mu}_1}$ and $\tau_n >0$.  Then in this case $v_p(C_{\tau_n}) + \alpha_2 v_p(B_{\tau_n} \setminus C_{\tau_n}) > v_p(C_{\tau_n}) + \alpha_1 v_p(B_{\tau_n} \setminus C_{\tau_n})$ because from Proposition \ref{join_dependency_of_tail_functions} it follows that $v_p(B_{\tau_n} \setminus C_{\tau_n})>0$ if $\tau_n \in \mathcal{Y}_{\bar{\mu}_1} \setminus \left\{0 \right\}$.   Again the result in this case follows from (\ref{eq_pr_main_ams&ergodic_7}). Mixing these two scenarios and using the monotonic property of the tail functions $(\phi_{\bar{\mu}_1}(\cdot),  \phi_{v_p}(\cdot) )$, we can prove the strict monotonic property of $(f_{\bar{\mu}_1}(\cdot))$ in $(0,1]$ if $0\notin \mathcal{Y}_{\bar{\mu}_1}$   and, the strict monotonic property of $(f_{\bar{\mu}_1}(\cdot))$  in  $(0,1] \setminus [1- \bar{\mu}_1(\left\{0 \right\}), 1]$ if  $0\in \mathcal{Y}_{\bar{\mu}_1}$.  
	
	{\bf Proof of iii)}: From the fact that $(f_{\bar{\mu}_1}(\cdot))$ is strictly monotonic in $r\in (0,1-\bar{\mu}_1(\left\{ 0 \right\}))$ (proof of ii)) and 
	the conditions on iv) (proved above), we have that  $f_{\bar{\mu}_1}(r) \in (0,1)$ if, and only if, $r\in (0,1-\bar{\mu}_1(\left\{ 0 \right\}))$. This suffices 
	to show that ${f}_{\bar{\mu}_1}^{-1}((0,1))= (0,1-\bar{\mu}(\left\{0 \right\}))$.
 	
	{\bf Proof of v)}:  This part comes directly from the continuity of $f_{\bar{\mu}_1}(\cdot)$ in $(0,1)$ and the limiting values 
	of $f_{\bar{\mu}_1}(\cdot)$ (i.e., $f_{\bar{\mu}_1}(1)=0$ and $\lim_{r \longrightarrow 0} f_{\bar{\mu}_1}(r)=1$). 
\end{proof}

\section{Proof of Proposition \ref{pro_tail_function_properties}}
\label{proof_pro_tail_function_properties}
\begin{proof}
The statement in i) follows  from the definition of $\phi(\cdot)$ and the statement ii) comes from the continuity of a measure under a monotone sequence of events converging to a limit \cite{varadhan_2001}. The left continuous property of $\phi(\cdot)$ and the fact that $\phi^+_m(\tau)= \phi_m(\tau) - m(\left\{ \tau \right\} \cup \left\{ -\tau \right\} )$ (stated in iii)) follow mainly from the continuity of a measure  \cite{varadhan_2001}.
\end{proof}

\section{Proof Proposition \ref{join_dependency_of_tail_functions}}
\label{proof_join_dependency_of_tail_functions}
\begin{proof}
	The proofs of these two points derive directly from the definition of the tail function and the construction of $v_p$ from $\bar{\mu}_1$.  More precisely, both results derive from the observation that these two measures are almost mutually absolutely continuous in the sense that for all $B\in \mathcal{B}(\mathbb{R})$ such that $0\notin B$, $\bar{\mu_1}(B)=0$ if, and only if, $v_p(B)=0$. In fact,  for all $B\in \mathcal{B}(\mathbb{R})$ such that $0\notin B$, $v_p(B)=\int_{B} \frac{|x|^p}{\left| \left| (x^p)  \right|\right|_{L_1(\bar{\mu}_1)}} d\bar{\mu}_1(x)$  and, conversely,  $\bar{\mu}_1(B)=\int_{B} \frac{\left| \left| (x^p)  \right|\right|_{L_1(\bar{\mu}_1)}}{|x|^p} d v_p(x)$.
\end{proof}

\bibliography{main_jorge_silva}				

\end{document}